\documentclass[letter]{aa}

\usepackage[sectionbib]{bibunits}
\usepackage{graphicx} 
\usepackage[export]{adjustbox}
\usepackage{array}
\usepackage{multirow}%
\usepackage{amsmath,amssymb,amsfonts}%

\usepackage{amsthm}%
\usepackage{mathrsfs}
\usepackage{xcolor}%
\usepackage{textcomp}
\usepackage{manyfoot}%
\usepackage{booktabs}%
\usepackage{hyperref}
\usepackage{txfonts}

\usepackage{natbib}
\bibpunct{(}{)}{;}{a}{}{,}
\begin{document}

\title{A patchy CO$_2$ exosphere on Ganymede revealed by the James Webb Space Telescope}

\author{Dominique Bockel\'ee-Morvan \inst{1}
\and Olivier Poch \inst{2}
\and Fran\c{c}ois Leblanc \inst{3}
\and Vladimir Zakharov \inst{1}
\and Emmanuel Lellouch \inst{1}
\and Eric Quirico \inst{2}
\and Imke de Pater \inst{4,5}
\and Thierry Fouchet \inst{1}
\and Pablo Rodriguez-Ovalle \inst{1}
\and Lorenz Roth \inst{6}
\and Fr\'ed\'eric Merlin \inst{1}
\and Stefan Duling \inst{7}
\and Joachim Saur \inst{7}
\and Adrien Masson \inst{1}
\and Patrick Fry \inst{8}
\and Samantha Trumbo \inst{9}
\and Michael Brown \inst{10}
\and Richard Cartwright \inst{11}
\and St\'ephanie Cazaux \inst{12}
\and Katherine de Kleer \inst{10}
\and Leigh N. Fletcher \inst{13}
\and Zachariah Milby \inst{10}
\and Audrey Moingeon \inst{2}
\and Alessandro Mura \inst{14}
\and Glenn S. Orton \inst{15}
\and Bernard Schmitt\inst{2}
\and Federico Tosi \inst{14}
\and Michael H. Wong \inst{4} }

\institute{
LESIA, Observatoire de Paris, Université PSL, Sorbonne Universit\'e, Universit\'e Paris Cité, CNRS, 92195, Meudon, France, 
\email{Dominique.Bockelee@obspm.fr}
\and Univ. Grenoble Alpes, CNRS, IPAG, 38000 Grenoble, France
\and LATMOS/CNRS, Sorbonne Universit\'e, UVSQ, Paris, France
\and Department of Astronomy, University of California, 22 Berkeley, CA 94720, USA
\and Department of Earth and Planetary Science, University of California, 22 Berkeley, CA 94720, USA
\and Space and Plasma Physics, KTH Royal Institute of Technology, Stockholm, Sweden
\and Institute of Geophysics and Meteorology, University of Cologne, Albertus Magnus Platz, 50923 Cologne, Germany
\and University of Wisconsin, Madison, WI, 53706, USA
\and Department of Astronomy \& Astrophysics, University of California, San Diego, La Jolla, CA 92093, USA
\and Division of Geological and Planetary Sciences, Caltech, Pasadena, CA 91125, USA
\and Johns Hopkins University Applied Physics Laboratory, 11001 Johns Hopkins Rd. Laurel, MD 20723, USA
\and Faculty of Aerospace Engineering, Delft University of Technology, Delft, The Netherlands
\and School of Physics and Astronomy, University of Leicester, University Road, Leicester, LE1 7RH, UK
\and Istituto Nazionale di AstroFisica – Istituto di Astrofisica e Planetologia Spaziali (INAF-IAPS), 00133 Rome, Italy
\and Jet Propulsion Laboratory, California Institute of Technology, Pasadena, California 91109, USA}

\date{Received  / Accepted }

\abstract{Jupiter's icy moon Ganymede has a tenuous exosphere produced by sputtering and possibly sublimation of water ice. To date, only atomic hydrogen and oxygen have been directly detected in this exosphere.
Here, we present observations of Ganymede's CO$_2$ exosphere obtained with the James Webb Space Telescope. CO$_2$ gas is observed over different terrain types, mainly over those exposed to intense Jovian plasma irradiation, as well as over some bright or dark terrains. Despite warm surface temperatures, the CO$_2$ abundance over equatorial subsolar regions is low. CO$_2$ vapor has the highest abundance over the north polar cap of the leading hemisphere, reaching a surface pressure of 1 pbar.  From modeling we show that the local enhancement observed near 12 h local time in this region can be explained by the presence of cold traps enabling CO$_2$ adsorption. However, whether the release mechanism in this high-latitude region is sputtering or sublimation remains unclear. 
The north polar cap of the leading hemisphere also has unique surface-ice properties, probably linked to the presence of the large atmospheric CO$_2$ excess over this region. These CO$_2$ molecules might have been initially released in the atmosphere after the radiolysis of CO$_2$ precursors, or from the sputtering of CO$_2$ embedded in the H$_2$O ice bedrock. Dark terrains (regiones), more widespread on the north versus south polar regions, possibly harbor CO$_2$ precursors. CO$_2$ molecules would then be redistributed via cold trapping on ice-rich terrains of the polar cap and be diurnally released and redeposited on these terrains. Ganymede's CO$_2$ exosphere highlights the complexity of surface-atmosphere interactions on Jupiter's icy Galilean moons.}


\titlerunning{Ganymede's CO$_2$ exosphere}
\authorrunning{Bockel\'ee-Morvan et al.}
\maketitle

\section{Introduction}
Jupiter's icy satellites, Europa, Ganymede and Callisto, are known to have rarefied atmospheres. The surface composition of these moons is dominated by H$_2$O ice and non-ice components, possibly salts, hydrated minerals and organics, hosting volatiles such as CO$_2$ \citep{1996Sci...274..385C,mccord1998,Tosi2024}. Sublimation and weathering processes, such as sputtering by charged particles from Jupiter's magnetosphere and micro-meteoroids bombardment, lead to the formation of weakly bound atmospheres composed primarily of H$_2$O, O$_2$, OH, H, O, and CO$_2$ species. Because of strong telluric absorption by Earth's atmosphere, detection of atomic and molecular emissions from icy moon exospheres is difficult from ground-based facilities. Most of our knowledge comes from the detection of auroral O and H emission lines in the atmospheres of the three icy moons \citep{Hall98,Cunningham2015,Barth1997,Roth2017a,Roth2017b,deKleer2023}, with some constraints obtained on H$_2$O vapor content for Ganymede \citep{Roth2021}. So far, atmospheric CO$_2$ was only detected in the atmosphere of Callisto \citep{Carlson1999,Cartwright2024}. The maximum CO$_2$ column densities do not coincide with the subsolar region, nor the regions with the greatest solid-state CO$_2$ abundance on Callisto’s surface, suggesting that CO$_2$ gas may be partly sourced by outgassing from its crust \citep{Cartwright2024}.
Characterizing how icy moon exospheres are formed and sustained is pivotal for understanding surface-atmosphere interactions, geomorphological and chemical changes driven by erosion. 

Ganymede is the only known moon with an intrinsic magnetic field, resulting in a complex space plasma environment which has been explored by the in-situ flybys of  Galileo and Juno spacecraft \citep[e.g.,][]{Kivelson1996, Allegrini2022, Ebert2022, Clark2022}. The intrinsic magnetic field directs most of the external Jovian magnetospheric plasma in a way that it primarily interacts with the moon's surface where Ganymede's mini-magnetosphere has open field lines, i.e., around the polar regions \citep{poppe2018,liuzzo2020,Greathouse2022}. This results in specific surface properties with respect to shielded equatorial latitudes, such as the formation of H$_2$O ice-rich patches at the polar caps \citep{khurana2007, ligier2019, stephan2020, king2022} with higher amounts of amorphous H$_2$O ice \citep{ligier2019,DBM2024}, radiolytically produced H$_2$O$_2$ \citep{Trumbo2023}, and CO$_2$ possibly trapped in amorphous H$_2$O ice \citep{DBM2024}. In addition, asymmetries between the north/south polar caps, and leading/trailing hemispheres are observed \citep{ligier2019,deKleer2021,Trumbo2023,DBM2024}. 
 Here we present the first detection of CO$_2$ in the exosphere of Ganymede, achieved using the James Webb Space Telescope (JWST), and we link the observed highly hetereogeneous CO$_2$ exosphere to surface properties and processes. This paper follows the investigation of Ganymede's surface properties from the same JWST data set \citep[][ hereafter Paper I]{DBM2024}.

\section{JWST observations of Ganymede's exosphere}
\begin{figure*}[ht]
\centering
\begin{minipage}{14cm}
\includegraphics[width=4cm]{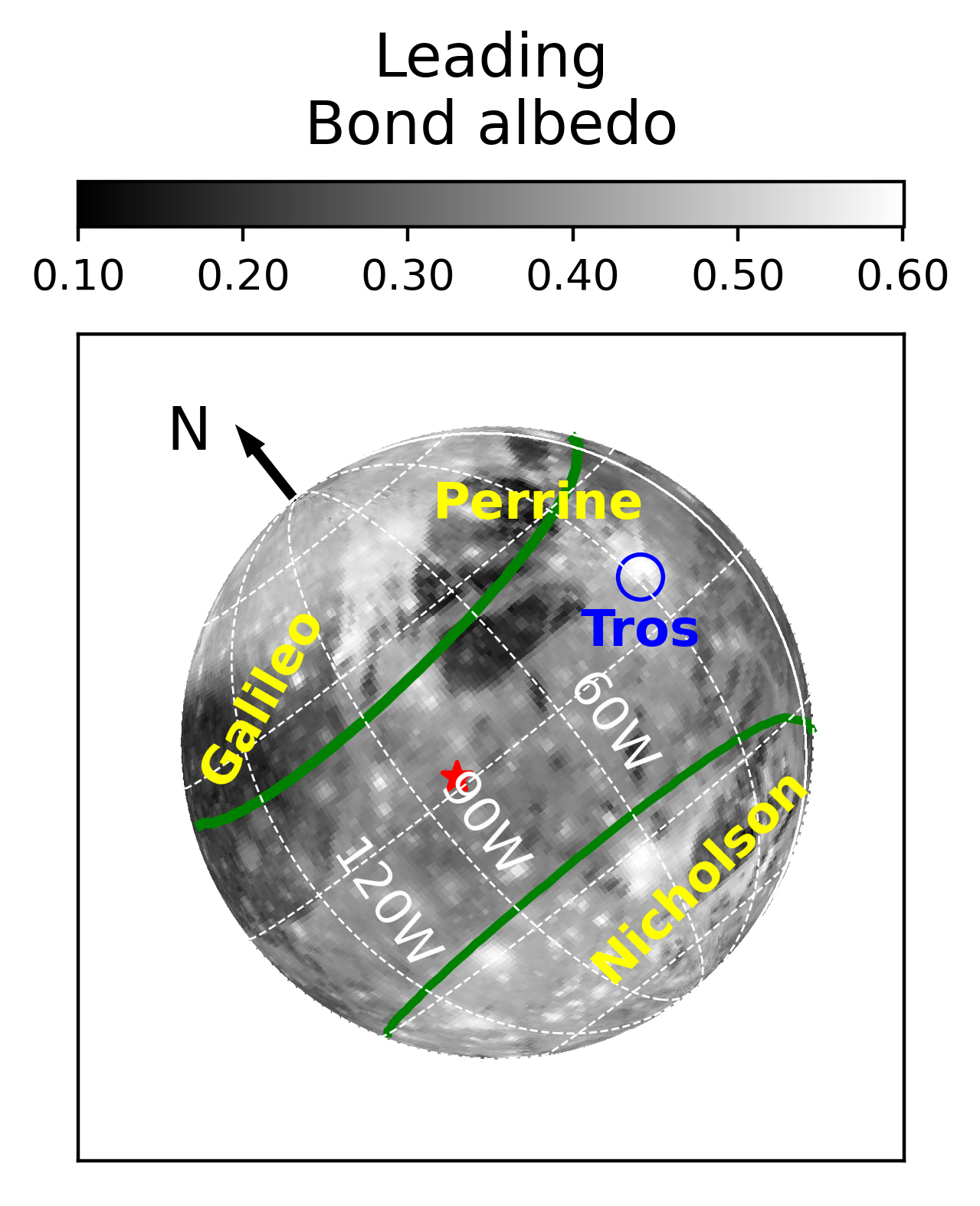}
\includegraphics[width=4.13cm]{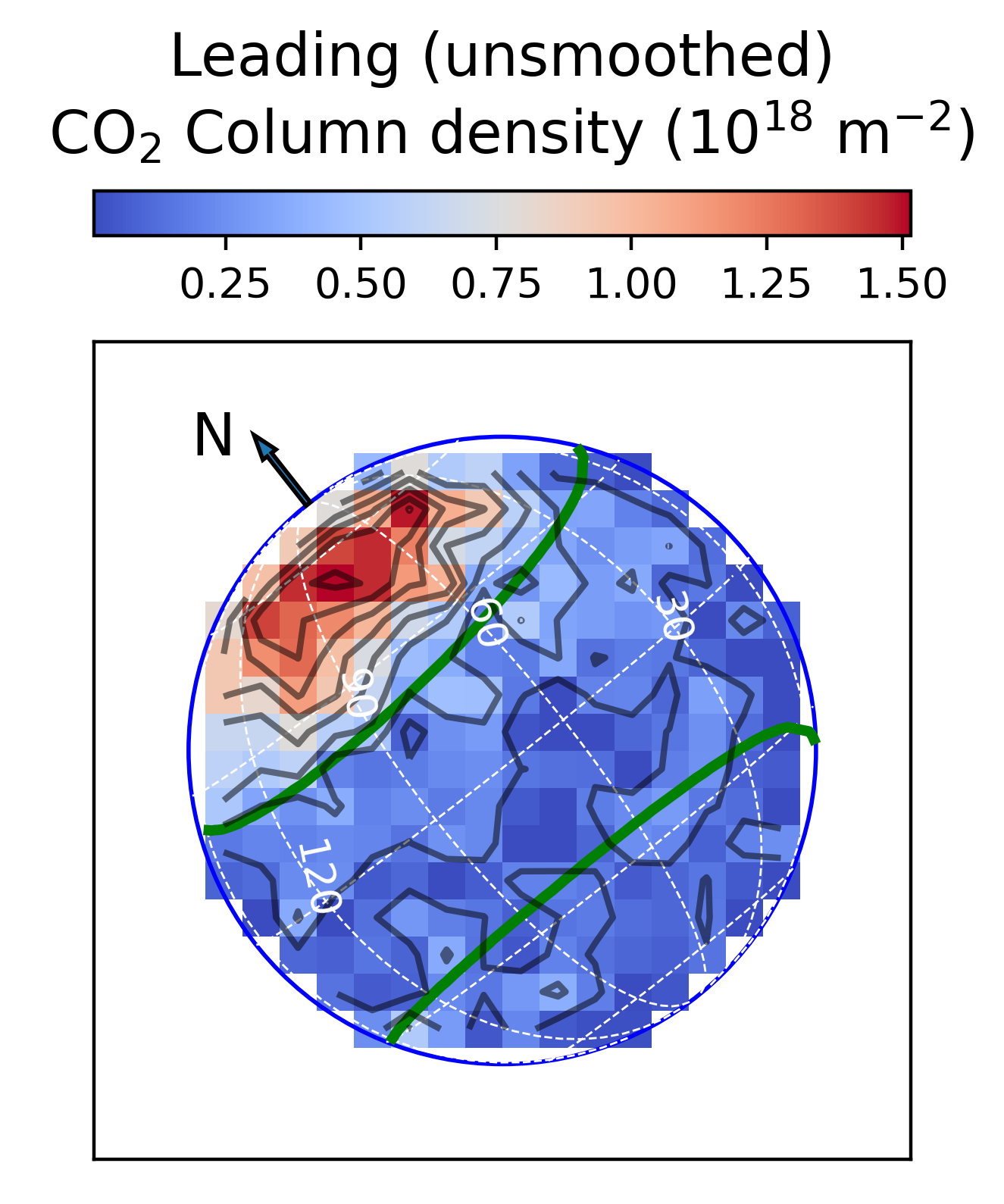}\hfill 
\includegraphics[width=6cm]{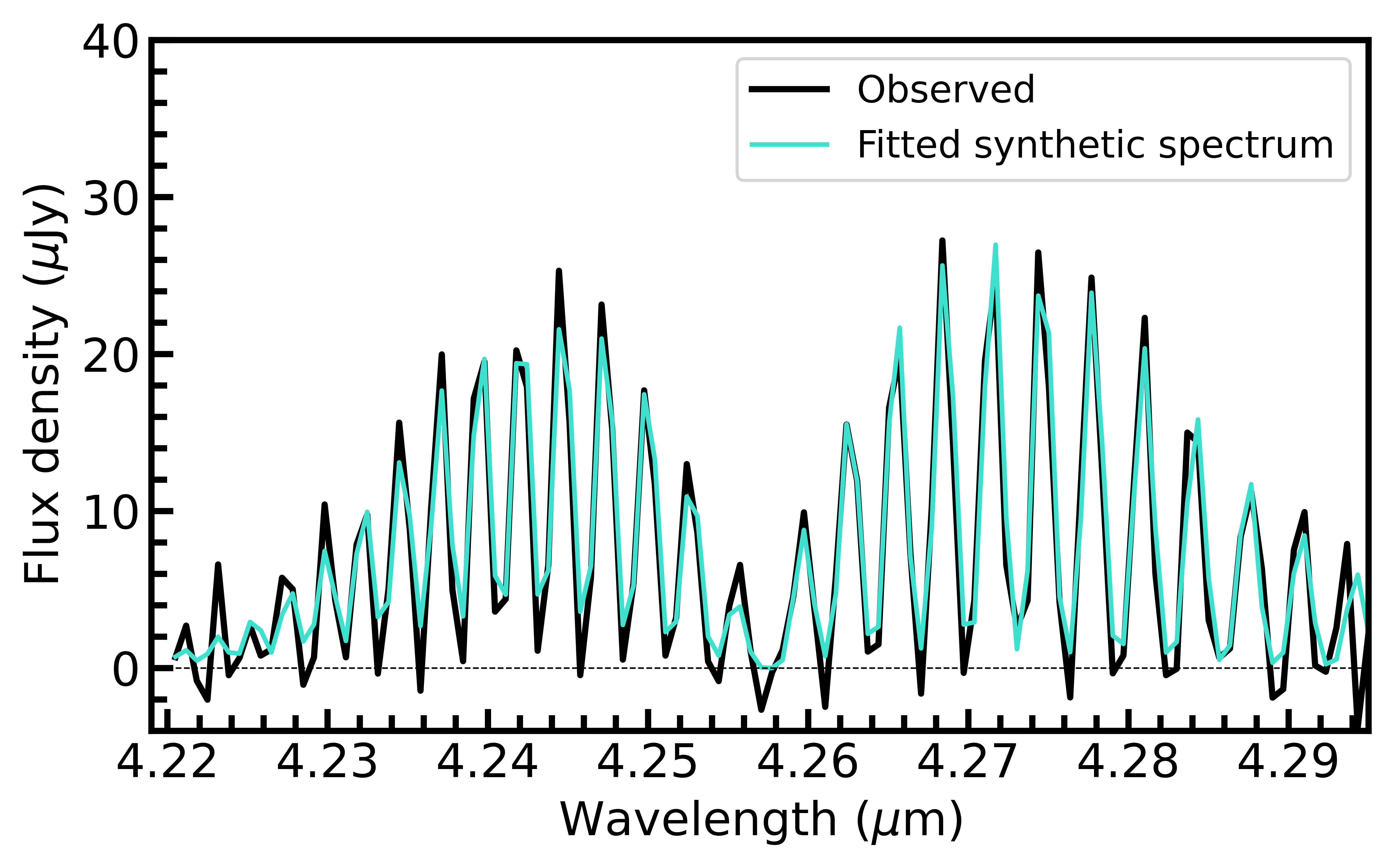}\hfill
\end{minipage}
\begin{minipage}{14cm}
\includegraphics[width=4cm]{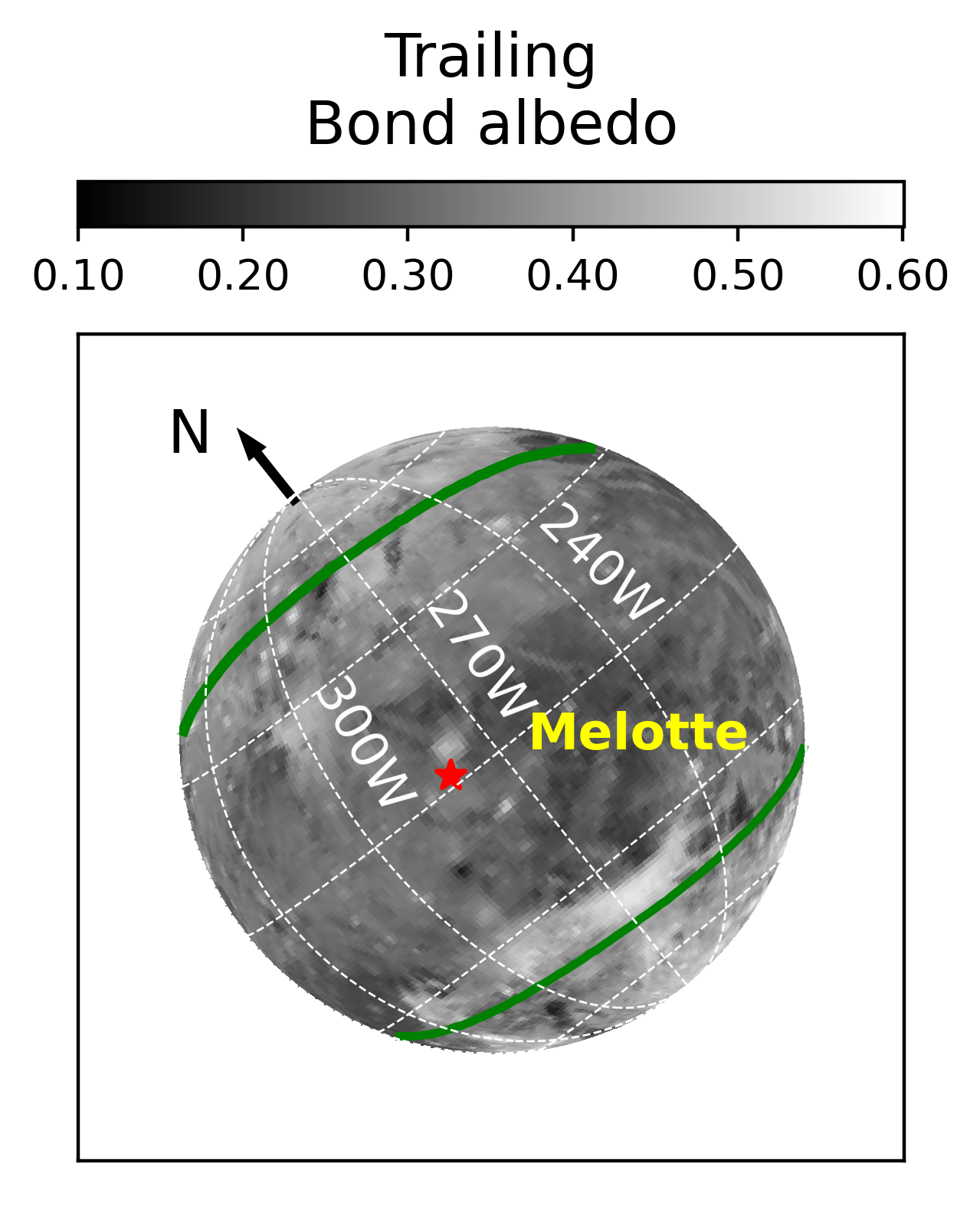}
\includegraphics[width=4.13cm]{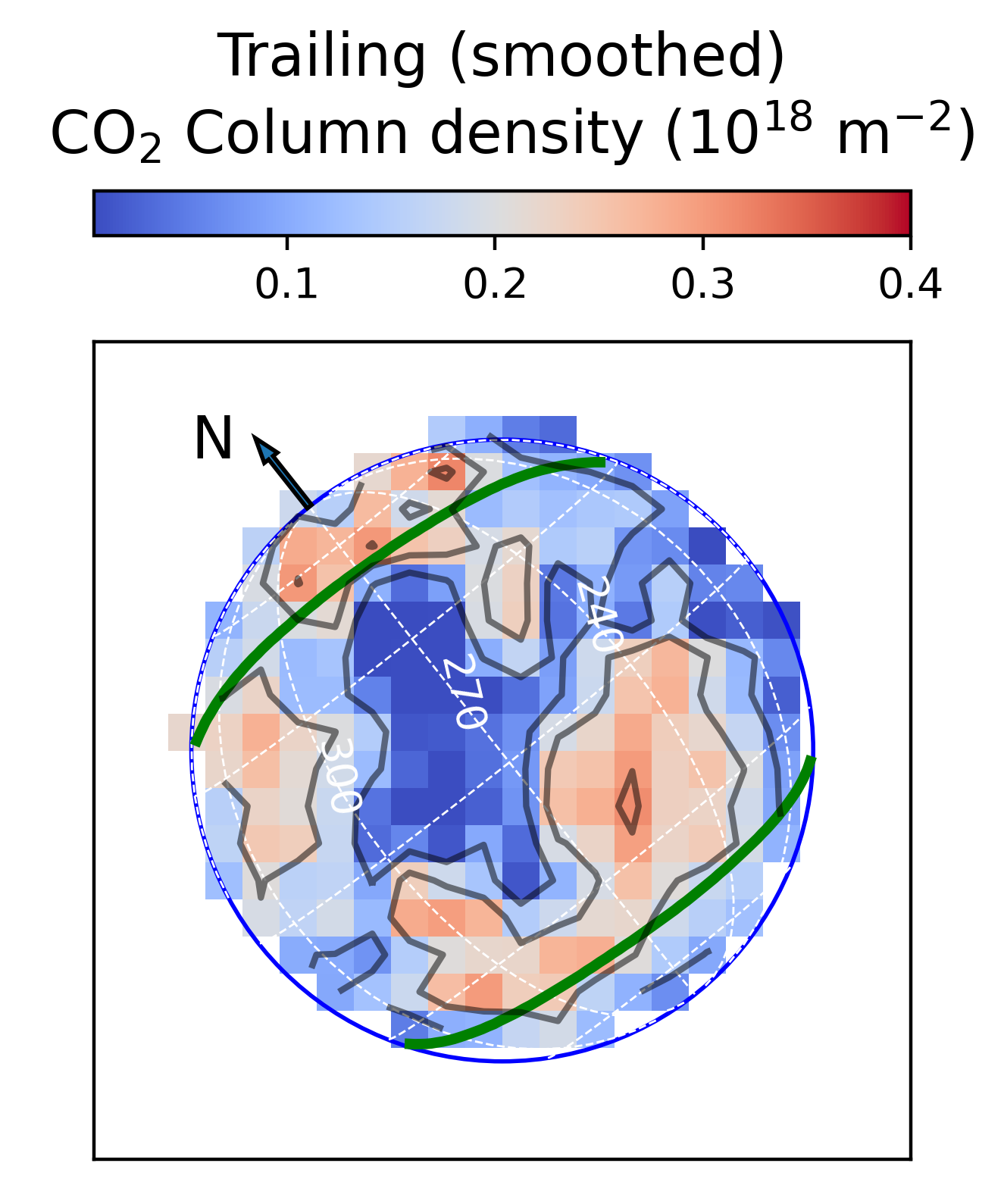}\hfill 
\includegraphics[width=6cm]{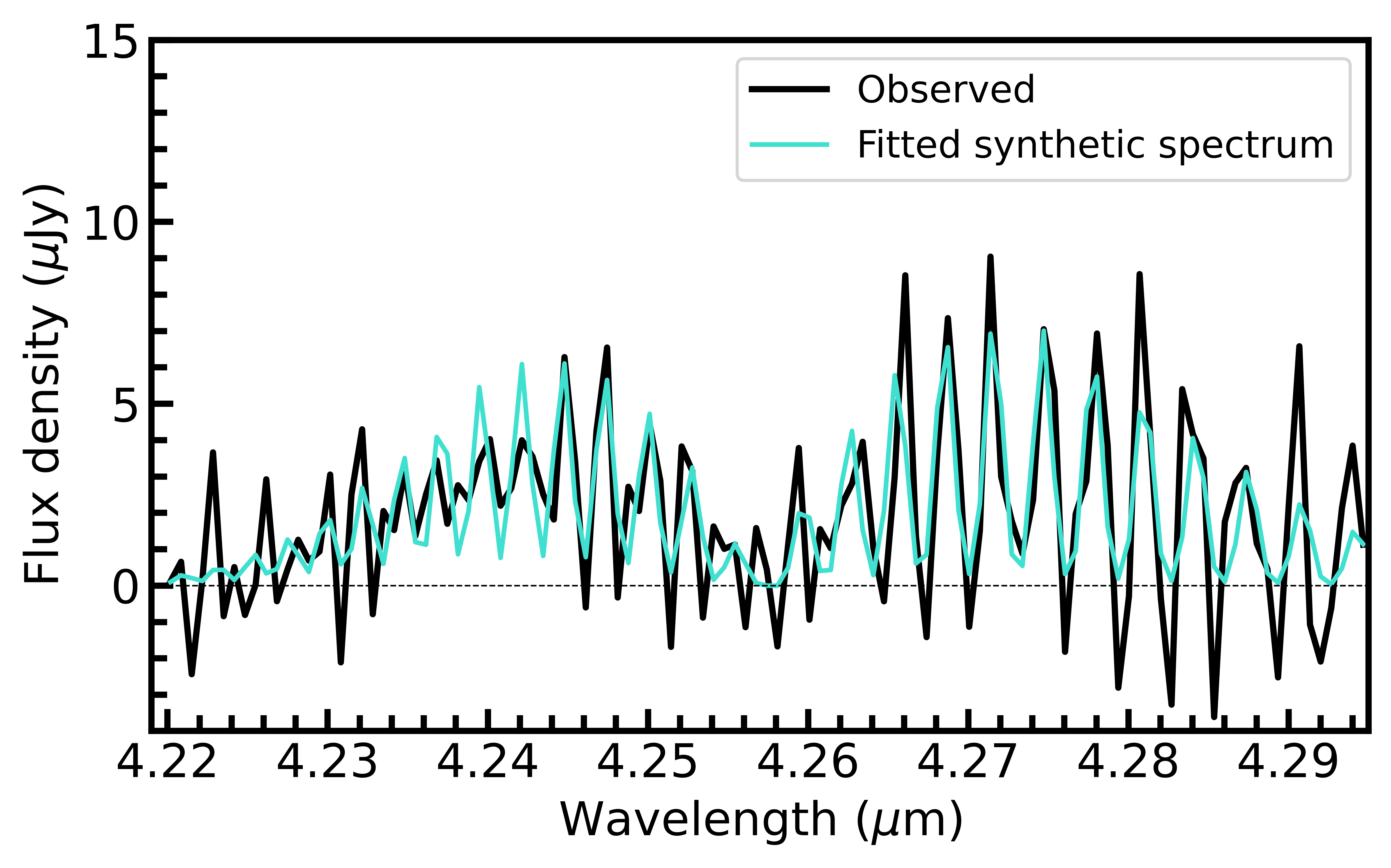}\hfill
\end{minipage}
\caption{CO$_2$ in Ganymede's exosphere. Top and bottom rows are for the leading and trailing sides, respectively. The first, second and third columns show: 1) Bond albedo maps derived by \citet{deKleer2021} from Voyager-Galileo mosaic; 2) Line-of-sight CO$_2$ column density maps inferred from spectral modeling (Appendices~\ref{app:B}--\ref{app:D}); trailing data were smoothed using a 3$\times$3 boxcar filter; color scales for the leading and trailing sides differ, and indicated above the plots; pixel sizes are 0.1$\times$0.1'' and the PSF is $\sim$ 0.19'' (FWHM); CO$_2$ maximal emission in the leading hemisphere (based on central contour) is at 81$^\circ$W, 51$^\circ$N ($\sim$12h local time); correcting for the line of sight, the maximum vertical column density is at 72$^\circ$W, 45$^\circ$N (12.6 h local time), Figs. ~\ref{leading-surface}A,~\ref{fig:vertical-colum-CO2}; 3) CO$_2$ gaseous emission spectra obtained after removing the continuum emission from Ganymede's surface, averaged over latitudes 45--90$^{\circ}$N for leading (top), and 30--60$^{\circ}$S for trailing (bottom); best fit synthetic spectra are shown in cyan, with a fitted rotational temperature of 108$\pm$8 K for the leading side, and a fixed rotational temperature of 105 K for the trailing side; the $y$ scale is $\mu$Jy per pixel. The green lines in the maps show the open-closed-field line boundary at the time of the JWST observations \citep[Appendix~\ref{app:J},][]{Duling2022}. Ganymede was north of and within the plasma sheet at the time of the leading and trailing sides observations, respectively. The subsolar point at the time of the JWST observations is shown by a red star in the Bond albedo maps. \label{fig:exosphere-combined}}
\end{figure*}

\begin{figure}[h]
\centering
\includegraphics[width=8cm]{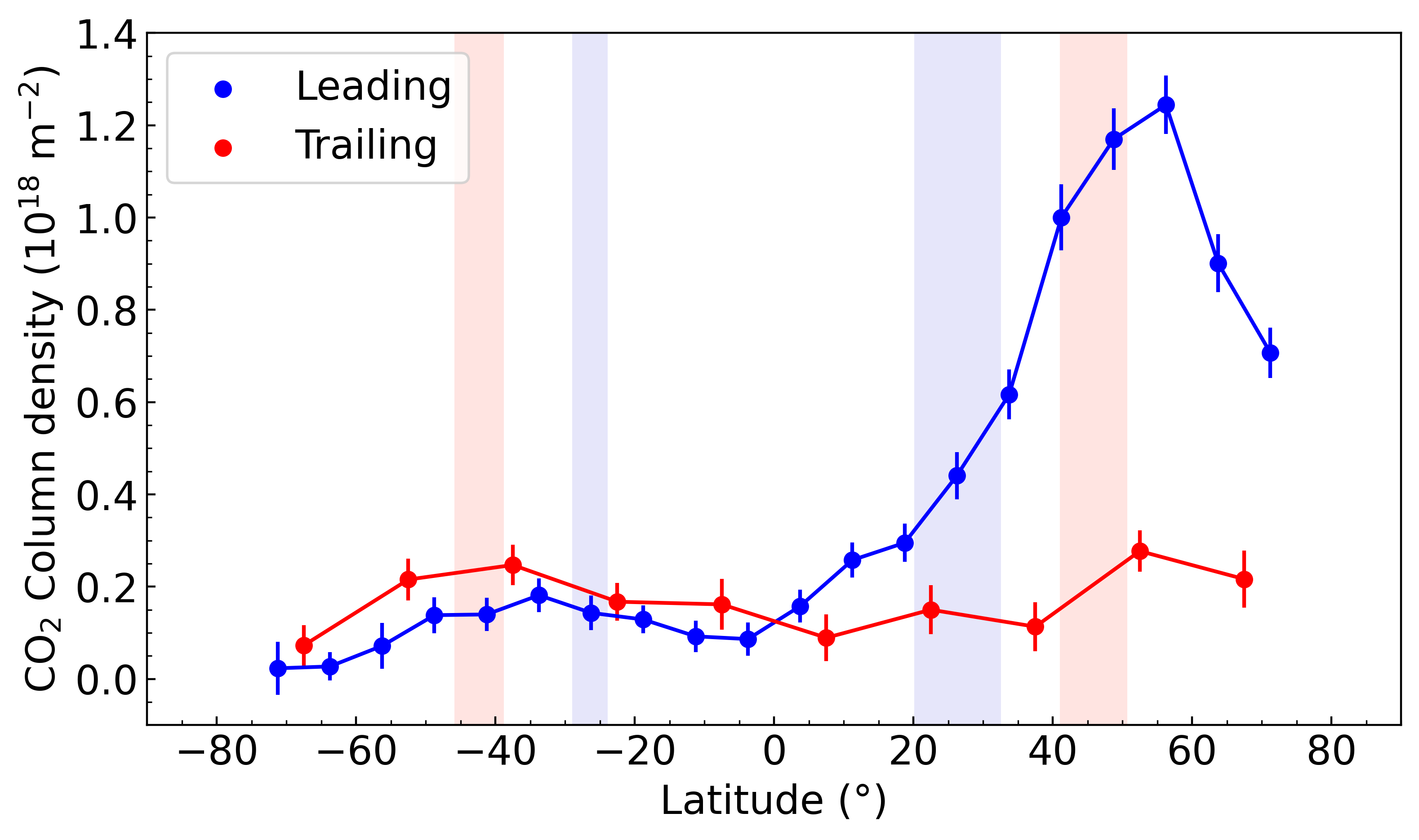}
\caption{Variation of CO$_2$ gas line-of-sight column density with latitude. Blue and red symbols refer to the leading and trailing sides, respectively. Column densities  were derived from spectra that have been averaged in latitude bins of 7.5$^\circ$ (leading, Fig.~\ref{sup:CO2-spectra-leading}) and 15$^\circ$ (trailing).  The blue (resp. pink) vertical domains show the latitude range of the open-closed-field-line boundaries (OCFB) for the leading and trailing sides, respectively, restricted to longitudes of 10--130$^{\circ}$ W (leading) and  210--330$^{\circ}$ W (trailing). \label{collat-distribution} }
\end{figure}

 Observations undertaken with the NIRSpec/IFU instrument provided spatially resolved (0.1'' pixel size, with $\sim$190 pixels across Ganymede's disk) spectra of the leading and trailing sides of Ganymede in the 2.9--5.2 $\mu$m range at high spectral resolution ($R$ $\sim$ 3000) (Paper I and Appendix~\ref{sec:sup-reduc}).  Ro-vibrational emission lines of the CO$_2$ $\nu_3$ band at 4.26 $\mu$m were detected within the broad solid-state CO$_2$ absorption band (Figs~\ref{fig:exosphere-combined},~\ref{spectra}). We used several data processing techniques to extract the CO$_2$ gas signal and best evaluate the confidence level of the detection for weak signals (Appendices~\ref{app:B}, \ref{app:C}). CO$_2$ column densities were inferred using a non-LTE excitation model (Appendices~\ref{app:D}, \ref{app:E}). The distributions of column densities for the two hemispheres are shown in  Fig.~\ref{fig:exosphere-combined}. Figure~\ref{collat-distribution} presents the dependence on latitude, from the analysis of spectra after averaging pixels over ranges of latitude. 

The distribution of CO$_2$ gas shows strong regional variations (Fig.~\ref{fig:exosphere-combined}) and is at odds with expectations that the peak surface 
location of the exosphere would be at the dawn terminator due to condensation on the surface at night and early morning re-evaporation \citep{Steckloff2022}. The CO$_2$ exosphere is most prominent over the north polar regions of the leading hemisphere, peaking at 81$^{\circ}$W, 51$^{\circ}$N (12 h local time), with a column density along the line of sight of (1.5 $\pm$ 0.11) $\times$ 10$^{18}$ m$^{-2}$ corresponding to a pressure at the surface of 1 pbar. The rotational temperature of CO$_2$ measured in this region (107$\pm$5 K, Fig.~\ref{fig:sup-Trot}) constrains the gas kinetic temperature in Ganymede's exosphere (Appendix~\ref{app:H}). 
A point-spread function (PSF) correction to the CO$_2$ column density map suggests that the decrease poleward of 50$^{\circ}$N is real (Appendix~\ref{app:M}). At southern latitudes of the leading hemisphere, and on the trailing hemisphere, the CO$_2$ exosphere is on average at least five-times less dense (Fig.~\ref{collat-distribution}). Low column densities are measured at or near equatorial latitudes for both hemispheres. The trailing hemisphere displays a north/south asymmetry, with the exosphere extending over a broader range of latitudes in the southern hemisphere. Noticeable in Fig.~\ref{fig:exosphere-combined} (see also Fig.~\ref{sup-gas-distribution}) is a CO$_2$ gas enhancement in a large region around (30$^{\circ}$W, 25$^{\circ}$N), encompassing the Tros crater (27$^{\circ}$W, 11$^{\circ}$N). CO$_2$ excess is also present at around 30$^{\circ}$S on the leading side, which corresponds to the position of the expected southern open-closed-field-line boundary (OCFBs, Appendix~\ref{app:J}).

Exospheric H$_2$O was unsuccessfully searched for in 5.5--7.1 $\mu$m spectro-imaging data acquired with the JWST MIRI/MRS instrument  (see Appendix~\ref{sec:H2O}).
Our 3-$\sigma$ upper limit on the H$_2$O column density for the subsolar region of the leading side (Table~\ref{tab:H2O}) is about an order of magnitude higher than the minimum of  6$\times$10$^{18}$ H$_2$O/m$^2$ required to explain UV HST data of atomic oxygen emission lines \citep{Roth2021}. On the other hand, for the trailing side, our derived upper limit for a 105 K atmosphere is slightly below the lower limit from HST \citep[3.6$\times$10$^{19}$ H$_2$O/m$^2$,][]{Roth2021}. Since HST constrains the H$_2$O/O$_2$ ratio and not directly the H$_2$O abundance, 
this could imply that the atmosphere is overall more dilute and that both H$_2$O and O$_2$ densities are lower than assumed in \citet{Roth2021}. This would contradict recent results that suggested a denser global atmosphere based on plasma measurements \citep{Carnielli2020,Waite2024}. Alternatively, a higher atmospheric temperature (e.g., 130 K, Table~\ref{tab:H2O}), as might be expected above subsolar regions, increases the JWST upper limit to values consistent with the HST lower limits.

\begin{table*}[]
    \caption{H$_2$O and CO$_2$ line-of-sight column densities in selected Ganymede's areas.}
    \begin{tabular}{lcccccc}
     \hline
     Region$^a$ &  H$_2$O band area (1-$\sigma$)$^b$  &  H$_2$O band area (1-$\sigma$)$^c$ &   $N({\rm H_2O})$ (3-$\sigma$)$^d$ &  $N({\rm H_2O})$ (3-$\sigma$)$^d$ & $N({\rm CO_2})^f$\\
     & $\lambda < 6.2~\mu$m & $\lambda > 6.2~\mu$m &$T_{\rm rot} = 105$ K & $T_{\rm rot} = 130$ K & $T_{\rm rot} = 105$ K \\
     & (W m$^{-2}$ sr$^{-1}$) & (W m$^{-2}$ sr$^{-1}$) & (m$^{-2}$) & (m$^{-2}$) & (m$^{-2}$)\\
     \hline\addlinespace[2mm]
     Leading CO$_2$ source & $<$ 3.1 10$^{-8}$ & $<$ 1.0 10$^{-7}$ & $<$ 2.0 10$^{19}$ & --  & \phantom{00}1.0 10$^{18}$\\
     Leading SZA$< 15^{\circ}$    & $<$ 4.7 10$^{-8}$ & $<$ 2.2 10$^{-7}$ & $<$ 6.8 10$^{19}$ & $<$ 1.7 10$^{20}$ & $<$ 1.9 10$^{17}$ \\
     Trailing SZA$< 15^{\circ}$  & $<$ 6.8 10$^{-8}$ & $<$ 3.4 10$^{-7}$ & $<$ 3.1 10$^{19}$ & $<$ 4.6 10$^{19}$ & $<$ 2.5 10$^{17}$\\ 
     \hline 
    \end{tabular}
    
    \footnotesize{$^a$ Using extracted spectra from either the subsolar region (Solar Zenith Angle SZA $< 15^{\circ}$, average of 8 pixels for MIRI, 27 pixels for NIRSpec) or from the region with large CO$_2$ gas emission  (leading, 40--65$^{\circ}$N, 46-100$^{\circ}$W, 7 pixels for MIRI, 15 pixels for NIRSpec). $^b$ Using the most intense 10--15 ro-vibrational lines expected in absorption in the 5.7--6.2 $\mu$m spectral range. $^c$ Using the most intense 10--15 ro-vibrational lines expected in emission in the 6.2-7.1 $\mu$m spectral range. $^d$ 3-$\sigma$ upper limits on H$_2$O line-of-sight column density combining upper limits obtained for the two spectral ranges ($<$ 6.2 $\mu$m and $>$ 6.2 $\mu$m) (Appendix~\ref{sec:H2O}). $^f$ Upper limits are 3-$\sigma$.}
    \label{tab:H2O}
\end{table*}

\section{Processes releasing CO$_2$ in Ganymede’s exosphere}
\begin{figure}
    \centering
    \includegraphics[width=8cm]{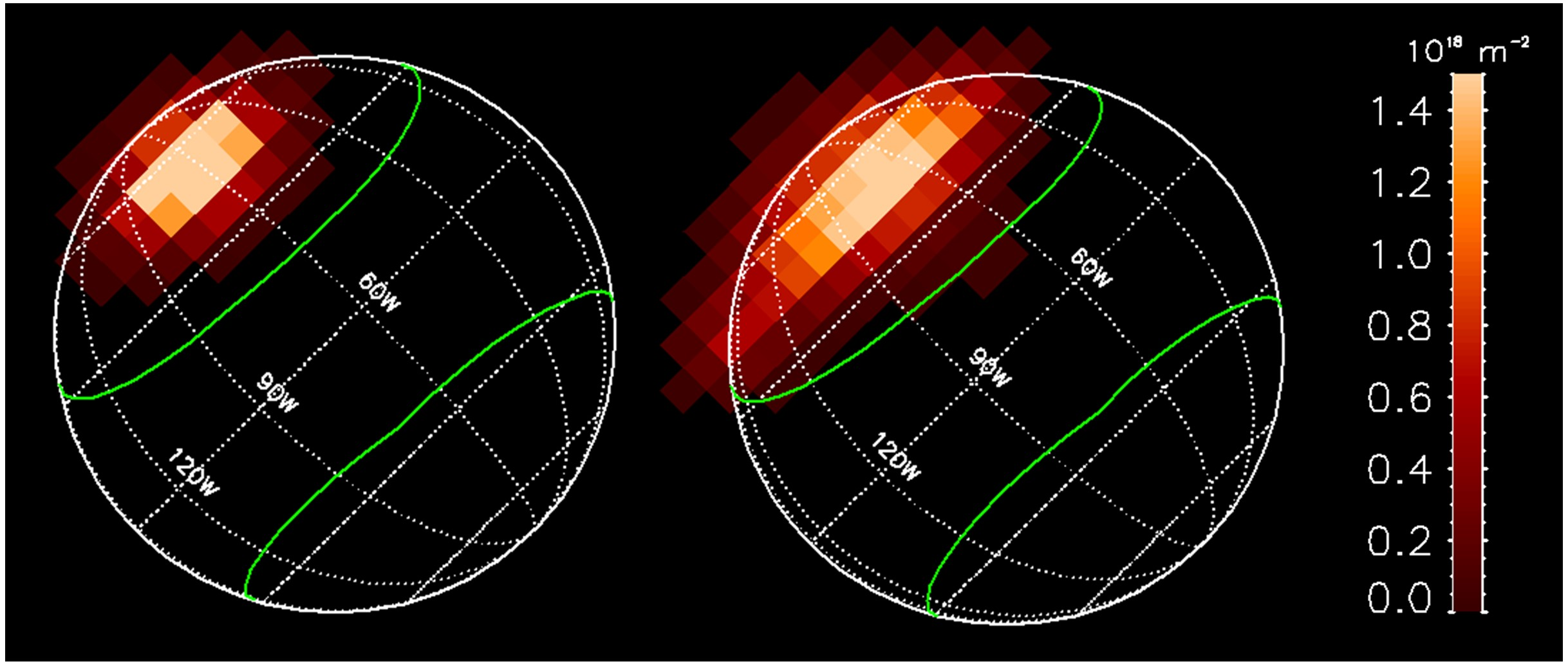}
\caption{Calculated line-of-sight column-density maps of the CO$_2$ exosphere of Ganymede above the leading side from the EGM model (in unit of 10$^{18}$ m$^{-2}$).  Left: CO$_2$ release associated with H$_2$O sublimation with a CO$_2$/H$_2$O relative abundance of 5 for an H$_2$O areal ice fraction of 50\% at latitudes $>$ 50$^{\circ}$N. Right: sputtering of H$_2$O ice with CO$_2$/H$_2$O = 0.01 at latitudes $>$ 40$^{\circ}$N; the result was multiplied by 382 to match the observations (Appendices~\ref{app:K}, \ref{app:L}).  The green lines display the boundary between open and close field lines (OCFB). The subsolar point is at 2.6$^{\circ}$N, 82$^{\circ}$W. \label{fig:model-exosphere}}
\end{figure}

Possible processes releasing CO$_2$ into Ganymede's exosphere include surface ice sublimation and sputtering by energetic particles. We investigated whether these mechanisms, acting either on H$_2$O ice containing CO$_2$ molecules or on pure CO$_2$ ice, could be distinguished from the observed properties of Ganymede's exosphere. For this purpose, we used the Exospheric Global Model \citet[EGM,][]{Leblanc2017}, a multi-species 3D Monte Carlo model that considers sources and sinks (photodestruction, surface sticking, gravitational escape) of such exospheres  (Appendix~\ref{app:K}). The simulations (Appendix~\ref{app:L}) were designed to explain to first order the CO$_2$ column density peak observed in the northern latitudes of the leading hemisphere, and the dichotomy between the trailing and leading hemispheres.

A key question to address is the localized character of the atmosphere. Mean surface temperatures, even in the polar regions (100--110 K, Fig.~\ref{fig:temp-distributiom}), are much warmer than the expected condensation temperature of pure CO$_2$ (73 K at 1 pbar pressure). Hence, the CO$_2$ atmosphere might have been expected to be more global, as shown by EGM calculations considering mean surface temperatures (Appendix~\ref{app:L:1}). This indicates that CO$_2$ interacts with the surface material much more strongly than expected in such a simplistic view. Similar conclusions were reached for O$_2$ gas at Ganymede \citep{Waite2024}, but also at Dione and Rhea \citep{Teolis2016}, based on inconsistencies on O$_2$ column densities between exospheric models and measurements. In those Saturn's moons, the O$_2$ source rates implied by the observations are 50 (Dione) - 300 (Rhea) times
less than expected from the known O$_2$ radiolysis yields from ion-irradiated pure water ice measured in the laboratory, and surface interactions (adsorption/diffusion) appear to control the exospheric structure, density, and seasonal variability. We note that for CO$_2$ at Ganymede, the prime evidence for strong surface/atmosphere interactions comes from the non-global character of the atmosphere, rather than the absolute CO$_2$ column densities (which remain difficult to explain, see below). 

To explain the atmospheric patchiness, Ganymede’s
surface may have properties increasing the effective binding and desorption energies of  O$_2$ and CO$_2$.  In addition to surface roughness producing cold traps, surface irradiation (creating defects) and microstructure (enabling diffusion, and re-adsorption on adjacent grains) could increase effective binding and desorption energies of adsorbates \citep{Yakshinskiy2000,Cassidy2015,Sarantos2020}. Such hypotheses were drawn to explain the distribution of alkali gases surrounding Mercury and the Moon. In the EGM simulations, the surface temperature model considers surface roughness, as constrained from JWST/MIRI brightness temperature maps (Paper I), and simulates the presence of local cold spots through a temperature distribution (Appendix~\ref{app:K}).  As shown in Appendix~\ref{app:L:1}, CO$_2$ diffusion is, to a large extent, controlled  by the ability of molecules to condense on cold traps, thereby explaining localized enhancements of the CO$_2$ exosphere at high latitudes.

Figure~\ref{fig:model-exosphere} shows simulations of Ganymede's CO$_2$ exosphere above the leading hemisphere, assuming that the release of CO$_2$ is induced by the sublimation (left panel), or the sputtering (right panel) of H$_2$O ice containing CO$_2$ molecules. Sublimation of CO$_2$ ice was also investigated (Fig.~\ref{fig:CO2profile_leading}c, d). In all three cases, the CO$_2$ column density peaks at the right latitude, as long as the source region covers the north polar cap (latitude $>$ 40--50$^{\circ}$N, longitude range 0--180$^{\circ}$), and follows a diurnal/longitudinal trend with a maximum at $\sim$ 13.1 to 13.4 h, slightly shifted from the maximum surface temperature (12.5 h, Fig.~\ref{fig:temp-distributiom}) and observed CO$_2$ peak (12 h). Sputtering explains the smooth diurnal variation of the CO$_2$ column density better than sublimation (Fig.~\ref{fig:CO2profile_leading}f). 

In our models in which CO$_2$ is released through sputtering of H$_2$O ice with 1\% CO$_2$ molecules, we had to multiply the sputtered flux from \citet{Leblanc2017} by a factor $\sim$ 380 to match the observed peak column density. The need to increase the sputtered flux significantly might be related to the approach used to calculate this flux (which follows \cite{Leblanc2017}, see Appendix~\ref{app:K}), which consisted in using the yield definition of \citet{Cassidy2013} and a precipitating Jovian ion flux of 10$^6$ particles/cm$^2$/s , ignoring any sputtered component from electron impact. In fact, \citet{Carnielli2020} modeled the ion population in the ionosphere and concluded, based on electron measurements from the Galileo spacecraft, that ionospheric ions could be a significant source of ion precipitation, especially on the leading hemisphere \citep{Carnielli2020b}. Using measurements from the Juno spacecraft, \citet{Waite2024,Vorburger2024} concluded that low-energy electrons are an important sputtering agent. Another source of uncertainty is sputtering yields for production of CO$_2$ by ion and electron impacts, which are unconstrained because the relevant  experiments are sparse. Simulations investigating sputtering on the entire Ganymede's surface (Appendix~\ref{app:L:4}) show that it might be possible to explain the overall distribution of CO$_2$ exosphere by considering strong regional variations of surface properties.

A consequence of the factor of ~380 enhancement of the sputtered flux from \citet{Leblanc2017} is that to first order the O$_2$ column density in our model is multiplied by the same factor, bringing it to values $\sim$ 1.5$\times$10$^{17}$ O$_2$/cm$^2$, at odds with results from \citet{Leblanc2023} and \citet{Roth2021}, based on the atomic O line intensities in the UV. This result goes in the same direction as the ionospheric calculations of \citet{Carnielli2020b} and the post-Juno analyses of  \citet{Vorburger2024} and \citet{Waite2024}, who advocated for O$_2$ columns $\sim$20 times enhanced compared to previous estimates, but the discrepancy is here much larger, which at  face value could be taken as an argument against sputtering being at the origin of the CO$_2$ atmosphere.

In our sublimation models in which CO$_2$ gas is released in proportion with the H$_2$O sublimation flux, to reproduce the peak column density requires an unrealistic CO$_2$ abundance relative to water, three orders of magnitude larger than estimated for the surface ($\sim$1\% in mass, Paper I). Hence, this scenario cannot explain the CO$_2$ exospheric excess on the northern polar cap of the leading side. On the other hand, direct sublimation of CO$_2$ ice is a possible mechanism as only a very small amount of surface coverage (3 $\times$ 10$^{-14}$, Table~\ref{table:model}) is required to explain the peak column density, albeit with an expected diurnal variation more extreme than observed (Fig.~\ref{fig:CO2profile_leading}d). Regarding sub-solar regions and considering H$_2$O ice sublimation with an areal H$_2$O abundance of 20\% appropriate for the leading side \citep{ligier2019}, our model predicts a H$_2$O column density of 4.1$\times$10$^{19}$ m$^{-2}$, consistent with the JWST upper limit for Ganymede's leading hemisphere (Table~\ref{tab:H2O}), but a factor of 7 above the minimum value derived from HST data for this hemisphere \citep[6$\times$10$^{18}$ m$^{-2}$,][]{Roth2021}. 

In summary, processes releasing CO$_2$ in Ganymede's exosphere are not well understood. The smooth diurnal variation of the CO$_2$ column density favors sputtering, but explaining the measured column densities with this process requires further model developments taking advantage of the most recent magnetospheric data acquired by the Juno mission. We can anticipate that the interpretation of Callisto's CO$_2$ exosphere \citep{Cartwright2024}, which is one order of magnitude denser than Ganymede's CO$_2$ exosphere , will be similarly challenging.    

\begin{figure*}[h]
\centering
\begin{tabular}{l@{\hspace{0.3cm}}l@{\hspace{0.3cm}}l@{\hspace{0.3cm}}}
\includegraphics[scale=0.4,valign=b]{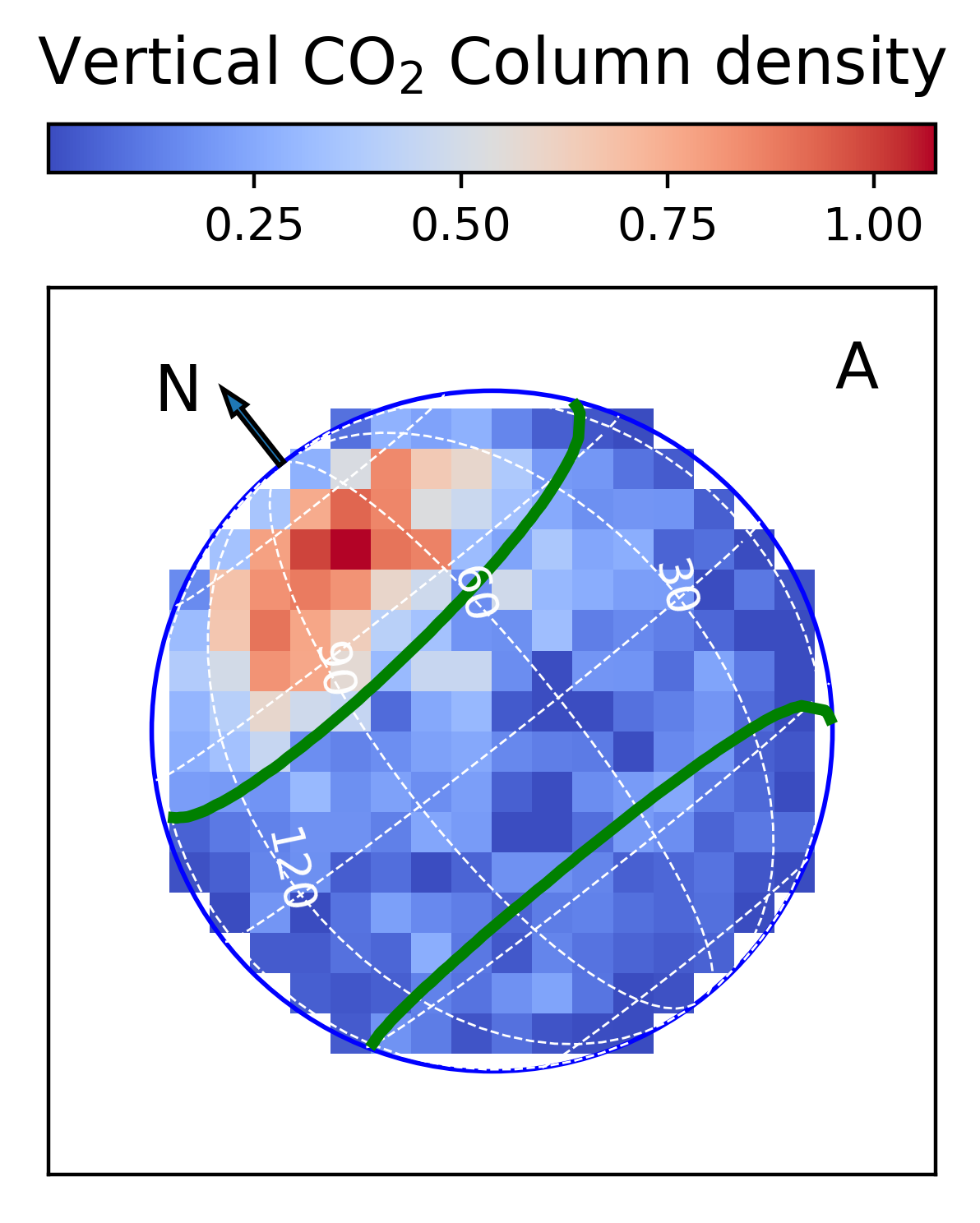}  & \includegraphics[scale=0.4,valign=b]{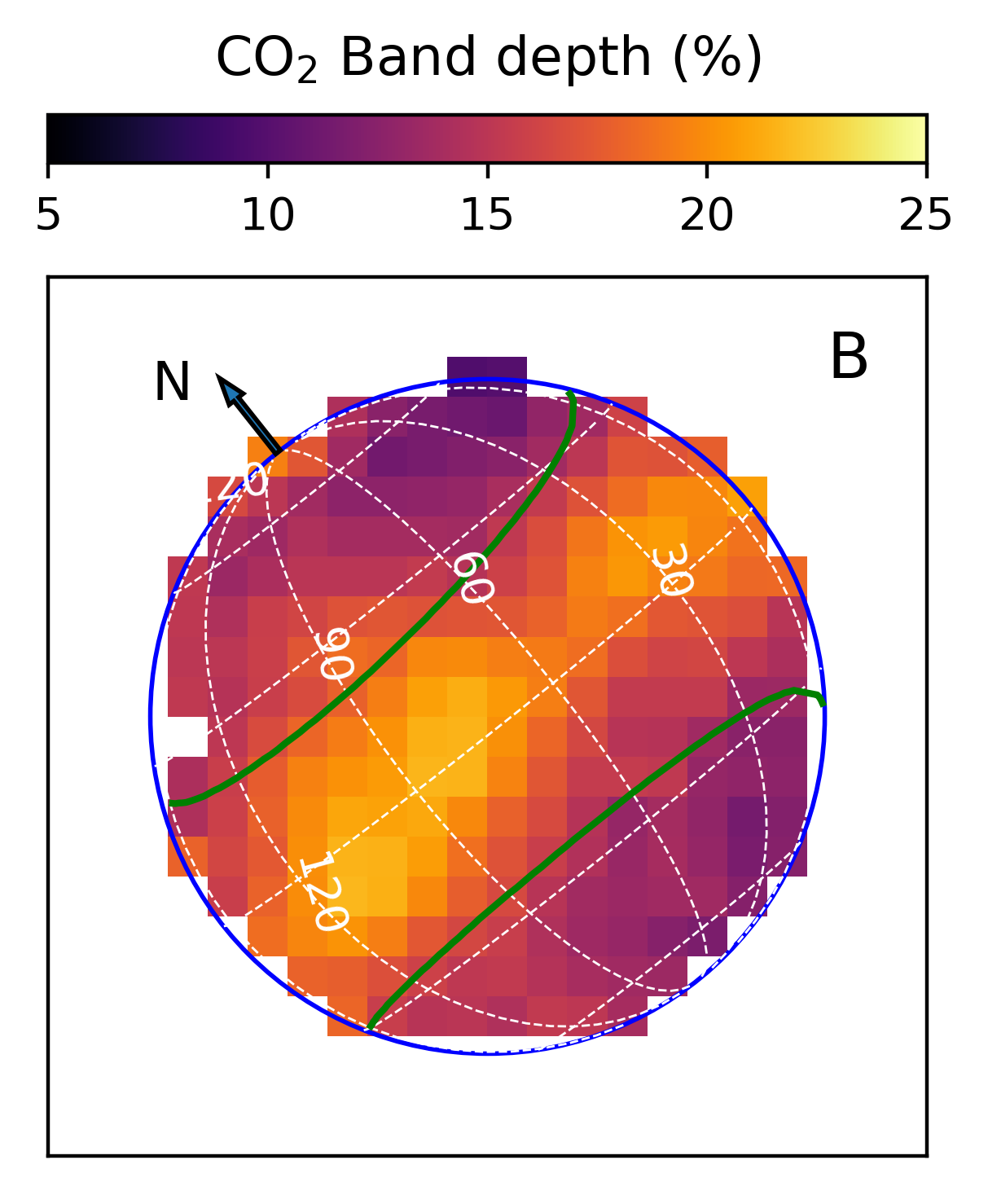} & \includegraphics[scale=0.4,valign=b]{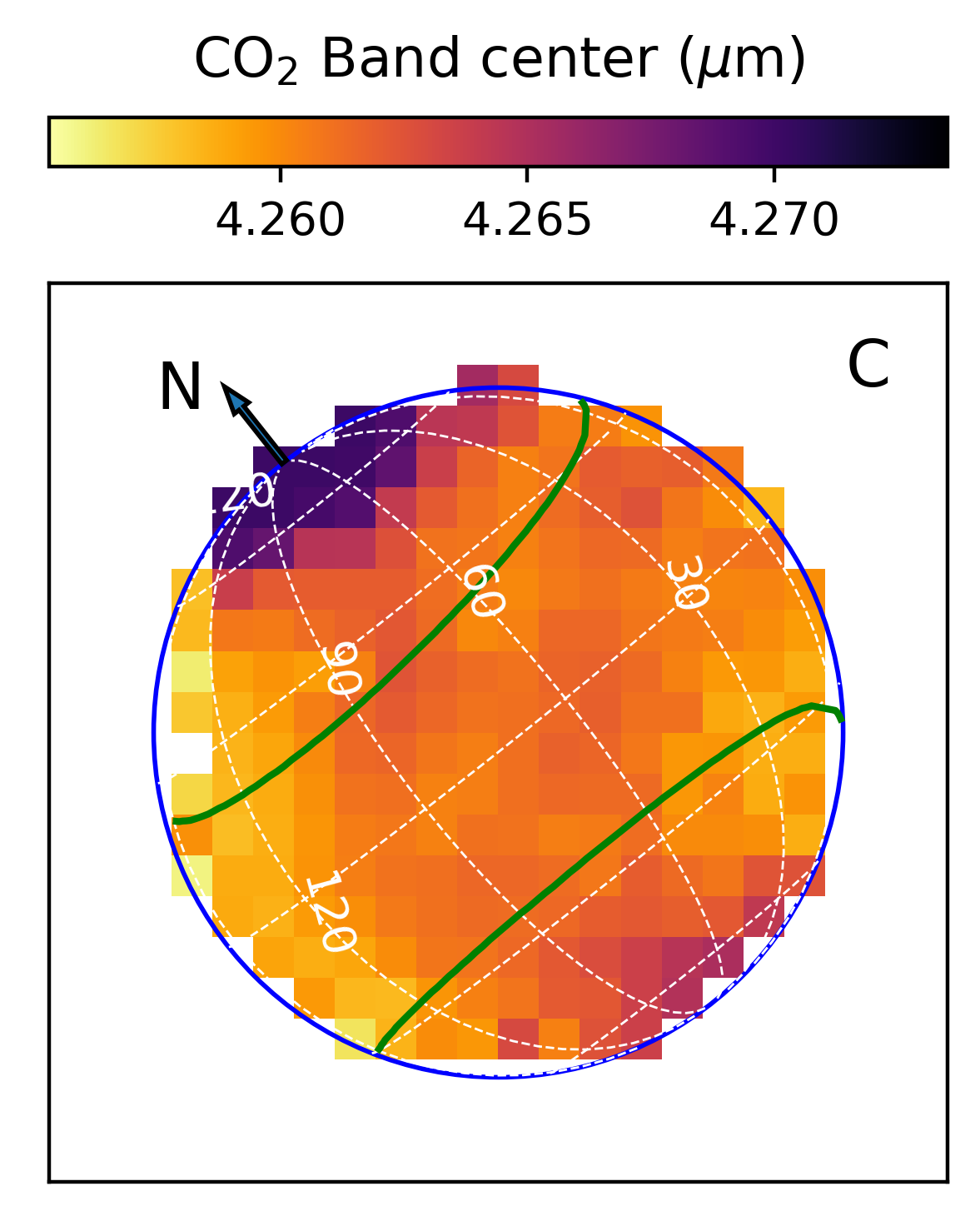}  \\\includegraphics[scale=0.4,valign=b]{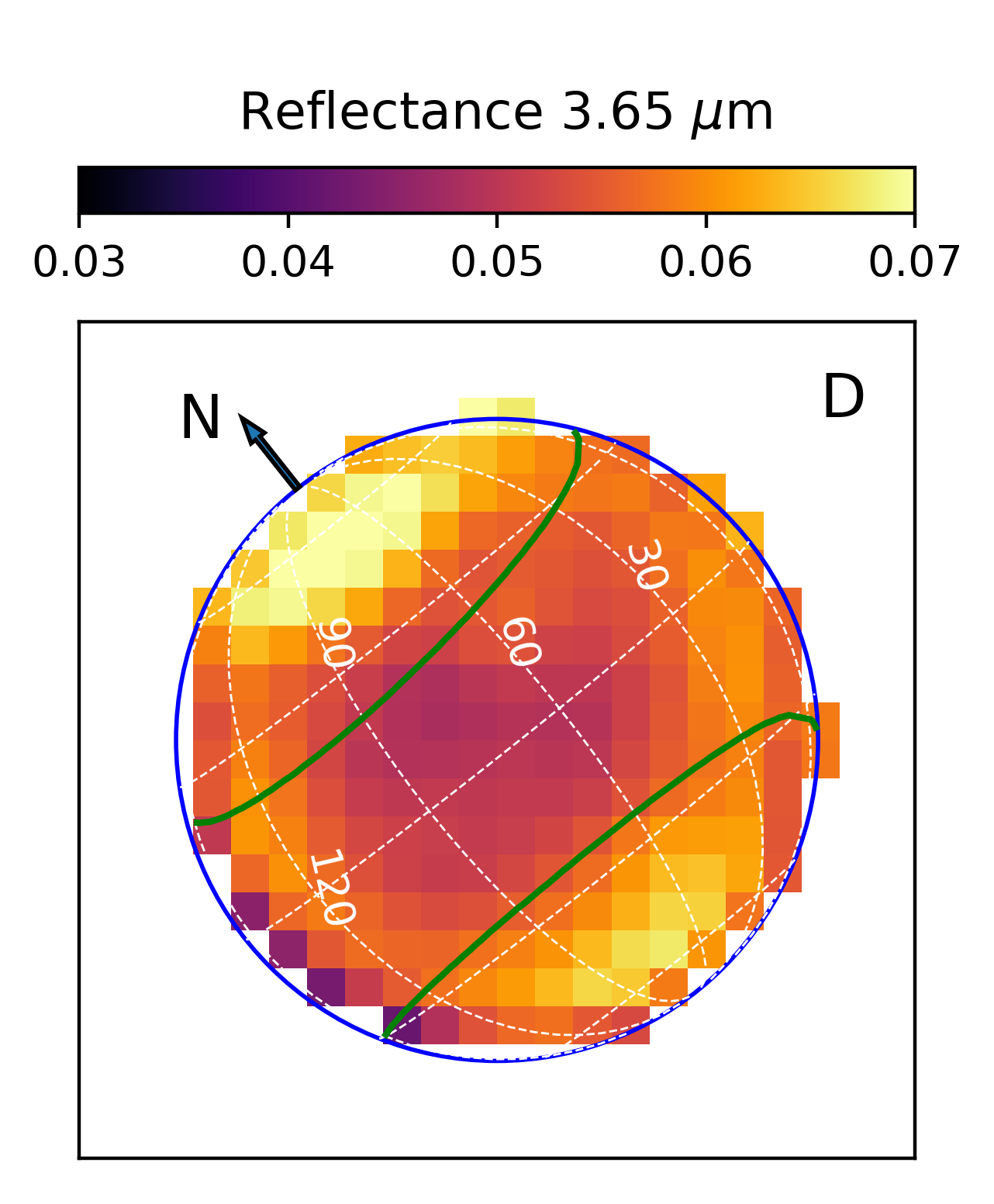} & 
\includegraphics[scale=0.4,valign=b]{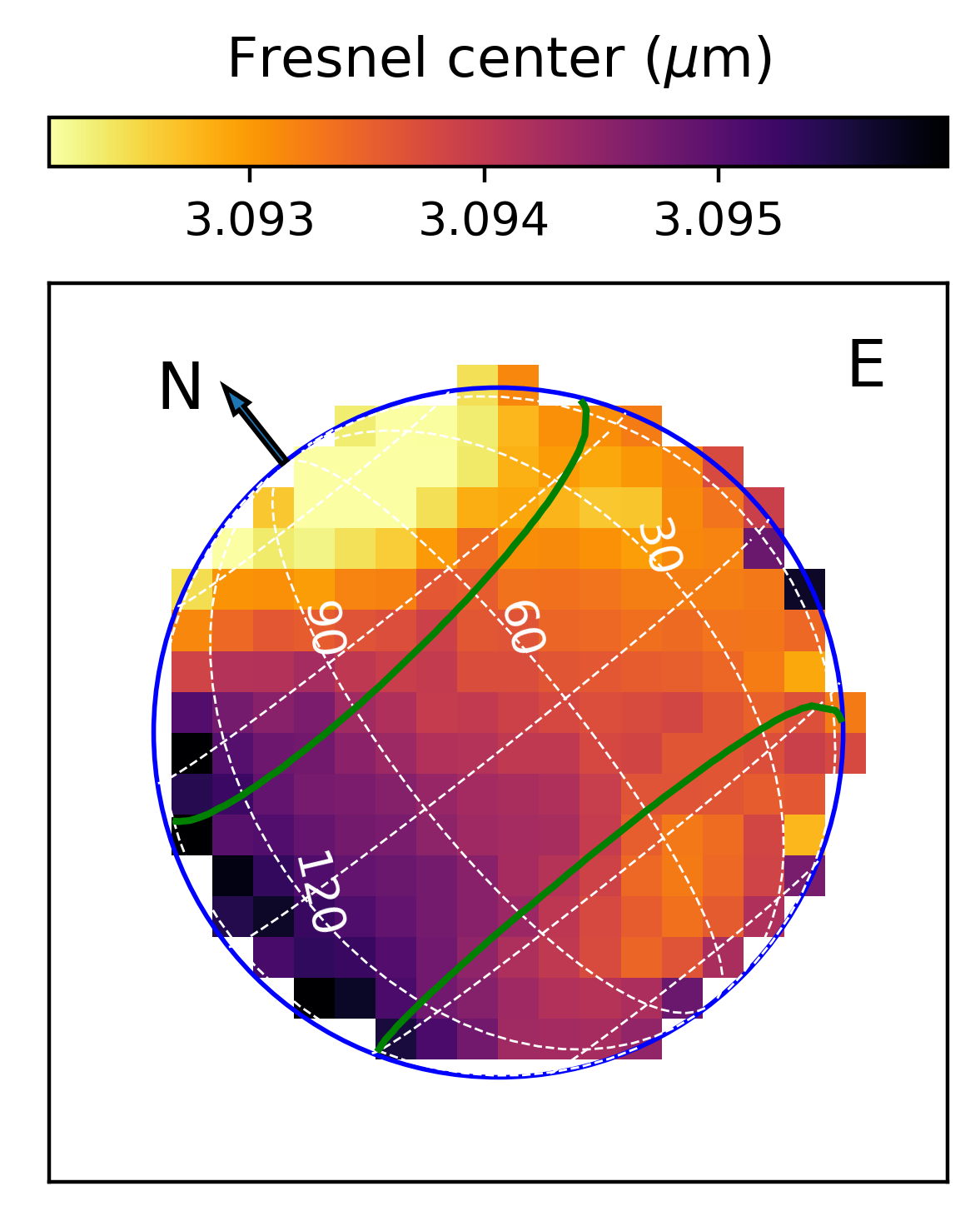} 
&\includegraphics[scale=0.4,valign=b]{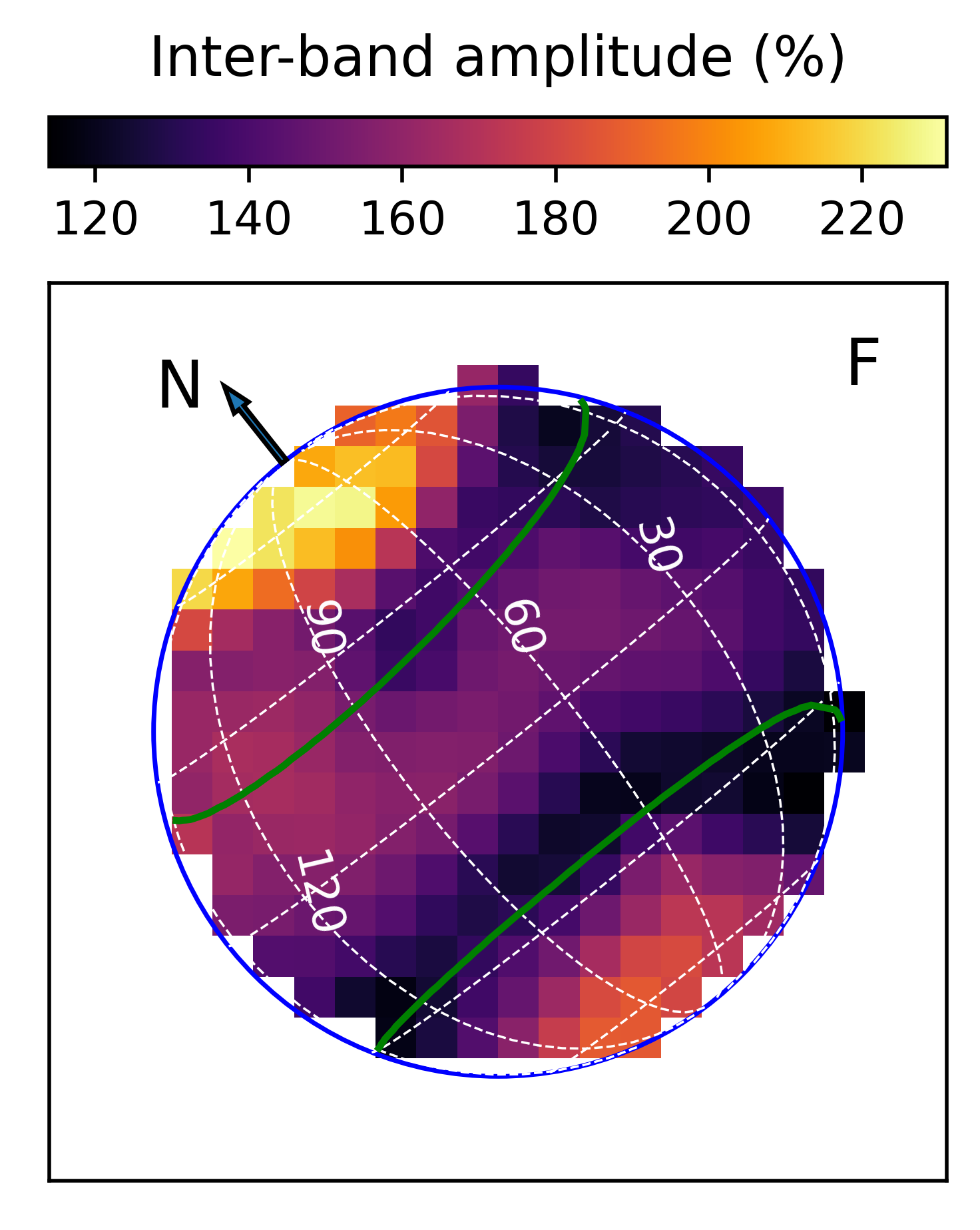} \\
\end{tabular}

\caption{Comparing CO$_2$ exosphere to surface properties on Ganymede's leading hemisphere. A) Vertical CO$_2$ gas column density (unit of 10$^{18}$ m$^{-2}$, this work, see also Fig.~\ref{fig:vertical-colum-CO2}); B) Depth of CO$_2$-solid absorption band (Paper I); C) Central wavelength of CO$_2$-solid absorption band (Paper I); D) Reflectance at 3.65 $\mu$m (Paper I); E) Central wavelength of H$_2$O Fresnel peak (Paper I); F) Relative amplitude of the maximum reflectance between 3.5 and 4 $\mu$m (H$_2$O interband amplitude, Paper I). The north pole of the leading hemisphere possesses the most red-shifted absorption band center of solid CO$_2$, consistent with CO$_2$ trapped in amorphous H$_2$O ice (Paper I). It also has the higher reflectance at 3.65 $\mu$m and H$_2$O interband amplitude, indicative of a higher density of facets in H$_2$O ice for the photons (i.e., smaller grains and/or more internal defects and/or higher micro-roughness/porosity), and the most blue-shifted central wavelength of the H$_2$O Fresnel peak due to a higher proportion of amorphous water ice \citep{mastrapa2009}. \label{leading-surface} }
\end{figure*}

\section{Linking Ganymede's CO$_2$ exosphere to surface properties}

The 4.26-$\mu$m absorption band of surface CO$_2$ is ubiquitous on Ganymede, and is caused by CO$_2$ under different physical states. However, the CO$_2$ gas column density does not correlate with the CO$_2$ surface distribution globally (Fig.~\ref{leading-surface}A, B ; see Appendix~\ref{app:N} and Figs~\ref{leadingtrailing-surface}A, C, D, F). Rather, the prominence of the CO$_2$ exosphere on the northern polar cap is associated with other surface properties.  

According to Galileo high-resolution images, Ganymede’s polar ''caps'' are actually made of discrete patches of optically thick ice, preferentially located on pole-facing slopes \citep{khurana2007}, likely formed by H$_2$O ice sputtering and subsequent re-deposition on these coldest locations \citep{khurana2007}. On both hemispheres, Ganymede’s north polar regions show spectral properties indicative of H$_2$O ice particles having a higher density of facets for the photons (i.e., smaller grains and/or more internal defects and/or higher micro-roughness/porosity) causing multiple scattering and a higher proportion of amorphous ice than the south polar regions \citep[][; Paper I]{denk2009, ligier2019} (Fig.~\ref{leading-surface}D,E,F).

Remarkably, these north/south polar asymmetries in spectral properties are most pronounced on the leading hemisphere. As shown in Fig.~\ref{leading-surface}, the fact that the peak column density of CO$_2$ gas is found over regions where water ice has the highest density of facets, the largest amorphous fraction, and the most red-shifted absorption band center of solid CO$_2$, indicative of CO$_2$ trapped in amorphous H$_2$O ice, suggests that all these properties are probably linked. They are co-located poleward of 40$^{\circ}$N, so they are probably specific to the ice-rich patches constituting the so-called ''polar cap''. The CO$_2$ exosphere is maximum over the polar cap, but it extends over all of the northern open-field-lines area. 

 During the JWST observation of the leading side, the southern hemisphere of Ganymede was facing towards the center of the plasma sheet, where the column density of plasma along Jupiter's magnetospheric field lines is larger than on the northern side of the plasma sheet. On the plasma sheet facing hemisphere, the auroral band of Ganymede is brighter than on the other hemisphere \citep{Saur2022,Greathouse2022, Milby2024}. The reason for the auroral asymmetry is not fully understood. It could be due to the larger plasma momentum and resultant larger magnetic stresses on hemispheres facing the plasma sheet center and/or asymmetric reconnection processes \citep[e.g.,][]{Saur2022,Milby2024}. The larger auroral brightness requires larger auroral electron fluxes of which the largest fraction will collide with the surface. Additionally, the hemisphere facing the center of the plasma sheet is facing larger fluxes of energetic ions and electrons. Integrating these electron and ion fluxes over a full Jovian synodic rotation period should however lead to similar fluxes on the northern and southern polar regions \citep{poppe2018, liuzzo2020}. The observations of higher density of CO$_2$ gas and of enhanced/specific surface properties on the northern hemisphere of the leading side are thus not consistent with what would be expected from either instantaneous or time-averaged plasma effects. Therefore, the specifics of the north polar regions of the leading hemisphere are likely an inherent property of Ganymede’s surface.

The north and south polar caps mainly differ in the nature of their underlying terrains. Galileo Regio, the largest patch of the darker and more cratered terrains on Ganymede, encompasses much of the leading north polar latitudes, while the leading south polar latitudes have fewer of such dark cratered terrains (Fig.~\ref{fig:sup-Patterson} from \citet{Patterson2010}). The low-albedo material, concentrated in topographic lows by sublimation and mass wasting \citep{prockter1998}, may be a remnant of Ganymede’s formation building blocks and/or may have been deposited by comet-like bodies \citep{zahnle1998,bottke2013}, so it could contain CO$_2$ precursors (organic and inorganic carbon-bearing components), whose radiolysis and/or disaggregation by energetic particles may produce and/or release CO$_2$.  As shown by laboratory experiments, the radiolysis of complex organic matter \citep{gomis2005,raut2012} or carbonates \citep{costagliola2017} in the presence of H$_2$O forms CO$_2$. In addition, the disaggregation of carbonaceous chondrite-like material \citep{yuen1984} or the radiolysis of some of their inorganic carbon-bearing components \citep[carbonates and other minerals,][]{Nakamura2023} could also release or produce CO$_2$. This CO$_2$ production may be specifically enhanced in the northern open-field-lines area of the leading hemisphere because they host the largest extent of dark cratered terrains than the southern ones (Fig.~\ref{fig:sup-Patterson}).

However, the peak in CO$_2$ column density is not only over the fraction of Galileo Regio poleward of the OCFB, but over the water ice polar cap (Fig.~\ref{leading-surface}A, D, F ; Fig.~\ref{leadingtrailing-surface}B), covering diverse terrain types (Fig.~\ref{fig:exosphere-combined}). Moreover, the peak of the CO$_2$ vertical column density is at 72$^\circ$W, 45$^\circ$N  (12.6 h local time), on the boundary between Galileo Regio and the bright terrain Xibalba Sulcus (Fig.~\ref{fig:exosphere-combined}, Fig.~\ref{fig:sup-Patterson}).
Therefore, if the CO$_2$ is initially produced on the dark terrains, it should migrate and accumulate over the polar cap on the long term, before being diurnally released and redeposited over the polar cap as is possibly observed. This redistribution might also occur if the CO$_2$ is initially produced from other sources, for example by a relatively recent resurfacing event (impact, mass movement) that would have exposed to the surface CO$_2$ or CO$_2$ precursors from the sub-surface and/or from the impactor. Notably, several impact craters with bright ejecta are present over the part of Xibalba Sulcus showing maximum CO$_2$ column density \citep{collins2014}, and the ice bedrock of this relatively recent region has been suspected of containing significant CO$_2$ based on its geomorphology \citep{moore1999}. If this CO$_2$ exosphere is permanent, geological mass-wasting events \citep{moore1999, pappalardo2004} and possibly micro-meteoritic gardening, may regularly expose new CO$_2$ or CO$_2$ precursors to the surface, maintaining the CO$_2$ exosphere over the long-term. The release of CO$_2$ gas from on-going volcanic activity seems unlikely given the surface age \citep[0.5--1 Ga with large uncertainties,][]{zahnle1998, showman2004}, but gravity anomalies were identified around this region \citep{casajus2022}.

Produced CO$_2$ may then preferentially co-deposit with H$_2$O and accumulate on high-latitude cold traps, potentially explaining the red-shift of the CO$_2$ absorption band with latitude (Fig.~\ref{leading-surface}C). The maximum CO$_2$ column densities and the H$_2$O ice having the highest density of facets both peak around a longitude at the maximum solar illumination and maximum surface temperature at this latitude (Paper I), suggesting a diurnal process releasing CO$_2$ from the ice in the atmosphere (Fig.~\ref{leading-surface}A, D, F). According to our analyses, sputtering appears to have a temperature dependence that is the most consistent with the observation (Fig.~\ref{fig:CO2profile_leading}). At mid-day, the maximum temperature of the ice enhances sputtering and thermal stress that may generate more ice facets, resulting in surface micro-roughness or internal cracks, which could further enhance CO$_2$ release \citep[][and references therein]{baragiola2003}. Later in the day and night, re-deposition and/or molecular movements induced by energetic ions might fill in these pores or cracks, decreasing the density of facets and trapping the CO$_2$ again.

\section{Summary}
In summary, the north/south polar asymmetry in the distribution of CO$_2$ gas of the leading hemisphere could be explained by the larger extent of dark terrains over the northern polar region, providing a larger initial source of CO$_2$ produced by radiolysis of organic or inorganic precursors. The existence of other initial sources specific to this region (impact, mass movement, cryo-volcanism) cannot be excluded, but lack compelling evidence. After its initial production, CO$_2$ may migrate and accumulate on cold traps of the polar cap and be diurnally released and redeposited, explaining the co-location of the northern polar atmosphere with the H$_2$O and CO$_2$ surface properties.  Whether the release mechanism in this high-latitude region is sputtering or sublimation remains unclear. Outside of the open-field-line areas, CO$_2$ gas is located above various terrain types, including the dark terrain Melotte and some other terrains having more (or smaller) grains of H$_2$O ice or H$_2$O-bearing minerals/salts (Fig.~\ref{leadingtrailing-surface}B, E). This spatial distribution suggests the existence of several mechanisms producing and releasing CO$_2$. Future investigations of Ganymede from JWST and space missions, together with further  models and experiments dedicated to sputtering processes, are needed to unravel the origin of Ganymede's patchy CO$_2$ exosphere.  

\begin{acknowledgements}
This work is based on observations made with the
NASA/ESA/CSA James Webb Space Telescope. The data were obtained from the Mikulski Archive for Space Telescopes at the Space Telescope Science Institute,
which is operated by the Association of Universities for Research in Astronomy, Inc., under NASA contract NAS 5-03127 for JWST. These observations are associated with program 1373, which is led by co-PIs Imke de Pater and
Thierry Fouchet and has a zero-exclusive-access period.
D.B.-M, E.Q., E.L., T.F., and O.P. acknowledge support from the French Agence Nationale de la Recherche (program PRESSE, ANR-21-CE49-0020-01). I.dP and M.H.W. were in part supported by the Space Telescope Science Institute grant nr. JWST-ERS-01373. L.F. was supported by STFC Consolidated Grant reference ST/W00089X/1; for the purpose of open access, the author has applied a Creative Commons Attribution (CC BY) licence to the Author Accepted Manuscript version arising from this submission. Some of this research was carried out at the Jet Propulsion Laboratory, California Institute of Technology, under a contract with the National Aeronautics and Space Administration (80NM0018D0004).
\end{acknowledgements}

\bibliography{biblio}

\begin{appendix}

\section{JWST observations and data reduction}
\label{sec:sup-reduc}
\noindent
NIRSpec/IFU observations of the leading and trailing sides of Ganymede were obtained as part of the ERS programme \#1373 (PIs I. de Pater, T. Fouchet). These observations, acquired with the G395H/F290LP grating/filter pair, provided spatially resolved imaging spectroscopy in the range 2.86--5.28 $\mu$m over a 3''$\times$3'' field of view with 0.1''$\times$0.1'' spatial elements (310 $\times$ 310 km at Ganymede), and a nominal spectral resolution of $R$ $\sim$ 2700. The estimated full width at half maximum of the point spread function (PSF) is $\sim$0.19'' (Appendix~\ref{app:M}). Detailed information on these observations is provided in \cite{Trumbo2023} and Paper I, focused on the analysis of solid state spectral features from CO$_2$, H$_2$O, and H$_2$O$_2$. For the data reduction we followed the procedure adopted in \cite{DBM2024} (Paper I). The updated JWST pipeline version 1.12.5 and context file version $jwst\_1148.pmap$ were used. Correction for the 1/f noise was done as explained in \cite{Trumbo2023} and Paper I. 

ERS~\#1373 comprised also observations of the leading and trailing sides of Ganymede using the Mid-Infrared Instrument/medium resolution spectroscopy (MIRI/MRS), which are described in  Paper I. These observations, made with the four IFU channels, provided spatially-resolved unsaturated spectra in the 4.9--11.7 $\mu$m range. Channel 1 (4.9--7.65 $\mu$m) covers the $\nu_2$ vibrational band (and weaker $\nu_2$+$\nu_3$-$\nu_3$ and $\nu_2$+$\nu_1$-$\nu_1$ hot-bands) of H$_2$O in vapor phase from which the H$_2$O content in Ganymede's exosphere can be studied. For Channel 1, the spaxel (aka pixel in main text) size is 0.13'' and the spectral resolution is $\sim$ 3700. The data were re-reduced using most recent JWST pipeline version 1.11.3, and context file $jwst\_1119.pmap$, and processed as in Paper I. 

Ganymede spectra are crowded with solar lines. For the study of solid-state features in NIRSpec spectra, the output of the JWST pipeline, calibrated in radiance units (MJy/sr), were divided by the solar spectrum \citep{Hase2010} at the spectral resolution of NIRSpec, giving data in units of radiance factor $I/F$ (Paper I). To obtain spectra in radiance units and corrected from solar lines, the data in $\textit{I/F}$ units were multiplied by the solar continuum. In the spectral 4.2--4.3 $\mu$m region where strong ro-vibrational lines of the CO$_2$ $\nu_3$ band are present, solar lines are not numerous and  much fainter than in nearby spectral regions. Nevertheless, we payed special attention to solar-line removal as gaseous emission lines from Ganymede are faint. We determined that solar lines present in the 4.4--4.6 $\mu$m range are best removed when applying a correction factor of $\sim$ 0.87 to the nominal spectral dispersion provided by JWST documentation (i.e., increasing the wavelength-dependent spectral resolution by 1.15). We used this factor (giving $R$ = 3365 at 4.2--4.3 $\mu$m) in subsequent analyses, including for producing synthetic line profiles.

In MIRI spectra, the most intense ro-vibrational lines from the H$_2$O $\nu_2$ band are expected between 5.6 and 7.4 $\mu$m (Fig.~\ref{fig:sup-H2O}). In this spectral region, both reflected light and thermal emission from Ganymede's surface contribute to the continuum, especially at the lowest wavelengths where the two components have similar intensities (Paper I). Therefore, spectra were corrected from solar lines by isolating the reflected-light component, and applying the method used for NIRSpec data.

\section{Extraction of CO$_2$ gaseous lines}
\label{app:B}
\begin{figure}[h!]
\centering
\begin{minipage}{9cm}
\includegraphics[width=4.cm]{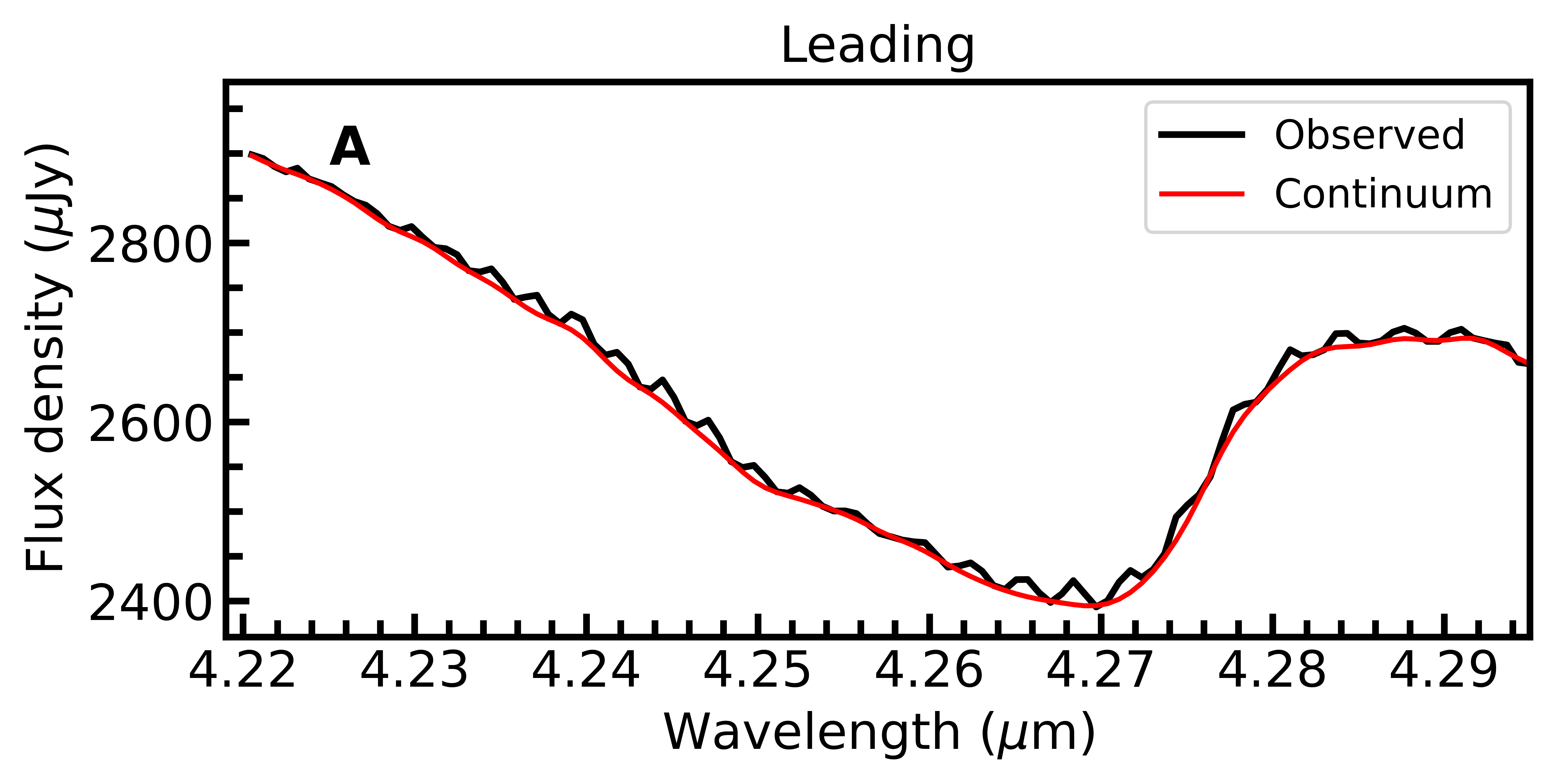}
\includegraphics[width=4.cm]{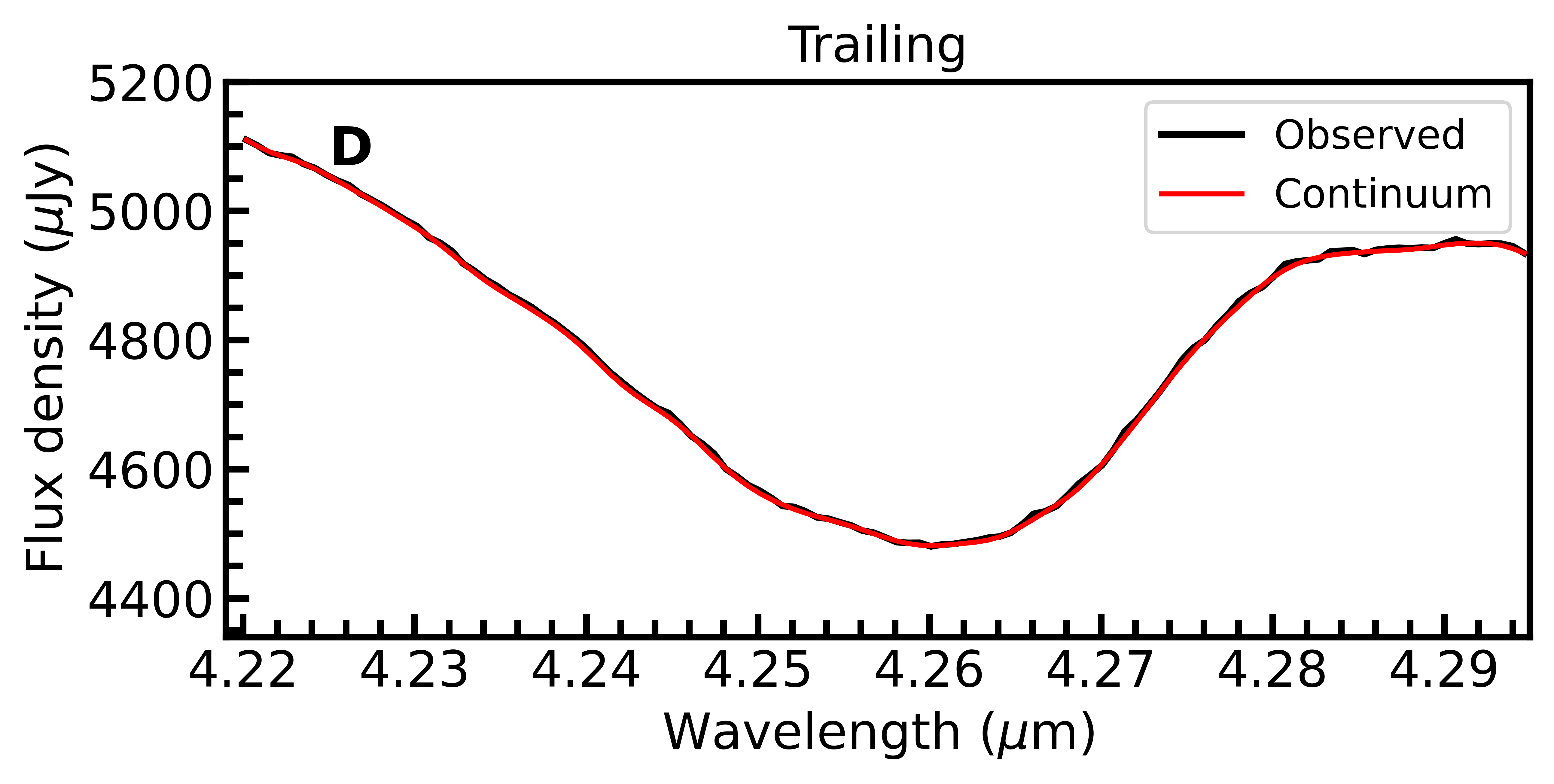}\hfill
\end{minipage}
\begin{minipage}{9cm}
\includegraphics[width=4.cm]{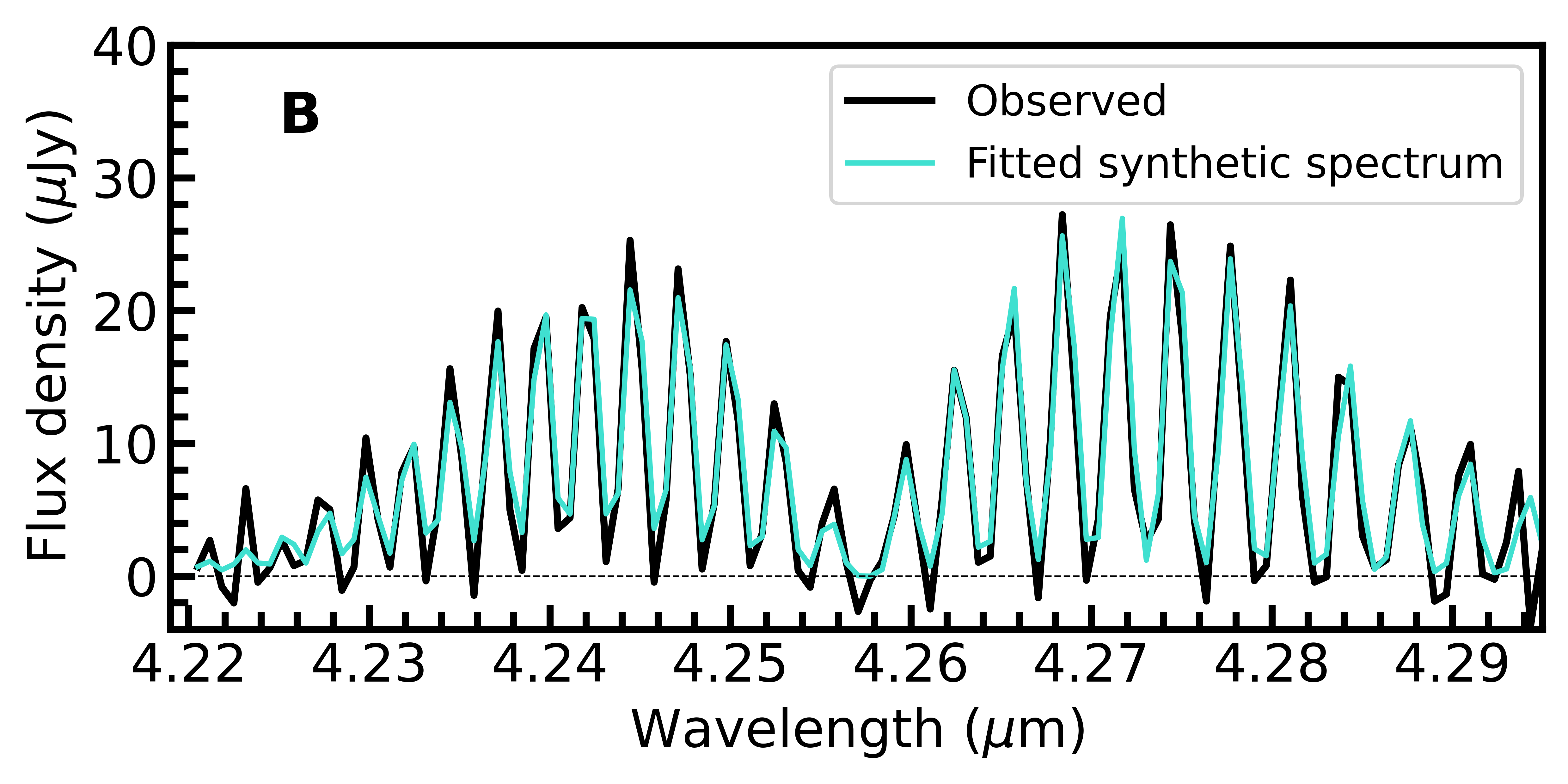}
\includegraphics[width=4.cm]{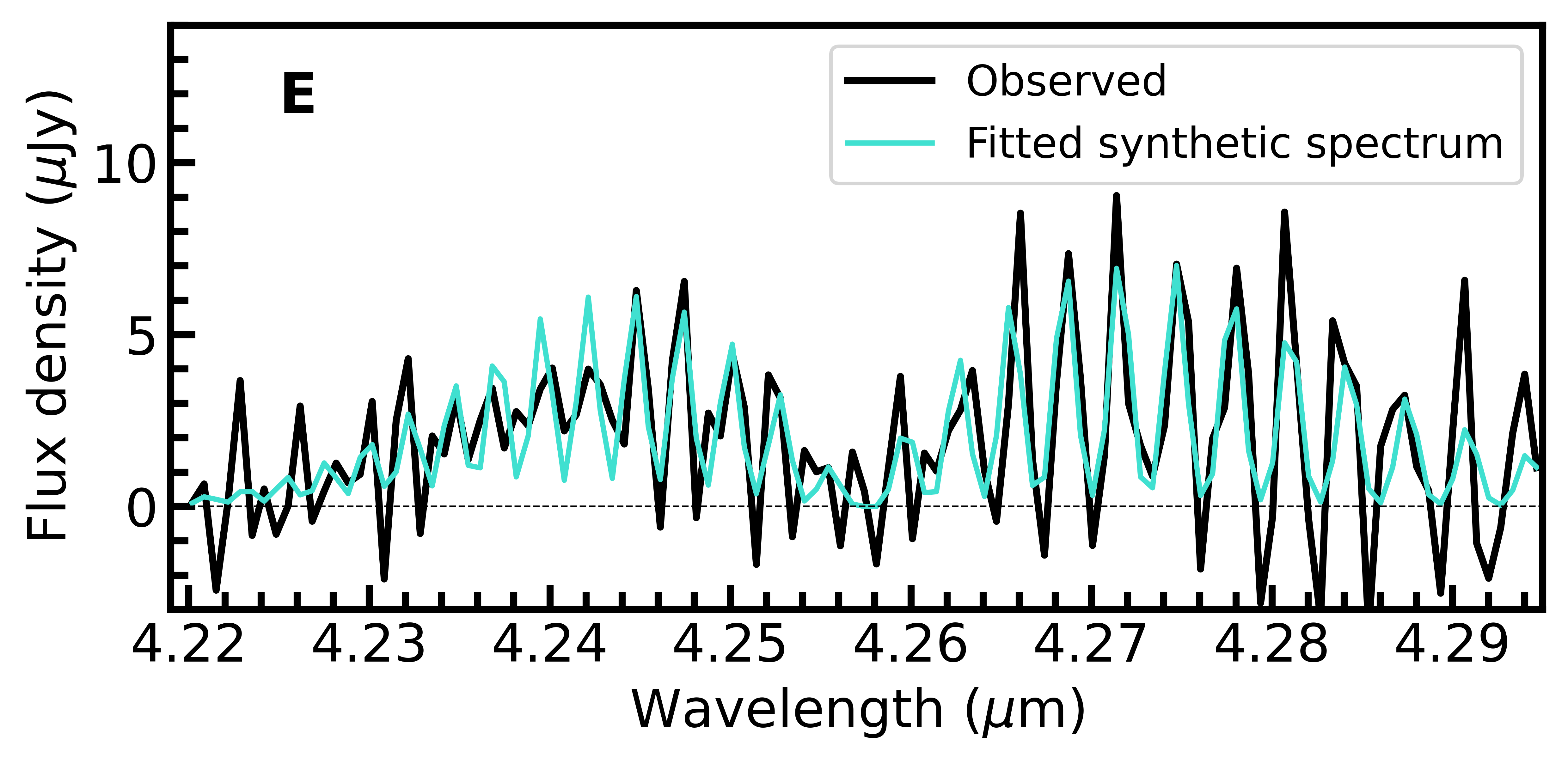}\hfill
\end{minipage}
\begin{minipage}{9cm}
\includegraphics[width=4.cm]{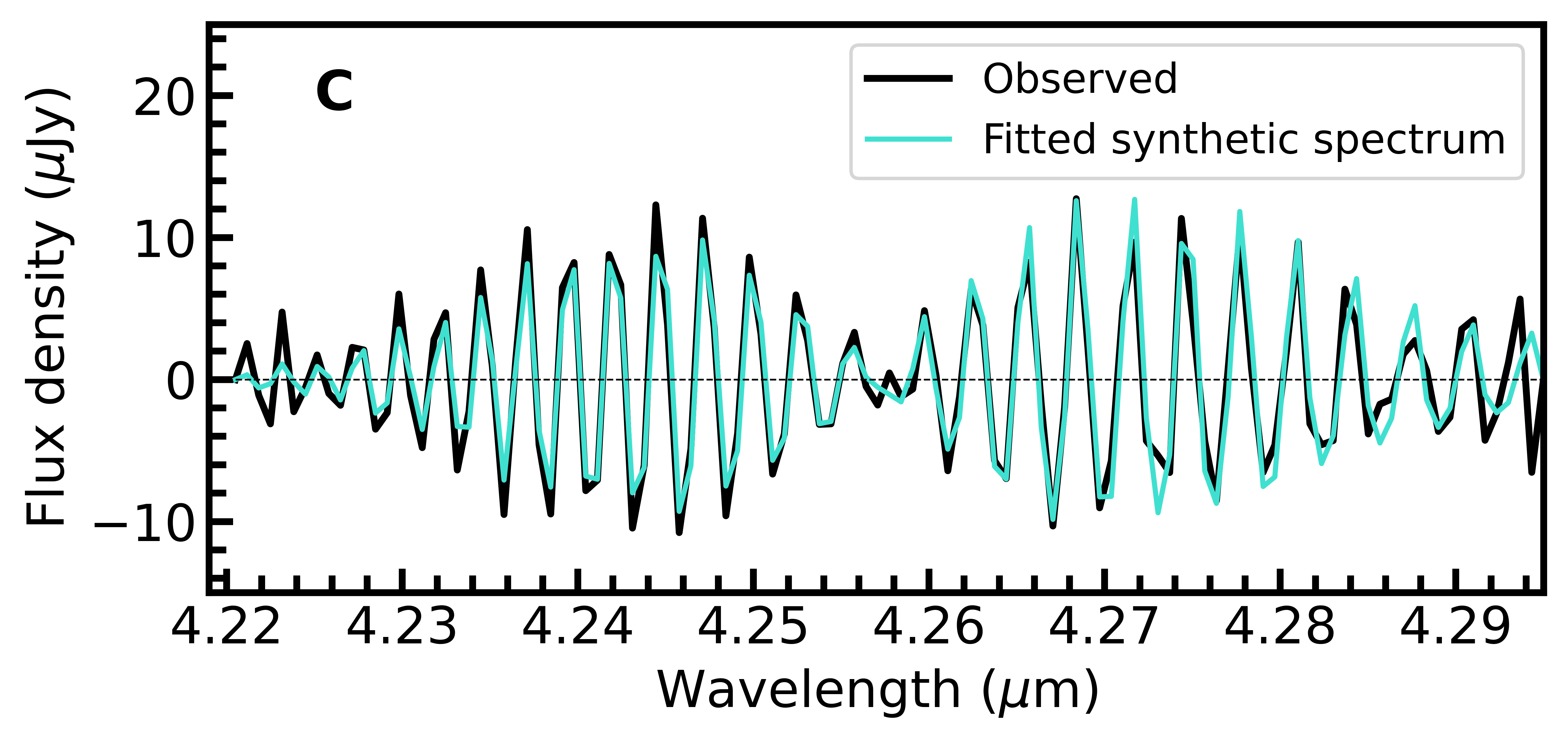}
\includegraphics[width=4.cm]{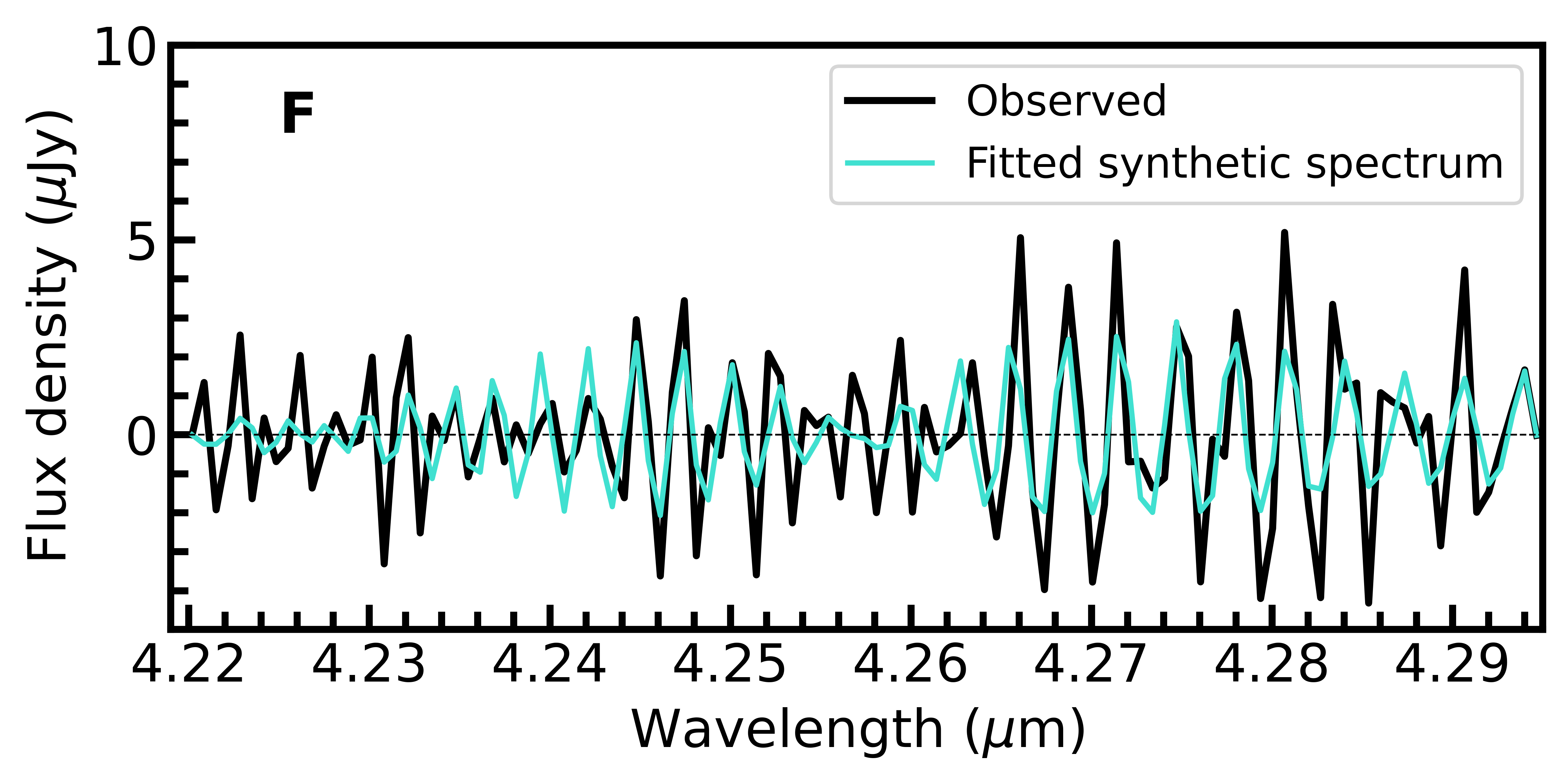}\hfill
\end{minipage}
\caption{CO$_2$ $\nu_3$ band gas spectra from Ganymede's exosphere. Left (A--C): North polar cap of the leading side of Ganymede (averaged spectra for latitudes $>$ 45$^{\circ}$ N); Right (D--F): Southern hemisphere of the trailing side (latitudes 30--60$^{\circ}$ S). A, D) Observed spectra showing both the CO$_2$ $\nu_3$ absorption band from CO$_2$ in solid state, and ro-vibrational emission lines of gaseous CO$_2$; B, E) CO$_2$ gaseous emission spectra obtained after removing the continuum emission shown in red in panels A and D (Appendix~\ref{app:B}). C, F) Residual CO$_2$-gas DIFF spectra obtained by removing the continuum obtained from low-pass filtering (Appendix~\ref{app:B}). Best fit synthetic spectra are shown in blue, with a fitted rotational temperature of 108$\pm$8 K for the leading side, and a fixed rotational temperature of 105 K for the trailing side. In all plots, vertical-axis unit is $\mu$Jy per pixel (1 pixel = 0.1''$\times$ 0.1''). \label{spectra}}
\end{figure}

\noindent
The $\nu_3$ bands of CO$_2$ in gaseous form and in solid state lie at the same wavelengths. The solid-state absorption band of CO$_2$ shows strong variations in shape and intensity on the surface of Ganymede (Paper I). Hence, to isolate the weak ro-vibrational emission lines from CO$_2$ gas from the broad absorption band, we developed specific tools, which were tested on synthetic spectra. We restricted the analysis to the 4.220--4.295 $\mu$m range where the strongest CO$_2$ gaseous lines and only weak solar lines are present.

In a first step, the solid-state contribution was estimated by applying low-pass filtering with a Butterworth filter. The optimum cutoff frequency  that preserves best the gaseous signatures was determined by applying the method to synthetic spectra combining the Ganymede CO$_2$-solid absorption band and CO$_2$-gas fluorescence emission. 
In a second step, the residual CO$_2$ gas signature (called DIFF) was obtained by subtracting this estimated solid-state signal from the observed spectra.  Two examples of residual DIFF spectra are shown in Figs~\ref{spectra}C, F. This method does not allow  retrieval of the correct shape of the CO$_2$ gaseous band. We show in  Figs~\ref{spectra}B, E (and Fig.~\ref{fig:exosphere-combined}) CO$_2$ gas spectra from Ganymede displaying the expected ro-vibrational structure of the CO$_2$ $\nu_3$ band for fluorescence emission. They were obtained through several iterations, by computing the envelope of the residual DIFF signal and adding the bottom part of the envelope to the solid-state signal extracted from low-pass filtering. From synthetic spectra processed in the same manner, we found that this third step produced an overestimation of the strength of the CO$_2$ gas signature, especially for faint signals at the limit of noise. Hence, analyses were made on DIFF spectra.

In order to evaluate the significance of detection of gas emissions, we computed the Pearson correlation coefficient $C_p$ between the DIFF spectra and a forward model. The forward model consists in a DIFF spectrum  computed by applying the same treatment as for the data to a synthetic spectrum obtained by combining the Ganymede average CO$_2$ absorption band observed on the leading side (obtained from low-pass filtering) and a CO$_2$ fluorescence spectrum at 105 K convolved to the instrumental spectral resolution $R$ = 3365. The Pearson correlation coefficient $C_p$ was computed for each individual spaxel on the leading side. Due to the faintness of the CO$_2$ gas emission lines on the trailing side, the trailing data were smoothed using a 3$\times$3 boxcar filter.  $C_p$ ranges from --0.22 to 0.87 on the leading side, and from --0.11 to 0.50 on the trailing side (left panels of Fig.~\ref{sup-gas-distribution}). Except for the northern regions of the leading Ganymede disk, $C_p$ values do not exceed 0.5. Hence the confidence level of the detection of the CO$_2$ exosphere is rather low for several regions, calling for the use of other detection criteria (cross-correlation technique). 

\begin{figure}[h]
\includegraphics[width=9cm]{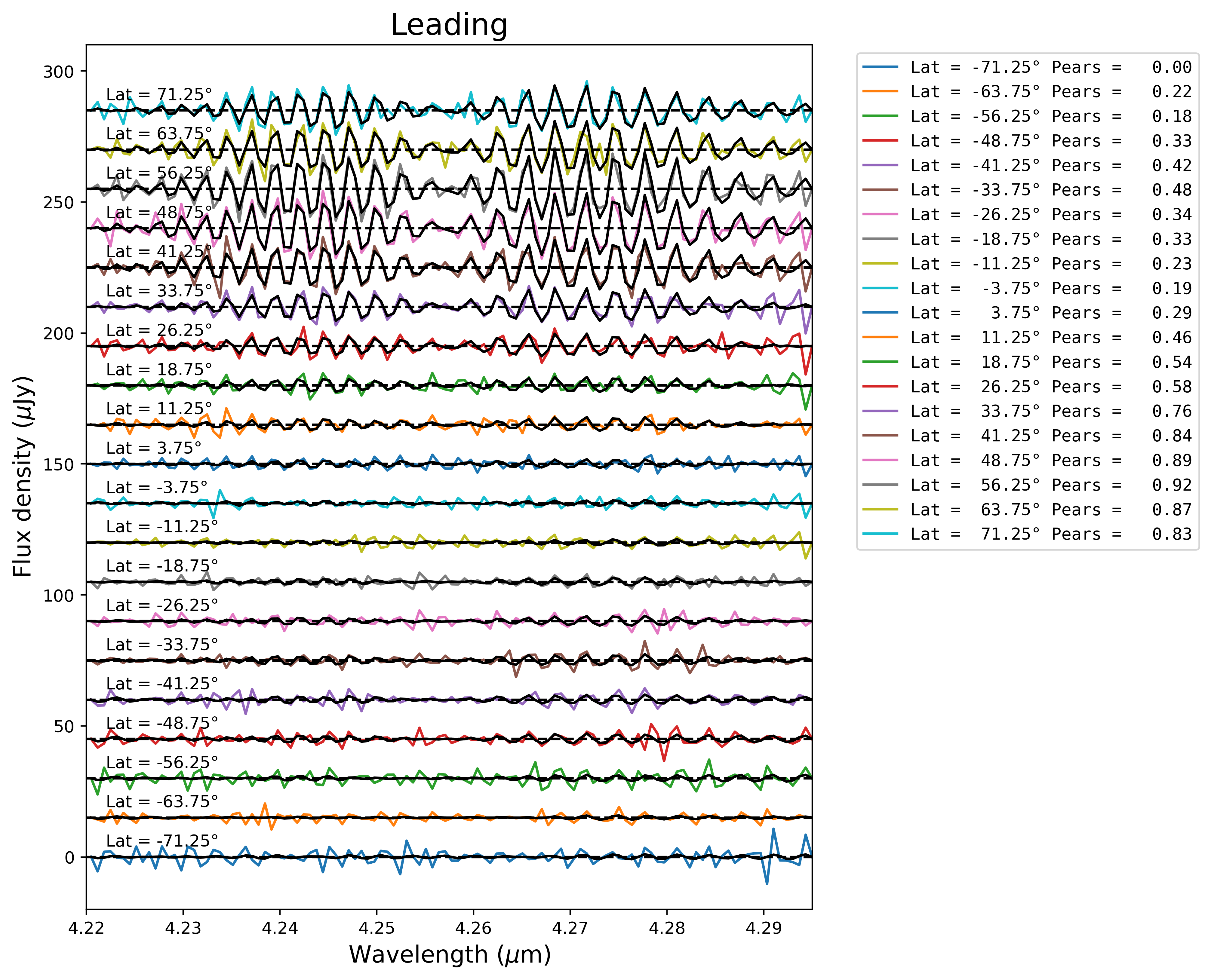}
\caption{DIFF spectra on the leading hemisphere as a function of latitude. Data were averaged over latitude bins of 7.5$^\circ$ and treated as explained in Appendix~\ref{app:B}. Fitted fluorescence DIFF spectra are shown in black. The Pearson correlation coefficient for each spectrum is given in the legend. Vertical-axis unit is the flux density per pixel. \label{sup:CO2-spectra-leading}}
\end{figure}

\section{Cross-correlation technique}
\label{app:C}
\begin{figure}[h]
\centering
\includegraphics[width=8cm]{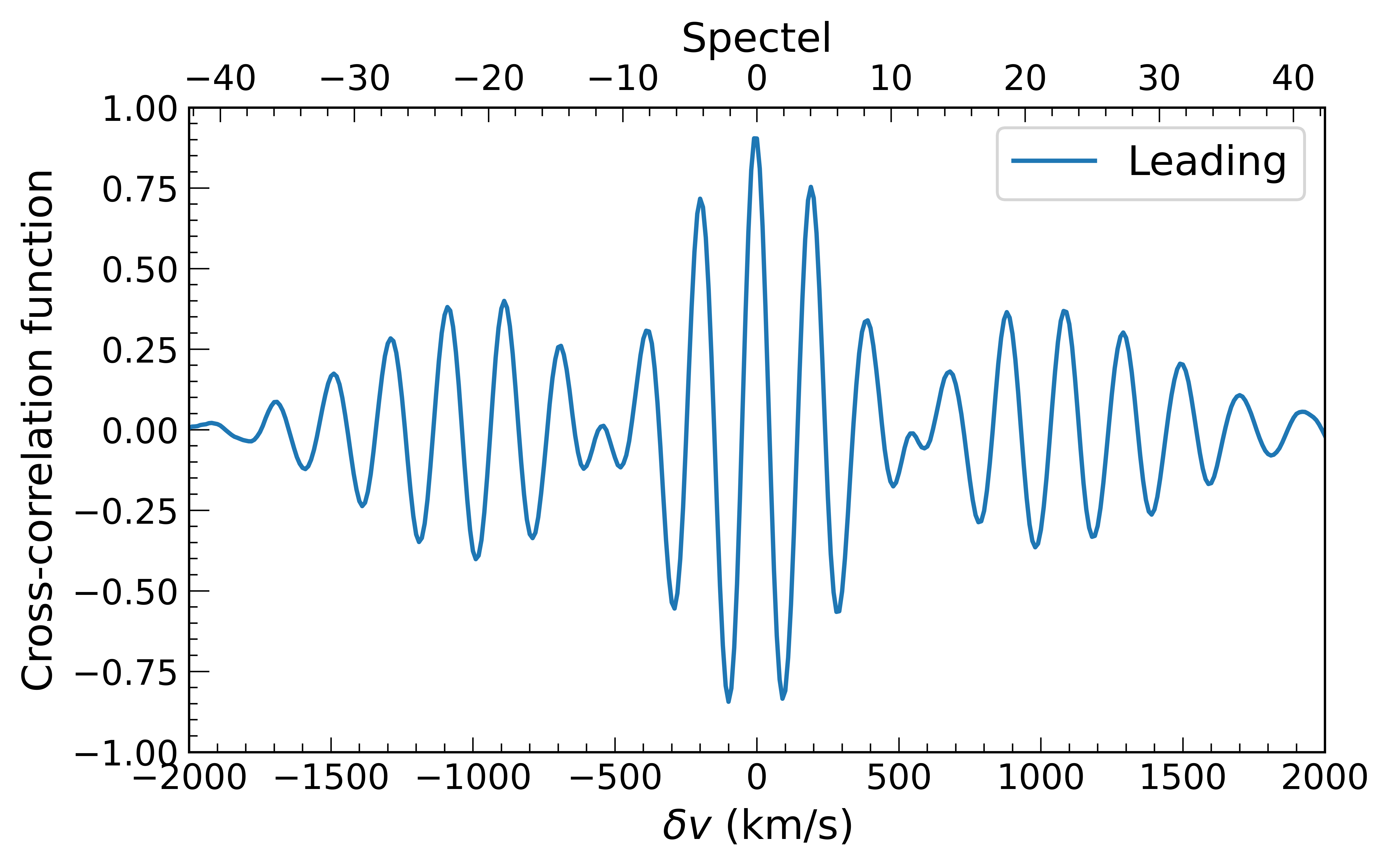}
\includegraphics[width=8cm]{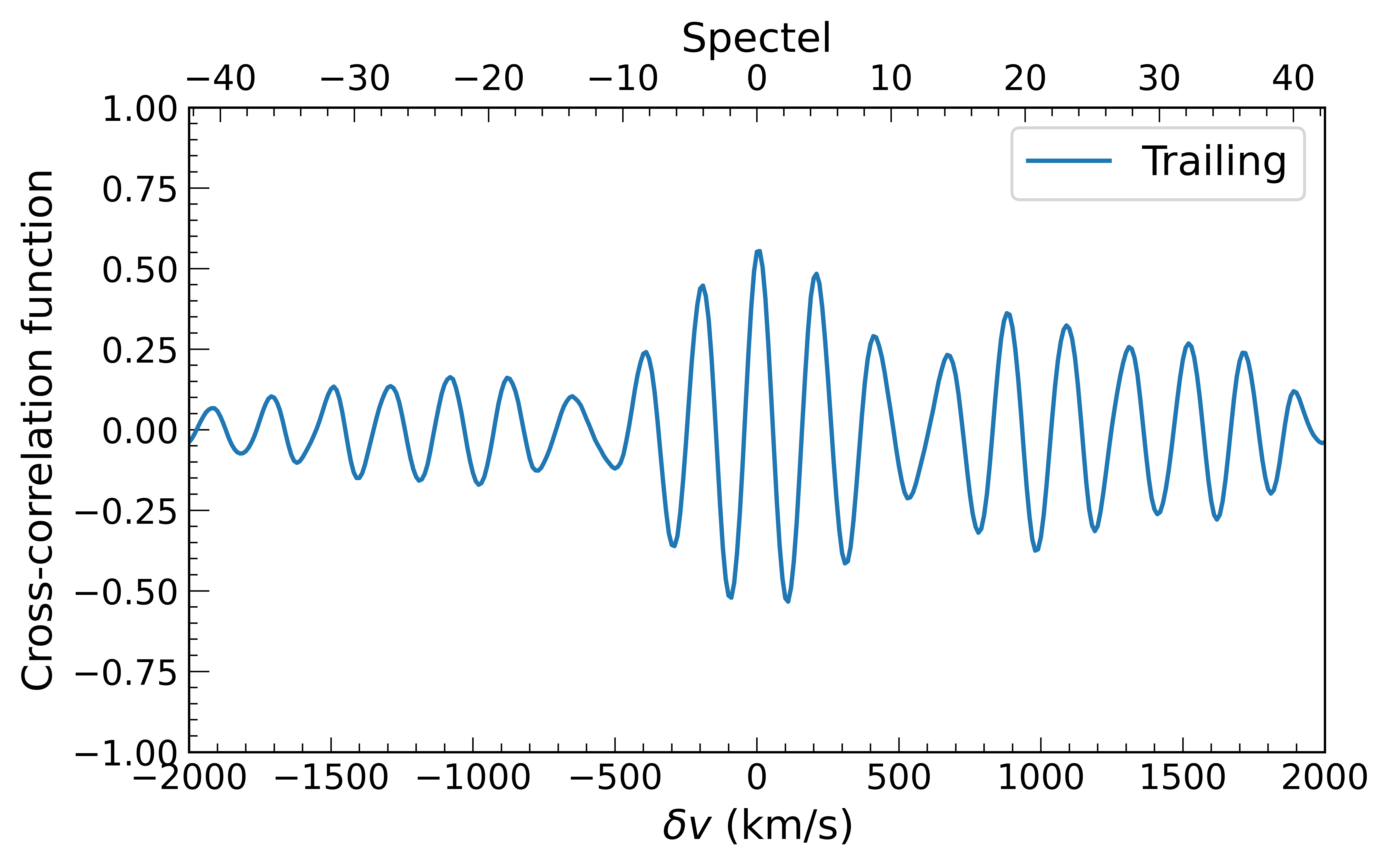}
\caption{Cross-correlation functions (CCFs). Top: from averaged data at latitudes $>$ 45$^\circ$N on the leading hemisphere. Bottom: from averaged data at latitudes 30--60$^\circ$S on the trailing hemisphere. The forward model for computing the cross-correlation is a fluorescence CO$_2$ spectrum. In both cases the CCF peaks at $\delta v$ = 0, indicating CO$_2$ exosphere detection. The maximum of the CCF is lower for the trailing side due to a fainter CO$_2$ signal.  \label{sup:CCF} }
\end{figure}

\begin{figure*}[h]
\centering
\includegraphics[width=16cm]{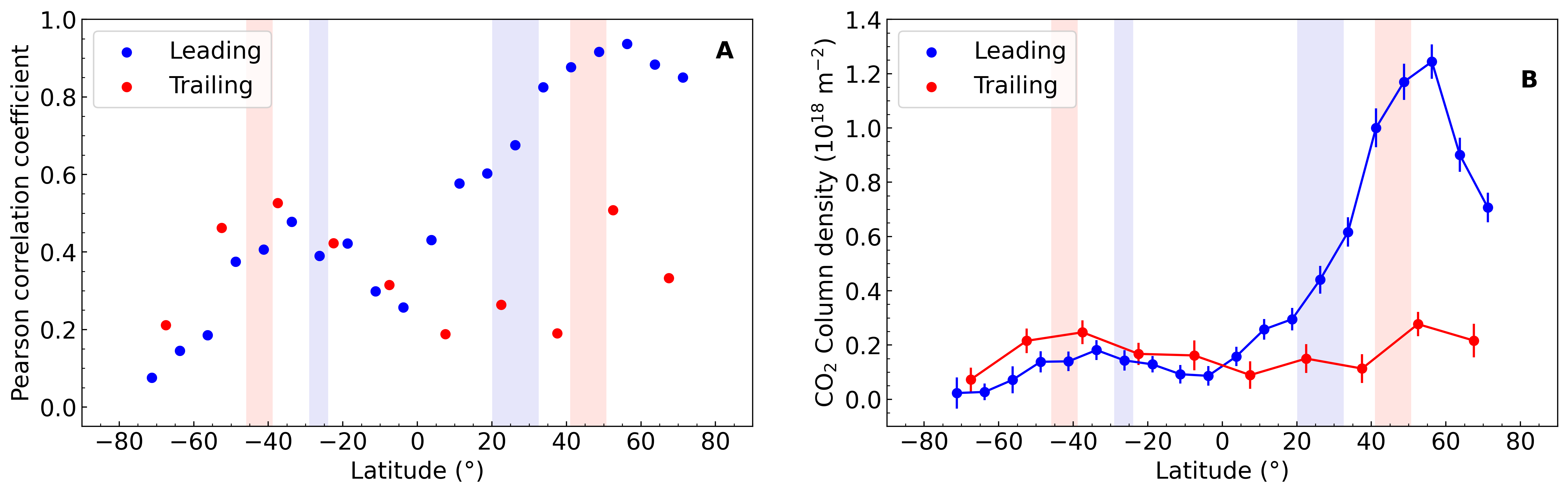}
\includegraphics[width=16cm]{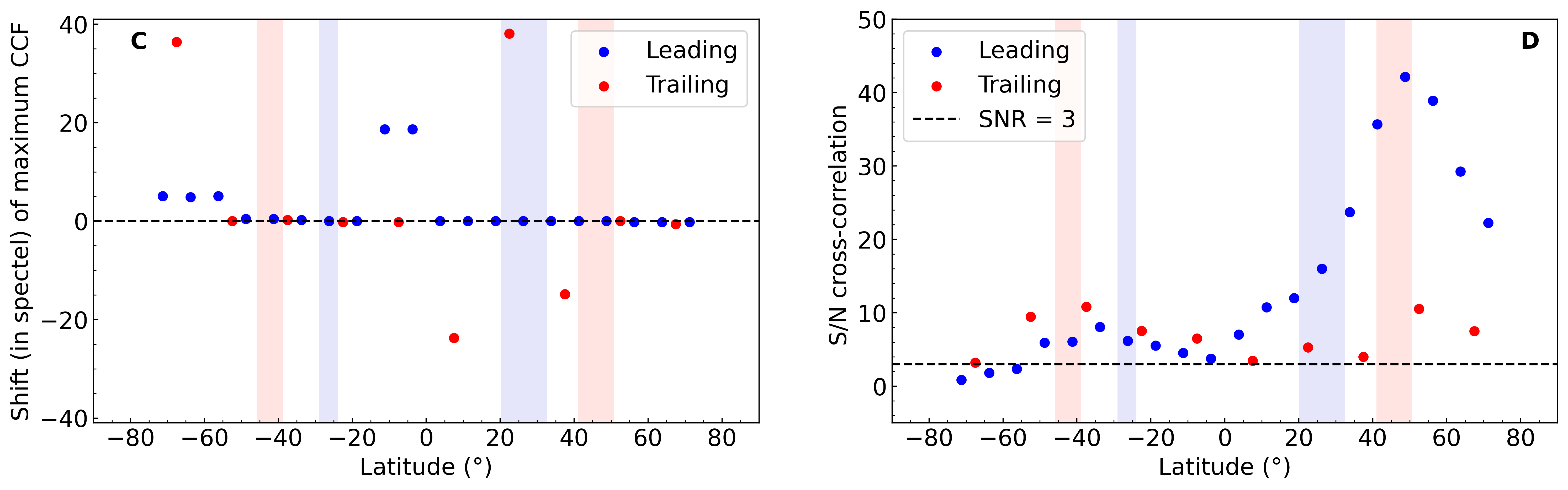}
\caption{Study of 4.22--4.295 $\mu$m spectra averaged over latitude bins. A) Pearson correlation coefficient between residual CO$_2$ DIFF spectra and a synthetic DIFF spectrum which uses a CO$_2$ fluorescence synthetic spectrum at $T$ = 105 K as input. B) CO$_2$ line-of-sight column density retrieved from the fit of CO$_2$ DIFF spectra. C) shift, in fraction of spectel, of the maximum of the cross-correlation function (CCF) between CO$_2$ DIFF spectra and the synthetic DIFF spectrum; secure detection is indicated when the shift is close to zero. D) S/N of the cross-correlation function at shift = 0. The S/N is obtained by computing the CCF$_{noise}$ obtained by using as input a simulated noisy spectrum, with the r.m.s deduced from the ERR entry in the Level 3 data cube, and scaled in $\sqrt{n}$, where n is the number of averaged spaxels (S/N = CCF/CCF$_{noise}$ at shift = 0). In all plots, blue and red symbols correspond to the leading and trailing sides, respectively. The blue (resp. pink) vertical domains show the latitude range of the open-closed-field line boundaries for the leading and trailing sides, respectively, restricted to longitudes of 10--130$^{\circ}$ W (leading) and  210--330$^{\circ}$ W (trailing).  \label{lat-distribution} }
\end{figure*}

\noindent
We used the cross-correlation technique to obtain additional criteria for confirming weak  CO$_2$ detections. This method is widely used, e.g., to search for molecular signatures in exoplanet spectra \citep{Snellen2010,Malin2023}. We computed the cross-correlation function (CCF) between the Ganymede DIFF spectra and the forward model over a velocity range (-3000, 3000) km/s (in total 127 spectral resolution elements, aka spectels) using velocity steps $\delta v$ spaced by 10 km/s (0.21 spectels). Cross-correlating the forward model with itself, the maximum of the autocorrelation function is obviously at $\delta v$ = 0. Because of the periodicity in frequency of the CO$_2$ ro-vibrational lines, which are equally spaced every 0.003 $\mu$m ($\sim$ 4 spectels), the autocorrelation function and CCFs present strong secondary peaks (reaching 80 \% of the maximum for the closest secondary peaks) spaced by the corresponding value in velocity units (Fig.~\ref{sup:CCF}). The  criterion used for confident CO$_2$ exosphere detection is that the maximum of the CCF stands close to $\delta v$ = 0, namely is shifted by at most one spectel element. 

We estimated the signal-to-noise ratio (S/N$_{\rm CCF}$) of the CCF at $\delta v$ = 0 to quantitatively measure the confidence level of the detections. For that purpose, we generated synthetic spectra adding random Gaussian noise to the Ganymede average CO$_2$ absorption band observed on the leading side obtained from low-pass filtering. As there is no possibility of estimating the noise level from the acquired spectra themselves (noise-like features are dominated by residuals in solar lines subtraction), we used the error cube given in the Level 3 hyperspectral data cubes to set the r.m.s., and assumed that it scales as $\sigma$/$\sqrt{n}$, when $n$ spaxels are averaged. The generated synthetic spectra were then processed as for the Ganymede data, and  the resulting DIFF spectra were cross-correlated with the forward model to obtain a cross correlation function CCF$_{\rm noise}$ for a spectrum containing only noise. We then measured the standard deviation of the CCF$_{\rm noise}$ curve. Eighty random-noise synthetic spectra were processed in this way, to derive a representative standard deviation $\sigma_{\rm CCFnoise}$ from the median of the values obtained for each shot. For processed Ganymede spectra, S/N$_{\rm CCF}$ is obtained by dividing the CCF at $\delta v$ = 0 by $\sigma_{\rm CCFnoise}$. 

The cross-correlation technique was applied on data averaged over latitude bins (Figs.~\ref{collat-distribution}, \ref{sup:CCF},~\ref{lat-distribution}).  Spectra with a Pearson correlation coefficient  $C_p$ $>$ 0.3 all display a CCF with S/N$_{\rm CCF}$ $>$ 5, and a maximum shifted by less than 1 spectel element. From those criteria, the CO$_2$ exosphere is detected with good confidence both in the northern and southern hemispheres of the leading and trailing sides of Ganymede. On the other hand, the S/N (and inferred CO$_2$-gas signal) is low near the equator for both hemispheres, indicating a more tenuous CO$_2$ exosphere in these regions. The decrease of the CO$_2$ signal observed at the most polar latitudes could be related to PSF blurring since spaxels probing extreme polar latitudes are near the limb of Ganymede disk. However, this decrease is still observed after deconvolution with modeled NIRSpec PSFs (Fig.~\ref{sup:figPSF1}).

\section{SMART-EVE excitation/radiative transfer model}
\label{app:D}
\noindent
A non-LTE Stochastic Modeling of Atmospheric Radiative Transfer-Exospheric Vibrational Excitation (SMART-EVE) has been developed to calculate the ro-vibrational populations of the  (1) H$_2$O $\nu_2$ mode (010) at 6.25 $\mu$m, and (2) CO$_2$ $\nu_3$ mode (001) at 4.25 $\mu$m. SMART-EVE solves the locally defined statistical equilibrium equations (SEEs) for all the energy levels considered and the radiative transfer equations (RTEs) for all the bands connecting these levels. Due to non-linearities arising from radiative transfer and/or collisional coupling, the resulting equation system is solved iteratively using the Accelerated Lambda iteration approach which alternates SEE calculations involving all the energy levels with RTE calculations involving all atmospheric layers.

The 1D model of the atmosphere is described by the kinetic gas temperature, assumed vertically uniform, and the gas density which follows hydrostatic equation. The model assumes that the atmosphere is illuminated by the Sun from the top, and by the surface thermal emission and the reflected solar and atmospheric radiance from below. The model parameters are: the column density and kinetic temperature $T_{\rm kin}$ of the atmosphere, the surface temperature $T_s$ and reflectance factor $Ref$, and the heliocentric distance. Electron-impact excitation of CO$_2$ is not considered, as most likely insignificant (Supplementary Information). 

The radiative processes considered are spontaneous and stimulated emissions, absorption of the upward thermal flux, incident solar, and reflected solar and atmospheric irradiance from the surface, as well as exchanges between layers. A single collisional process is considered for the vibrational state, its vibration-to-translation (V--T) relaxation/excitation in intermolecular collisions. However, vibrational de-excitation by collisions is insignificant in Ganymede's exosphere (see Appendix~\ref{app:G}).  It is assumed that rotational levels are at LTE at all altitudes with a rotational temperature $T_{\rm rot}$ = $T_{\rm kin}$.

The model is run from the surface up to 100 km, with 1-km thick layers. The spectral data are taken from the HITRAN database \citep{HITRAN}. We considered only (010)--(000) (H$_2$O) and (001-000) (CO$_2$) vibrational transitions, with a total number of lines of 1017 for H$_2$O and 129 for CO$_2$. The solar spectrum was taken from Paper I.

\begin{figure}[ht]
\includegraphics[width=9cm]{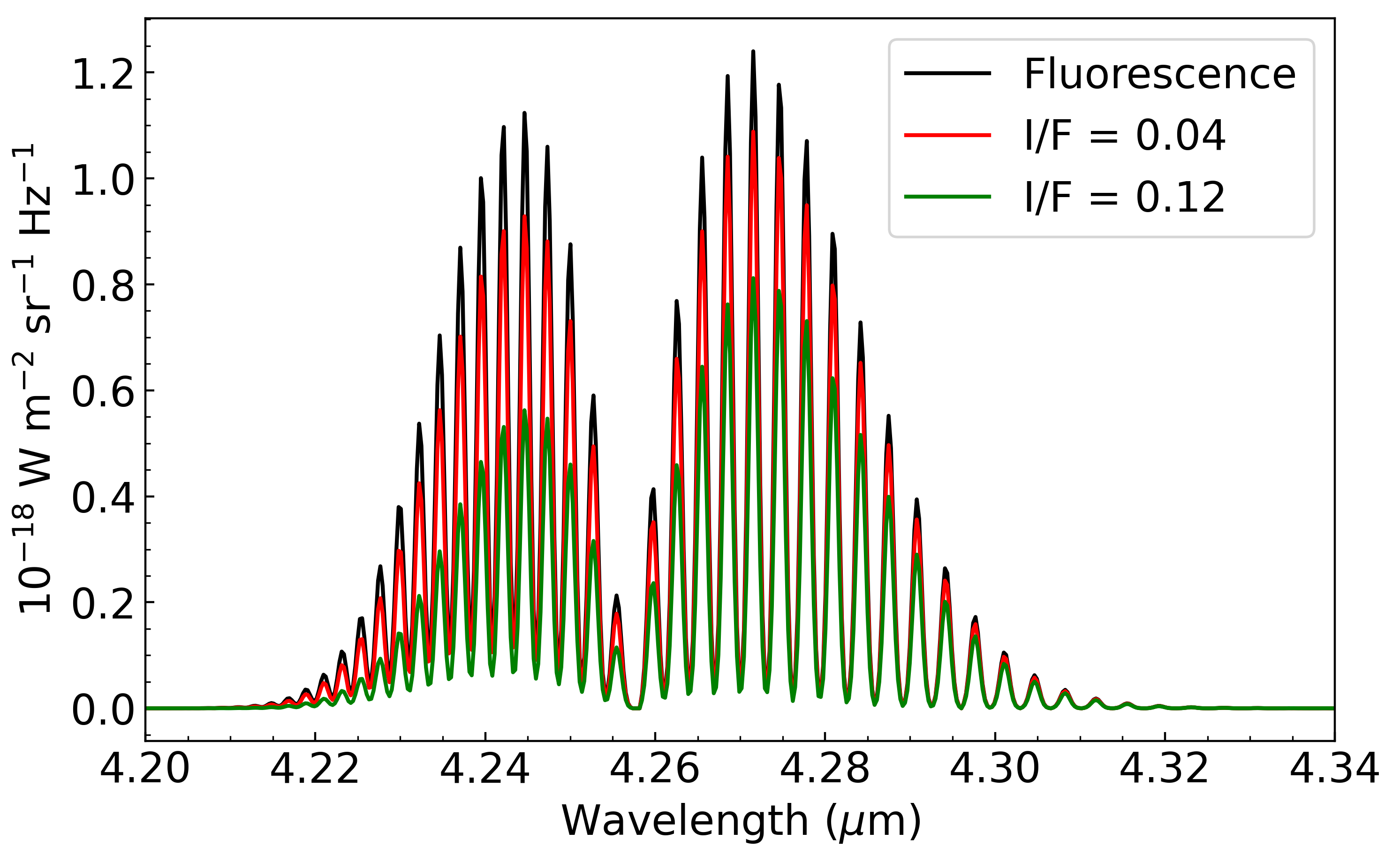}
\caption{CO$_2$ simulated nadir spectra. Input model parameters are: CO$_2$ column density $N$(CO$_2$) = 10$^{18}$ m$^{-2}$, surface temperature $T_s$ = 145 K, gas rotational temperature $T_{\rm rot}$ = 105 K, for $I/F$ values of 0.04 (red) and 0.12 (green). The fluorescence spectrum (black) corresponds to $I/F$ = 0 and  $T_s$ = 0 K. Ganymede's exosphere is described by hydrostatic equilibrium.   
\label{fig:sup-CO2modelspectra}}
\end{figure}

\section{Determination of CO$_2$ column density}
\label{app:E}
\noindent
CO$_2$ line-of-sight column densities were derived using a two-step approach. First, column densities were derived under the assumption of fluorescence equilibrium and optically thin lines. In the second step, a correction factor was applied, using prescriptions obtained from the SMART-EVE model described above. 

For CPU-time considerations, we used the Planetary Spectrum Generator (PSG) \citep{2018JQSRT.217...86V} for the first step. Optically thin CO$_2$ $\nu_3$ band fluorescence spectra at high spectral resolution (0.1 cm$^{-1}$) were generated (outputs for cometary atmospheres) and downloaded for a range of rotational temperatures in steps of 1 K (and fixed column density). As done for the forward model described above, they were combined with Ganymedes's solid-state CO$_2$ band and synthetic DIFF spectra were computed. This bank of synthetic spectra was utilized to fit the Ganymede DIFF spectra using the Levenberg-Marquardt algorithm (we used the $limfit$ Python package), with a normalizing factor as free parameter. The rotational temperature was set as a free parameter for the analysis of the high S/N spectra of the north hemisphere of the leading side (Fig.~\ref{fig:sup-Trot}), and fixed to 105 K elsewere (i.e., equal to the inferred value in leading north hemisphere, Appendix~\ref{app:H}). The CO$_2$ column density was derived from the inferred normalizing factor. For the uncertainty in the CO$_2$ column density, we used that provided by the $limfit$ package and derived from the covariance matrix.

Model simulations for the CO$_2$ $\nu_3$ band  (Fig.~\ref{fig:sup-CO2modelspectra}) show that spectral profiles from Ganymede's exosphere are expected to be less intense than in the assumption of cometary-like fluorescence emission (case $Ref$ = 0, $T_s$ = 0), which neglects absorption of surface reflected solar radiation by CO$_2$ gas, reflection of CO$_2$ gas emission on the surface, and surface thermal radiation. While this latter process is not significant at 4.25 $\mu$m, the other two processes affect the total band intensity of the $\nu_3$ band $BA$ according to:
\begin{equation}
\label{eq:eq1}
    BA = BA_{\rm 0} \times (1 - 4.2 \times I/F), \phantom{x} {\rm for~ N(CO_2) = 10^{17}~m^{-2} }
\end{equation}
\begin{equation}
\label{eq:eq2}
    BA = BA_{\rm 1} \times (1 - 3.6 \times I/F), \phantom{x} {\rm for~ N(CO_2) = 10^{18}~m^{-2} }
\end{equation}
\begin{equation}
\label{eq:eq3}
    BA = BA_{\rm 2} \times (1 - 2.1 \times I/F), \phantom{x} {\rm for~ N(CO_2) = 10^{19}~m^{-2} }
\end{equation}

\noindent 
where $I/F$ is the radiance factor on the surface, and  $BA_{\rm 0}$, $BA_{\rm 1}$, $BA_{\rm 2}$ are equal to 4.10$\times$10$^{-8}$, 
4.27$\times$10$^{-7}$, and 3.84$\times$10$^{-6}$ W m$^{-2}$ sr$^{-1}$, respectively. $BA_{\rm 0}$ is consistent with the value of 4.06$\times$10$^{-8}$ W m$^{-2}$ sr$^{-1}$ retrieved from PSG  \citep{2018JQSRT.217...86V} for optically thin cometary-like fluorescence emission at 4.95 au from the Sun with $N$(CO$_2$) = 10$^{17}$~m$^{-2}$. Eqs~\ref{eq:eq1}--\ref{eq:eq3} were obtained from multiple simulations fixing $T_s$ = 145 K and $T_{\rm rot}$ = 105 K, and varying $I/F$. 

We applied a correction factor  intermediate between Eqs~\ref{eq:eq1} and ~\ref{eq:eq2} (i.e., slope of --4.0 for the dependence with $I/F$) on the CO$_2$ column density inferred assuming fluorescence equilibrium, using radiance factors $I/F$ at $\sim$ 4.25 $\mu$m measured from JWST (Paper I).

\section{Electron impact excitation of CO$_2$}
\label{app:F}
We made estimations for electron-impact excitation of the CO$_2$ $\nu_3$ band using cross-sections from \citet{Itikawa2002}. Electron populations were assumed to follow a Maxwellian distribution around a mean temperature. For the total electron density and the temperature, values that explain the highest UV brightnesses (OI 1356 Å) of 1000 R \citep{Waite2024} measured for Ganymede were used. Specifically, we assumed an electron temperature of 20 eV and a high number density of 2500 cm$^{-3}$.  CO$_2$ emission from this process is found to be more than two orders of magnitude lower than fluorescence emission. The CO$_2$ $\nu_3$ band could be excited by much cooler electrons, well below 10 eV. However, information on these cold electrons is missing.  Cross-sections  for electron-impact excitation of the CO$_2$ $\nu_3$ band increase with decreasing electron energy \citep{Itikawa2002}. Using an electron temperature of 1 eV and the same number density, CO$_2$ emission from electron impact excitation is only two times higher than for 20 eV electrons. 

\section{CO$_2$ $\nu_3$-band collisional relaxation}
\label{app:G}
We have evaluated the role of de-excitation of the CO$_2$ $\nu_3$ band by collisions with H$_2$O, O$_2$, and CO$_2$ versus spontaneous emission. The result is that these processes are not significant in Ganymede's exosphere. The rate for collisional de-excitation of the CO$_2$ $\nu_3$ band via CO$_2$-H$_2$O collisions is 1.2$\times$10$^{-13}$ cm$^3$/s at 120 K \citep{Blauer1973}.
The H$_2$O number density is at most 1.4$\times$10$^{10}$ cm$^{-3}$ at the surface, derived from hydrostatic equilibrium for a water column density of 5$\times$10$^{20}$ m$^{-2}$ \citep{Roth2021}. This gives a collision rate of at most 1.7$\times$10$^{-3}$ s$^{-1}$, which is much lower than the spontaneous emission rate of the ro-vibrational levels (on the order of 400 s$^{-1}$). So the quenching is negligible.

The rates for de-excitation of CO$_2$ $\nu_3$ band via collisions with O$_2$ and CO$_2$ are much lower than for CO$_2$-H$_2$O collisions. So these collisional processes are still less significant.

\section{Rotational temperature of CO$_2$}
\label{app:H}

\noindent
For fluorescence emission, the relative intensities of the ro-vibrational lines of the CO$_2$ $\nu_3$ band are set by the population distribution in the ground vibrational state, described by a Boltzmann distribution at the rotational temperature $T_{\rm rot}$.  
Rotational temperatures of CO$_2$ derived on the northern latitudes ($>$30$^\circ$ N) of the leading hemisphere are shown in Fig.~\ref{fig:sup-Trot} and are on the order of 105--110 K (see Appendix~\ref{app:B} for details on how $T_{\rm rot}$ was derived). This is slightly lower than the surface temperature of Ganymede at these latitudes (from 120 to 140 K, Fig.~\ref{fig:temp-distributiom}). This rotational temperature possibly reflects the kinetic temperature of the exosphere at low altitudes where collisions with the major gas (H$_2$O or CO$_2$) are still efficient enough to thermalize CO$_2$ molecules. Alternatively, it might reflect the rotational energy of the CO$_2$ molecules when they left the surface, and be representative of the temperature of the surface where CO$_2$ molecules were released. The CO$_2$ molecule has no dipole moment, so radiative rotational decay within the ground vibrational state does not take place. The rotation temperature is expected to increase with residence time in the atmosphere  due to radiative decay from the excited vibrational states. However, one should mention that during their residence time in the exosphere (at most 18 h, which is the CO$_2$ lifetime set by electron-impact ionization), CO$_2$ molecules undergo at most 7 fluorescence cycles. Based on fluorescence calculations for cometary atmospheres, CO$_2$ molecules reach a warm fluorescence equilibrium only after about 3500 fluorescence cycles at 5 au from the Sun \cite{Crovisier1987}. In summary, the measured $T_{\rm rot}$ should reflect the thermal environment where last thermalizing collisions occurred, or the excitation state of the molecules when they left the surface.  

\begin{figure}[h]
\includegraphics[width=9cm]{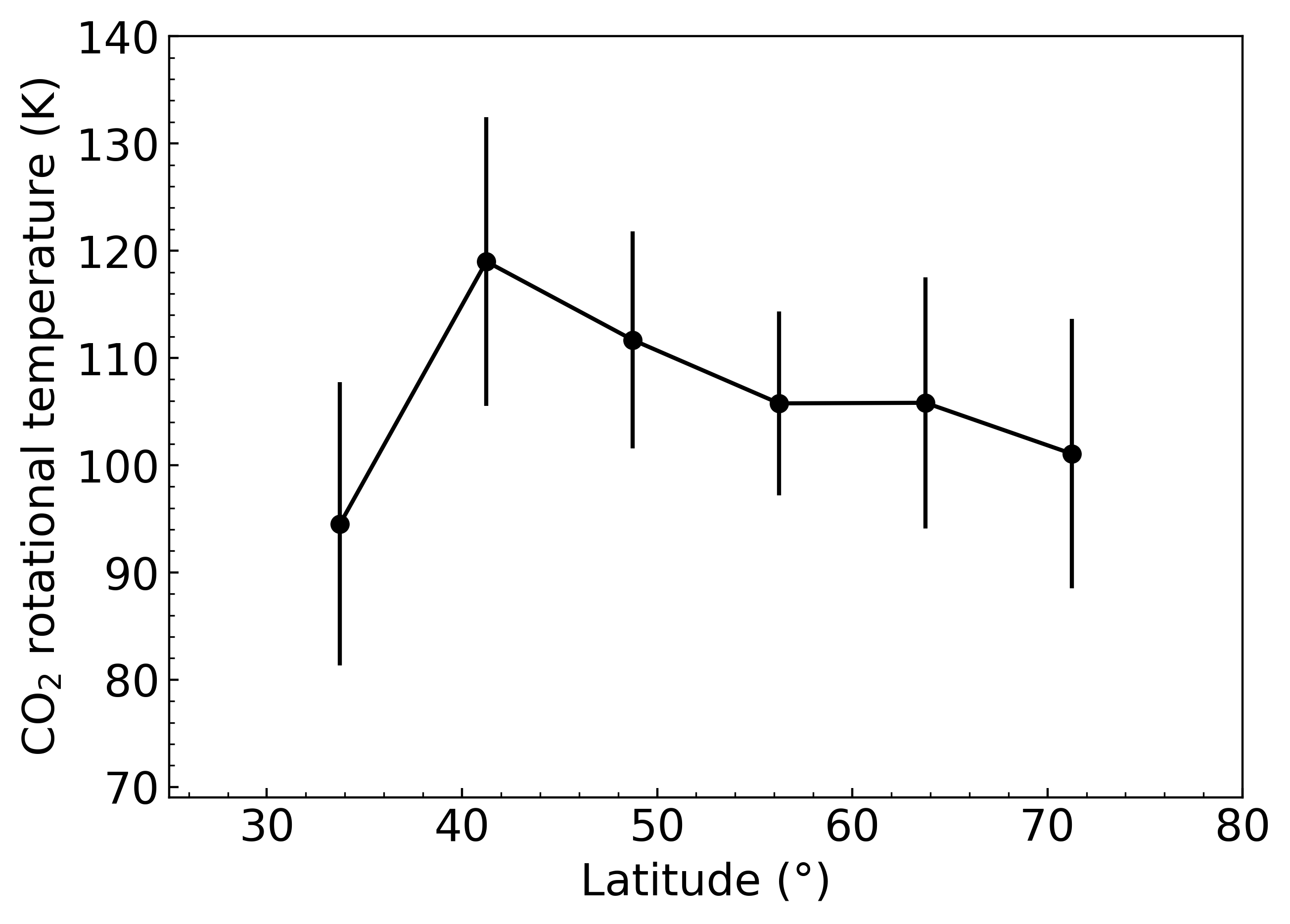}
\caption{CO$_2$ rotation temperature in the north hemisphere of the leading hemisphere. Spaxels within latitude bins of width 7.5$^\circ$ were averaged. Values were derived from the fitting of DIFF spectra using as model fluorescence emission with $T_{\rm rot}$ as a free parameter (Appendix~\ref{app:B}). The weighted mean value is $T_{\rm rot}$ = 107$\pm$5 K.  \label{fig:sup-Trot}}
\end{figure}

\section{H$_2$O analysis}
\label{sec:H2O}

\begin{figure}[ht]
\centering
\includegraphics[width=9cm]{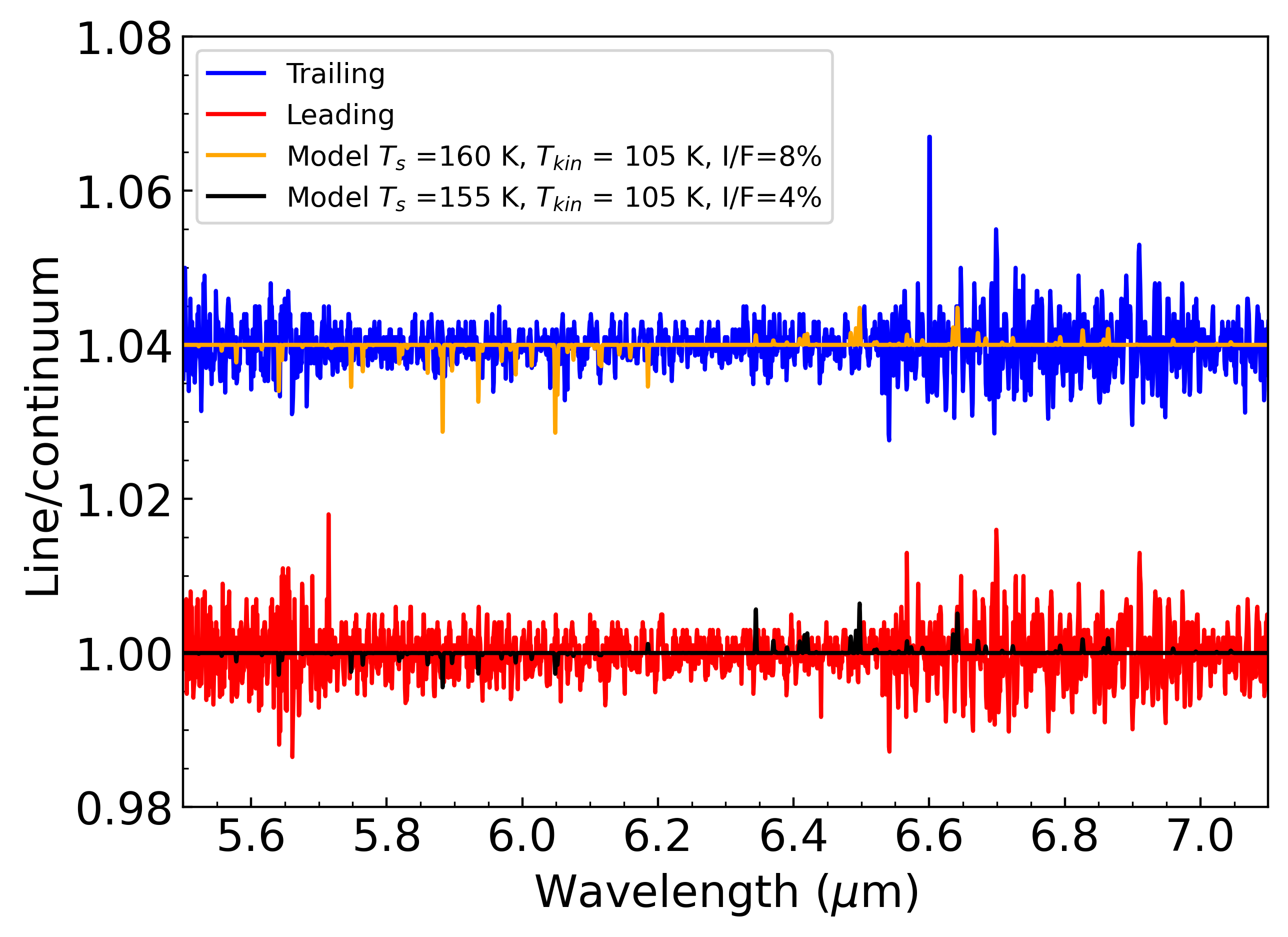}
\caption{Continuum-divided spectra of Ganymede observed with MIRI and synthetic H$_2$O spectra.  Spectra for the leading and trailing sides are shown in red and blue, respectively, with the spectrum of the trailing side shifted vertically. Spaxels for which the solar zenith angle is less than 15$^\circ$ at the center of the spaxel have been averaged. Synthetic spectra are superimposed, with input parameters indicated  in the legend (Appendix~\ref{sec:H2O}) and $N$(H$_2$O) =  10$^{20}$~m$^{-2}$. The Ganymede spectra do not show any hint of H$_2$O lines. 
\label{fig:sup-H2O}  }
\end{figure}
\noindent
We analysed MIRI/MRS Channel-1 spectra obtained by either 1) averaging spaxels around the subsolar point, namely eight spaxels for which the solar zenith angle (SZA) is less than 15$^\circ$ at the center of the spaxel; 2) averaging seven spaxels covering the region of the leading hemisphere where the CO$_2$ exosphere is prominent. Neither one shows any hint of the presence of water lines (Fig.~\ref{fig:sup-H2O}) and the non-detection of  H$_2$O was further confirmed by applying the cross-correlating technique using a forward model of a synthetic spectrum of H$_2$O computed with the SMART-EVE code. 

In the exosphere of Ganymede, the thermal radiation from the surface competes with the Sun's direct radiation for the excitation of the H$_2$O $\nu_2$ band at 6.2~$\mu$m (Paper I). In addition, in nadir viewing, absorption of the radiation from Ganymede's surface by the H$_2$O exosphere might compete with $\nu_2$ fluorescence emission, so that the band might be in absorption under certain conditions and a simple fluorescence model would not apply.
Therefore, we used the SMART-EVE model described above to derive upper limits on the H$_2$O column density ($N$(H$_2$O)).

Radiance factor values of 0.04 and 0.08 were assumed for the leading and trailing hemispheres, respectively (see Fig. 26 of Paper I).
The surface temperature was chosen such that the brightness temperature in the synthetic spectra matches the $T_{BB}$ value at 6.2 $\mu$m measured on the MIRI spectrum. For the "subsolar" spectra (SZA$< 15^{\circ}$), $T_s$ (and $T_{BB}$) are closed to 155 K (leading) and 160 K (trailing). For the spectrum extracted at the position of the CO$_2$ northern source, $T_s$ is
about 140 K. Synthetic spectra for the subsolar region are shown in Fig.~\ref{fig:sup-H2O}. Ro-vibrational lines at $\lambda$ $<$ 6.2 $\mu$m are expected in absorption whereas emission lines are expected at $\lambda$ $>$ 6.2 $\mu$m. As a matter of fact, the vibrational temperature of the $\nu_2$ band (mainly controlled by solar IR pumping) is $\sim$ 158--159 K, i.e. very close to the brightness temperature near 6.4 $\mu$m. The change from absorption to emission regimes is related to the fact that the vibrational temperature is close to the wavelength-dependent brightness temperatures near 6.2 $\mu$m.

Table~\ref{tab:H2O} presents measured 1-$\sigma$ uncertainties for the H$_2$O band area in the 5.7--6.2 and 6.2--7.1 $\mu$m spectral ranges, considering the 10--15 expected strongest lines (with intensities $> 0.2$ the intensity of the strongest line).   From the measured band areas in each wavelength window, we derived a 3-$\sigma$ upper limit for line-of-sight $N$(H$_2$O), using SMART-EVE model with appropriate parameters. The results were then combined. The final results are given in Table~\ref{tab:H2O} for two values (105 and 130 K) of the H$_2$O rotational temperature.

\section{Open-closed-field-line Boundary}
\label{app:J}
\noindent
The OCFB location is determined through magnetohydrodynamic modeling of Ganymede's magnetosphere similar to the method described in \citet{Duling2022}. Due to the variation of the upstream magnetic field and plasma density at Ganymede's position relative to the Jovian current sheet, the OCFB location can oscillate with an amplitude ranging between 2 to 6 degree latitude during Jupiters’s approximative 10-h rotation period \citep{Saur2015}. We modeled the OCFB analogous \citep{Duling2022} by adapting the upstream conditions to estimates for the times of the JWST observations. During the observation of the leading side, Ganymede was  above the center of the current sheet and we used 61 amu/cm$^3$ and (--11, --66, --79)nT for the upstream plasma mass density and magnetic field respectively. During the trailing side observation Ganymede was {\bf at the center} of the current sheet and we used 100 amu/cm$^3$ and (--18, --6, --79)nT. 

\section{CO$_2$, H$_2$O Ganymede exospheric model}
\label{app:K}
\noindent
We simulated the CO$_2$ exosphere using the Exospheric Global Model (EGM), a multi-species Monte Carlo model describing the fate of test particles in a gravitational field, interacting with a surface or an atmosphere and subject to sources of ionization and dissociation. EGM has been extensively used to model the exospheres of H$_2$O and related species (e.g., O$_2$, H) in various objects, in particular Ganymede \citep{Leblanc2017,Leblanc2023}. We considered two possible mechanisms of ejection of the CO$_2$ molecules from the surface: i) sputtering, i.e. ejection following bombardment of H$_2$O ice containing CO$_2$ molecules by the incident Jovian energetic ions and electrons; and ii) sublimation of the CO$_2$ molecules from Ganymede surface. We considered the release of CO$_2$ either from the sublimation of pure CO$_2$ ice or from the sublimation of H$_2$O ice containing CO$_2$ molecules. We took into account that CO$_2$ molecules re-impacting cold areas of the surface eventually recondense. The H$_2$O exosphere is also computed. The calculated images from the simulations (e.g., Fig.~\ref{fig:model-exosphere}) consider the orbital position of Ganymede around Jupiter at the time of the JWST observations and the viewing geometry of JWST observations (for the observations of the leading hemisphere, sub-observer coordinates were 2$^{\circ}$N, 72$^{\circ}$W). Convolution with a FWHM = 0.185'' PSF is applied (Appendix~\ref{app:M}). Line-of-sight CO$_2$ column densities averaged over latitude bins of 7.5$^\circ$ or 15$^\circ$ were computed for comparison with the data shown in Fig.~\ref{collat-distribution}. For the study of the CO$_2$ exosphere above the north polar cap of the leading hemisphere, we extracted the longitudinal variation of the CO$_2$ column density for latitudes in the range 42--62$^\circ$ N.      

{\bf Sublimation:}
For a CO$_2$ release associated with the sublimation of water ice, the release rate is in proportion with the H$_2$O sublimation rate (cm$^{-2}$ s$^{-1}$): 
\begin{equation}
F({\rm CO_2}) = f_c \times q_{\rm CO_2} \times 2.17~10^{32} \frac{e^{\frac{-U_0}{k_B T_s}}}{\sqrt{T_s}}. 
\label{Eq:sublimationCO2H2O}
\end{equation}

\begin{equation}
F({\rm H_2O}) =  q_{\rm H_2O} \times 2.17~10^{32} \frac{e^{\frac{-U_0}{k_B T_s}}}{\sqrt{T_s}}, 
\label{Eq:sublimationH2O}
\end{equation}

\noindent
where $U_0/k_B$ = 5950 K, and $q_{\rm H_2O}$ is the areal surface fraction of H$_2$O. The relative abundance of CO$_2$ in the sublimated gases (in number) is $q_{\rm CO_2}$/$q_{\rm H_2O}$.  The description of $F$(H$_2$O) follows \citet{Leblanc2023}. $T_s$ is the surface temperature. $f_c$ is a factor introduced to reproduce the CO$_2$ JWST data.

\noindent
For the sublimation of CO$_2$ ice, the sublimation rate (cm$^{-2}$ s$^{-1}$) is given by:
\begin{equation}
F({\rm CO_2}) = f_c \times \frac{N_{\rm tot} q_{\rm CO_2}}{\tau_0 \sqrt{T_s}} e^{\frac{-U_1}{k_B T_s}},
\label{Eq:sublimationCO2}
\end{equation}
\noindent
where $N_{\rm tot}$ = 10$^{18}$ cm$^{-2}$ s K$^{0.5}$ is determined from a fit of  the polynomial relation of CO$_2$-ice vapor pressure with temperature \citep{fray2009} and using $U_1/k_B$ = 2860 K (surface binding energy for CO$_2$ on H$_2$O ice, \citep{Sandford1990}) and $\tau_0$ = 3.45 10$^{-13}$ s \citep{Sandford1990}. $f_c$ is a factor introduced to reproduce the JWST data.

{\bf Sputtering:} The ejection of CO$_2$ molecules by sputtering is described by the efficiency by which CO$_2$ molecules are emitted from a surface when an incident ion or electron impacts the surface with a given energy. We hypothesized that CO$_2$ molecules are trapped in/on H$_2$O ice, so we assumed that the sputtering yield follows the same temperature dependence as for H$_2$O, and used the same definition as for H$_2$O \citep{Cassidy2013,Leblanc2023}:
\begin{equation}
Y({\rm CO_2}) = f_c \times Y_0 \times (1+Y_{00}\times e^{\frac{-U_{00}}{k_B T_s}}),
\label{eq:yield}
\end{equation}
\noindent
with $Y_0$ = 1200. In fact, we made the assumption that CO$_2$ molecules are released into the exosphere along with H$_2$, O$_2$, H$_2$O$_2$ and H$_2$O molecules ejected when pure H$_2$O ice is bombarded, therefore $U_{00}$ and $Y_{00}$ are set to be the same as for the bombardment of pure H$_2$O ice \citep{Fama2008}: $U_{00}$ = 0.06 eV (700 K), $Y_{00}$ = 220. As for the energy and angular distributions of the CO$_2$ molecules when ejected from the surface, we followed the approach used for sputtered O$_2$ in \citet{Leblanc2017} and assumed a Maxwell–Boltzmann energy distribution at the local surface temperature. Regarding the intensity and spatial distribution  of the Jovian ions impacting the surface, we assumed a given ion flux of 10$^6$ particles/cm$^2$/s derived from \citet{Cassidy2013} as in \citet{Leblanc2017}, impacting Ganymede’s surface only in the open-field-line regions.  Electron impacts are not considered. The flux of the CO$_2$ molecules released at a given position on Ganymede's surface is therefore the product of $Y$(CO$_2$) (Eq.~\ref{eq:yield}) times $q_{\rm CO_2}$ times the flux of impacting particles. The flux of H$_2$O follows the same equation, using $q_{\rm H_2O}$ instead. A multiplying factor $f_c$ is introduced in Eq.~\ref{eq:yield} with respect to \citet{Cassidy2013} and \citet{Leblanc2023} that is adjusted to reproduce the CO$_2$ column density measured by JWST. This factor is also applied to the flux of H$_2$O sputtered molecules. In our model, sputtering of water ice is assumed to release mainly H$_2$O molecules with a ratio H$_2$O/O$_2$ = 20 \citep{Leblanc2017,Cassidy2013}.

{\bf Surface adsorption:} To determine the fate of a CO$_2$ molecule re-impacting the surface, we define the CO$_2$ residence time at the surface as:

\begin{equation}
\tau = \tau_0 e^{\frac{U_1}{k_B T_s}}.   
\end{equation}  

\noindent
where $U_1$/$k_B$ = 2860 K is the binding energy for CO$_2$ adsorbed on H$_2$O ice, and $\tau_0$ = 3.45 10$^{-13}$ s \citep{Sandford1990}. We considered that when the CO$_2$ residence time is longer than the model time step (0.25 s), any particle hitting the surface gets trapped in the surface. We then calculated at each step and for each trapped particle a probability to be re-ejected as being equal to the ratio between the time step of the simulation and the residence time calculated from the surface temperature at the position of the particle. This probability is then compared to a random number between 0 and 1 and if larger than this random number, the particle is re-emitted into the exosphere. We checked that the results are not sensitive to the model time step.

\begin{figure*}[!ht]
\centering
\includegraphics[scale=1]{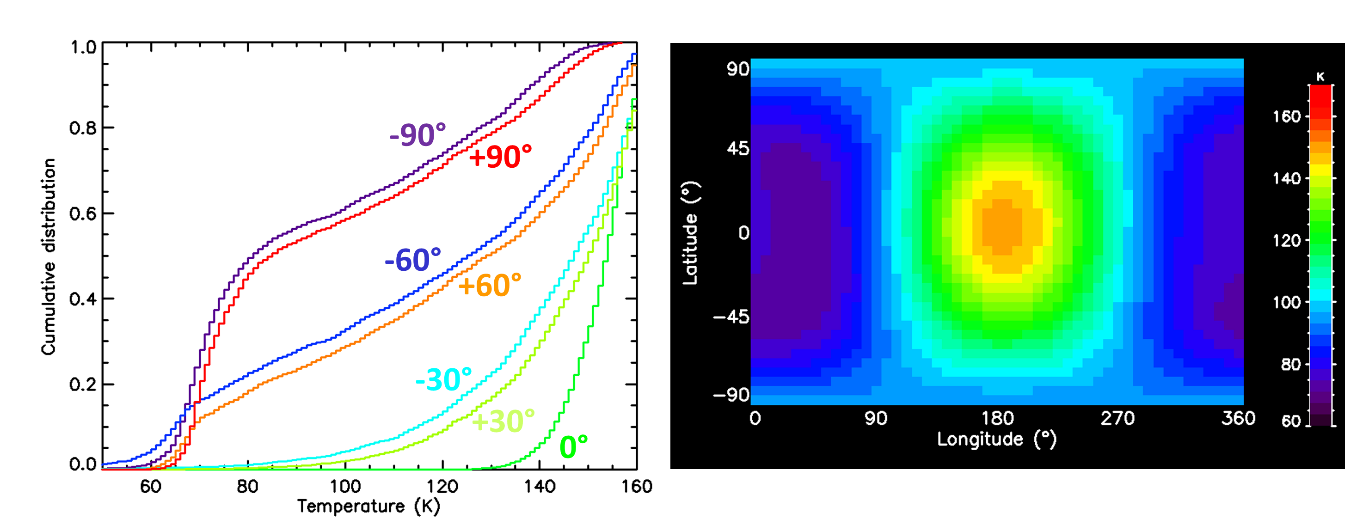}
\caption{Ganymede's surface temperature used in EGM model, representative of the leading hemisphere. Left panel: facet temperature distribution (cumulative probability) of the surface temperatures at 12 h local time for various latitudes indicated in the plot. Right panel: latitude/longitude map of the average surface temperature with the subsolar point being at a latitude of 2$^{\circ}$N (as for JWST observations, Paper I) and longitude of 180$^{\circ}$.\label{fig:temp-distributiom}}
\end{figure*}

{\bf Surface temperature:} In \citet{Leblanc2017,Leblanc2023}, the Ganymede's surface temperature was calculated using a 1-D heat conduction model.  Such description had some limitations, in particular it ignored surface roughness that leads to a distribution of facet temperatures (instead of a single temperature) at a given latitude, longitude and local time. \citet{DBM2024} (Paper I) showed that matching the JWST/MIRI brightness temperature maps, in particular the low-to-high latitude and the noon-to-dawn/dusk temperature contrasts, requires considering surface roughness effects. In the framework of a model for the distribution of slopes inherited from \citet{Hapke1984}, they found that the data could be fit by invoking mean slope angles $s$ =15$^{\circ}$--20$^{\circ}$ on the trailing side and 20$^{\circ}$--25$^{\circ}$ on the leading side, with some variations depending on the adopted surface albedo model. Here we adopted the following parameters, relevant to the leading side: $s$ = 25$^{\circ}$, Bond albedo = 0.30, thermophysical parameter $\Theta$ = 0.3 (i.e. thermal inertia $\Gamma$ = 22.5 SI units). We used a spatially constant Bond albedo to keep the number of free parameters tractable. Such rough temperature distributions were calculated on a 37 $\times$ 48 latitude $\times$ local time (or longitude) grid, i.e. with a 5$^{\circ}$ latitude and 0.5 h local time step. Figure~\ref{fig:temp-distributiom} (left) shows examples of cumulative facet  temperature distributions at noon local time and various latitudes, while the right panel shows the facet-averaged temperature map, where the maximum temperature is at 12.5 h local time. The multiplicity of temperatures at a given latitude/local time enables condensation in regions where it would not be expected without surface roughness. At the equator, the probability to find a surface element at a temperature smaller than 73 K (the theoretical condensation temperature of CO$_2$ at Ganymede atmospheric pressure of 1 pbar) is zero, even in presence of surface roughness. However, the probability of encountering temperatures lower than 73 K increases with latitude, to 1\% at +30$^{\circ}$, 12\% at  +60$^{\circ}$ and 30\% at +90$^{\circ}$.

For describing the temperature of the surface of the trailing side (used for the calculations shown in Appendix~\ref{app:L:4}), we adopted the following parameters, which fit at best the JWST/MIRI brightness temperature map of this hemisphere: $s$ = 20$^{\circ}$, Bond albedo = 0.20, thermophysical parameter $\Theta$ = 0.3. 

We stress that, as indicated in Paper 1, our thermal model describes roughness purely as a slope effect, and does not account for other more complex effects associated with topography, such as shadowing and self-heating due to scattering and reabsorption of solar and thermal radiation within craters, as done for investigating cold traps for  water ice on the Moon \citep{Hayne2021,Davidsson2021}. Applying such more advanced thermophysical models is left to future investigations.

\section{Simulated exospheres from EGM simulations}
\label{app:L}

\subsection{CO$_2$ gas spreading}
\label{app:L:1}
\begin{figure*}[h]
\centering
\includegraphics[scale=0.6]{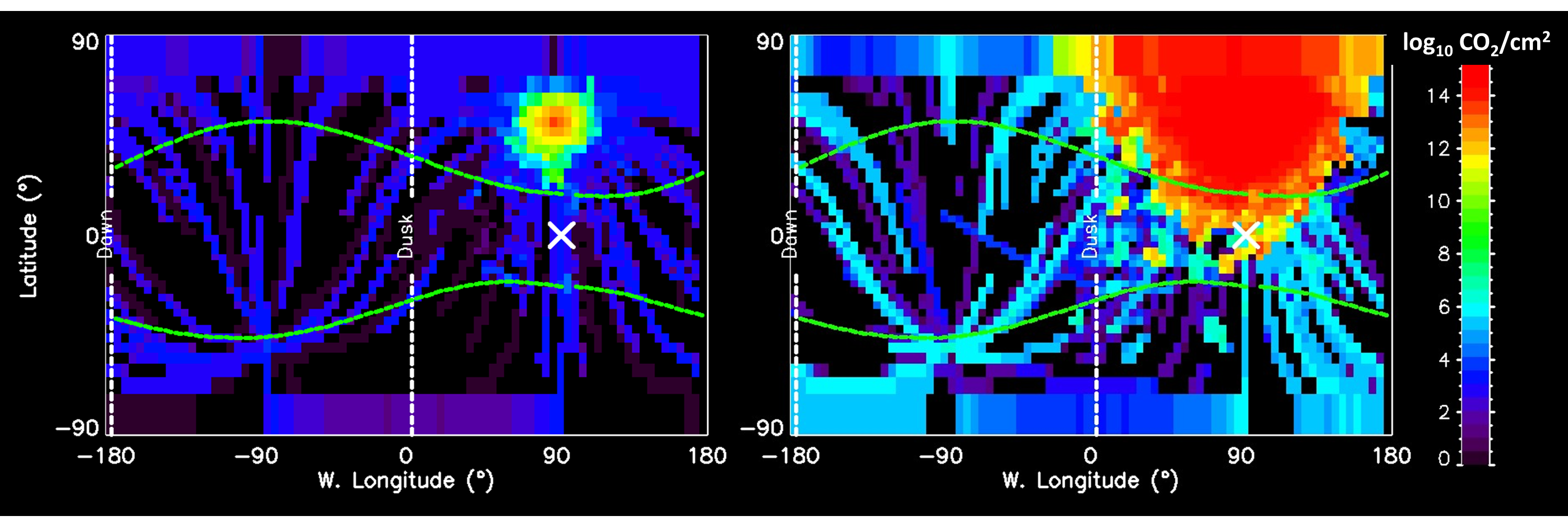}
\caption{Vertical CO$_2$ column density using a distribution of temperature at each location (rough surface, left panel) or the mean temperature (right panel). The simulations consider the sublimation of CO$_2$ ice from a 300$\times$300 km region at 52$^\circ$N, with $q_{\rm CO_2} \times f_c$ = 3 $\times 10^{-14}$.  \label{fig:CO2-vertical}}
\end{figure*}

A question to address is the localized character of Ganymede's CO$_2$ exosphere. Indeed, CO$_2$ does not condense efficiently at the typical Ganymede's surface temperatures, even in the polar regions (100-110 K), and therefore could spread out over the whole illuminated disk, possibly condensing only in the non-illuminated areas. To study the spreading of CO$_2$ molecules, we performed EGM simulations (Appendix~\ref{app:K}) assuming that CO$_2$ is released by the sublimation of CO$_2$ ice from a small surface area (300$\times$300 km) at 52$^\circ$N. The calculations were performed using a distribution of facet temperatures for each location, as computed with our thermal model with surface roughness (Appendix~\ref{app:K}), and, for comparison, using instead the facet-average temperature map (Fig.~\ref{fig:temp-distributiom}, right). Figure~\ref{fig:CO2-vertical} shows the vertical CO$_2$ column density in the two cases. It shows that the resulting atmospheric distribution is quite different for the rough surface. The migration distance is much smaller in this case, showing that the diffusion of CO$_2$ molecules is substantially controlled by the ability to condense on cold traps. These cold traps are probably the discrete patches of optically thick ice, preferentially located on pole-facing slopes, that constitute the polar cap \citep{khurana2007}. The limited horizontal spreading of CO$_2$ gas indicates that local column-density maxima are associated with local sources.

\begin{table*}[!h]
\caption{Line-of-sight column densities from EGM simulations.\label{table:model}}
\begin{center}
\begin{tabular}{|l|cccc|ccc|}
\hline
Models/Region & $q_{\rm CO_2}$$^{(a)}$ & $lat_c$$^{(a)}$ & $f_c$(CO$_2$)$^{(b)}$ & $N$(CO$_2$) & $q_{\rm H_2O}$ & $f_c$(H$_2$O) & $N$(H$_2$O) \\
& (\%) & ($^{\circ}$) & & (m$^{-2}$) & \% & & (m$^{-2}$) \\
\hline 
{\it H$_2$O-ice sublimation, with CO$_2$ release} &&&&&&&\\
Leading North source  & 1 & 50 & 260 & 9.0$\times10^{17}$ & 20 & -- & 1.4 $\times10^{18}$\\
Leading SZA$<$15$^{\circ}$ & 1 & 50 & 260 & 4.1$\times10^{10}$$^*$ & 20 & -- & 4.1$\times10^{19}$ \\
{\it CO$_2$ ice sublimation} &&&&&&&\\
Leading North source  & 1 & 50 & 3$\times10^{-12}$ & 8.8$\times10^{17}$ & -- & -- & -- \\
Leading SZA$<$15$^{\circ}$ & 1 & 50 & 3$\times10^{-12}$ & 4.4$\times10^{15}$$^*$ & -- & -- & -- \\
{\it Sputtering only} &&&&&&&\\
Leading North source  &1 & 40 & 382 & 9.9$\times10^{17}$ & 20 & 382 & 8.2$\times10^{18}$ \\
Leading SZA$<$15$^{\circ}$ &1 & 40 & 382 & 2.6$\times10^{15}$$^*$ & 20 & 382 & 2.6$\times10^{17}$ \\
\hline
\end{tabular}
\end{center}
$^{(a)}$ $q_{\rm CO_2}$ in a northern cap covering $lat_c$--90$^\circ$ in latitude and ranging from 0 to 180$^\circ$W in longitude. Elsewhere, $q_{\rm CO_2}$ is set to 0. 

$^{(b)}$ Correction factor to simulated CO$_2$ flux from the surface to reproduce the CO$_2$ line-of-sight column density peak in the latitude profile (Fig.~\ref{collat-distribution}).

$^{(*)}$ Values are low because $q_{\rm CO_2}$ = 0 for latitudes $<$ $lat_c$. 
\end{table*}

\subsection{CO$_2$ exosphere: Leading CO$_2$ source only}
\label{app:L:2}
\begin{figure*}[h!]
\centering
\includegraphics[scale=0.5]{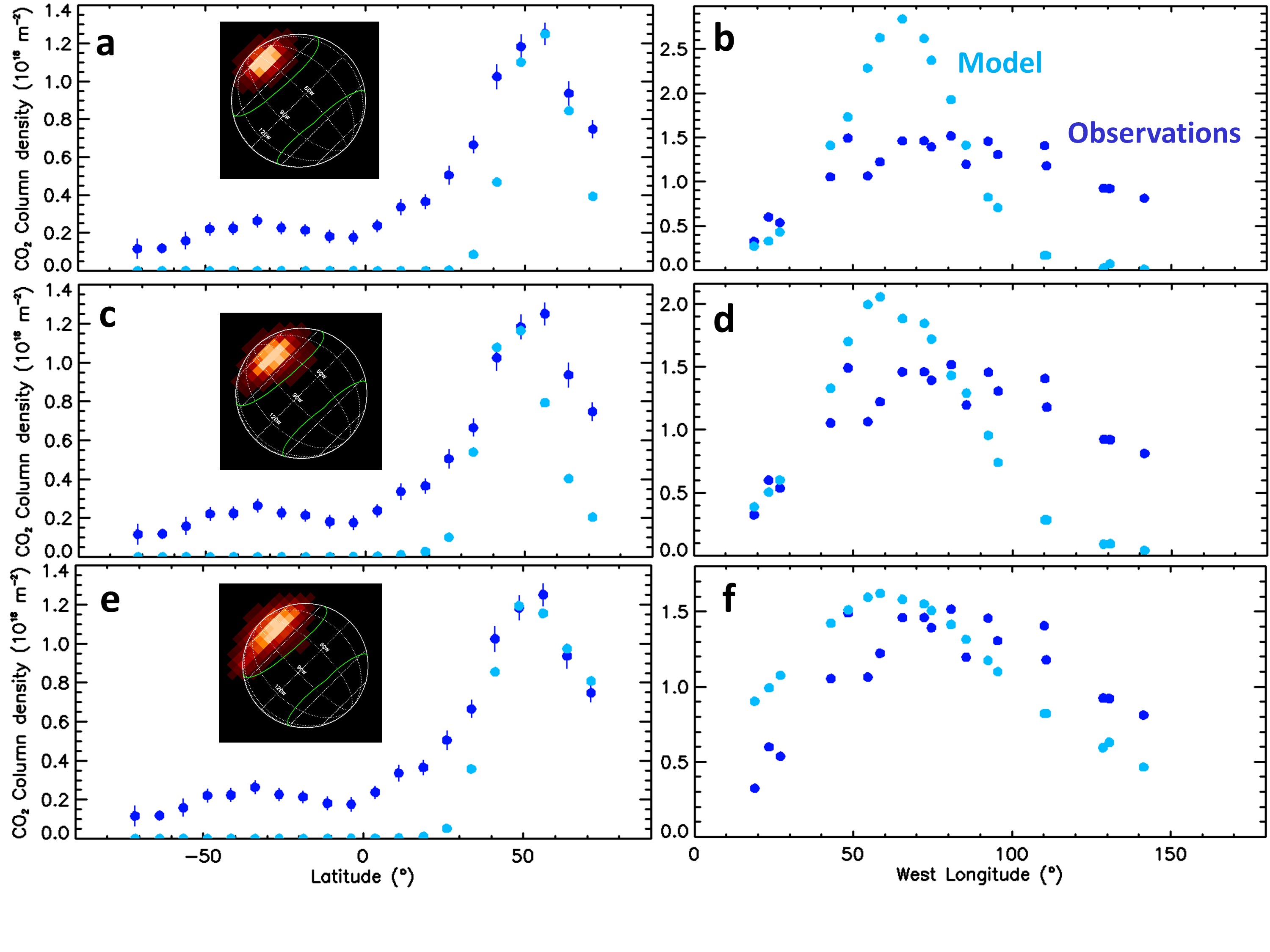}
\caption{CO$_2$ gas line-of-sight column density as a function of latitude (left) and longitude (right) for the leading side. Light blue symbols show model results from EGM and dark blue symbols refer to JWST observations.   Top panels (a, b): CO$_2$ release associated to H$_2$O sublimation with surface flux $f_c$ $\times$ $q_{\rm CO_2}$ = 2.6 at latitudes $>$ 50$^{\circ}$N and $q_{\rm CO_2}$ = 0 elsewhere. Middle panels (c, d): CO$_2$ release associated to CO$_2$ ice sublimation with surface flux $f_c$ $\times$ $q_{\rm CO_2}$ = 3.$\times$ 10$^{-14}$ at latitudes $>$ 50$^{\circ}$N and $q_{\rm CO_2}$ = 0 elsewhere. Bottom panels (e, f): sputtering only with surface abundance $q_{\rm CO_2}$ = 1\% at latitudes $>$ 40$^{\circ}$N, and $q_{\rm CO_2}$ = 0 elsewhere, and a multiplying factor to surface CO$_2$ flux $f_c$ = 382. In the left panels (a, c, e) column densities are averages in latitude bins of 7.5$^{\circ}$.  Longitudinal variations shown in right panels (b, d, f) consider latitudes in the range 42--62$^\circ$N (i.e., encompassing the northern region of the leading hemisphere with CO$_2$ gas enhancement). Calculated CO$_2$ column density maps for all three cases are shown in the left panels. \label{fig:CO2profile_leading}}
\end{figure*}

The most significant feature of Ganymede's CO$_2$ exosphere being the large excess in the northern hemisphere of leading side, our simulations (see Appendix~\ref{app:K} for model description) were designed to reproduce this feature. Specifically, we calculated line-of-sight column density maps $N$(CO$_2$), and extracted latitudinal and longitudinal profiles of $N$(CO$_2$) for comparison with observations. Model simulations were performed assuming  $q_{\rm CO_2}$ = 1\% in a northern cap extending from $lat_c$ to 90$^\circ$ in latitude and ranging from 0 to 180$^\circ$W in longitude. Elsewhere, $q_{\rm CO_2}$ was set to 0. 
The input value $q_{\rm CO_2}$ = 1\% was set as it is consistent within a factor of a few with the rough estimation of the CO$_2$ abundance at the surface based on the depth of the CO$_2$-solid 4.3 $\mu$m band (CO$_2$/H$_2$O $\sim$ 1\% in mass, Paper I).
The limiting latitude $lat_c$ and the correction factor $f_c$ to the CO$_2$ flux were adjusted to best reproduce the position in latitude and value of the column-density peak in the latitudinal profile of $N$(CO$_2$) shown in Fig.~\ref{collat-distribution}. 

Column density maps obtained for sputtering, CO$_2$ release associated with H$_2$O sublimation and sublimation of CO$_2$ ice (Appendix~\ref{app:K}) are shown in Fig.~\ref{fig:CO2profile_leading} (and in the main text Fig.~\ref{fig:model-exosphere} for the first two production mechanisms). EGM latitudinal profiles of line-of-sight column densities are compared to observations in the left panels of Fig.~ \ref{fig:CO2profile_leading}. Longitudinal variations for latitudes in the range 42--62$^\circ$ N (i.e., encompassing the northern region of the leading hemisphere with CO$_2$ gas enhancement) are compared  in the right-hand panels of Fig.~ \ref{fig:CO2profile_leading}. Table~\ref{table:model} lists model input parameters, and average column densities in the sub-solar region and so-called "Leading CO$_2$ source" defined in Table~\ref{tab:H2O}. For CO$_2$ release associated with H$_2$O sublimation (top panels), the latitudinal profile is reproduced for $lat_c$ = 50$^\circ$N and multiplying the flux of CO$_2$ molecules by $f_c$ = 260 (as $f_c$ $\times$ $q_{\rm CO_2}$ = 260$\times$0.01 = 2.6, this corresponds to a CO$_2$/H$_2$O relative abundance of 5 for an H$_2$O areal ice fraction of 50\%). Since the H$_2$O sublimation flux is highly dependent of the surface temperature (Eq.~\ref{Eq:sublimationH2O}) and is therefore strongly favored at low latitudes, $lat_c$ must be  close to the latitude at which the peak of the CO$_2$ column density is observed (Fig.~\ref{fig:exosphere-combined}). For the CO$_2$-ice sublimation case (middle panels), $lat_c$ is also close to 50$^\circ$N, and  $f_c$ =3 $\times 10^{-12}$. For the sputtering scenario (bottom panels), a limiting latitude $lat_c$ =  40$^\circ$N best reproduces the latitudinal trend of the column density (a too narrow distribution in latitude is obtained for $lat_c$ =  50$^\circ$N), and the flux of the sputtered CO$_2$ molecules had to be multiplied by $f_c$ = 382.

As shown in Fig.~\ref{fig:CO2profile_leading}, the sputtering only scenario (panels e and f) provides a  better fit of the spatial distribution of CO$_2$ exosphere than sublimation. Especially, for the sublimation scenarios, a strong variation with longitude (i.e., local time) is obtained whereas the diurnal variation is flatter and almost consistent with the observations for sputtering. This is essentially due to the difference in the temperature dependence of these mechanisms (Eqs~\ref{Eq:sublimationCO2H2O}, \ref{Eq:sublimationCO2}, \ref{eq:yield}). However, the peak of the CO$_2$ line-of-sight column-density distribution for sputtering is more shifted towards the afternoon ($\sim$13.4 h) than in the sublimation cases ($\sim$13.1 h) and for the observed peak ($\sim$12 h). 

\begin{figure*}[ht!]
\centering
\includegraphics[scale=1.0]{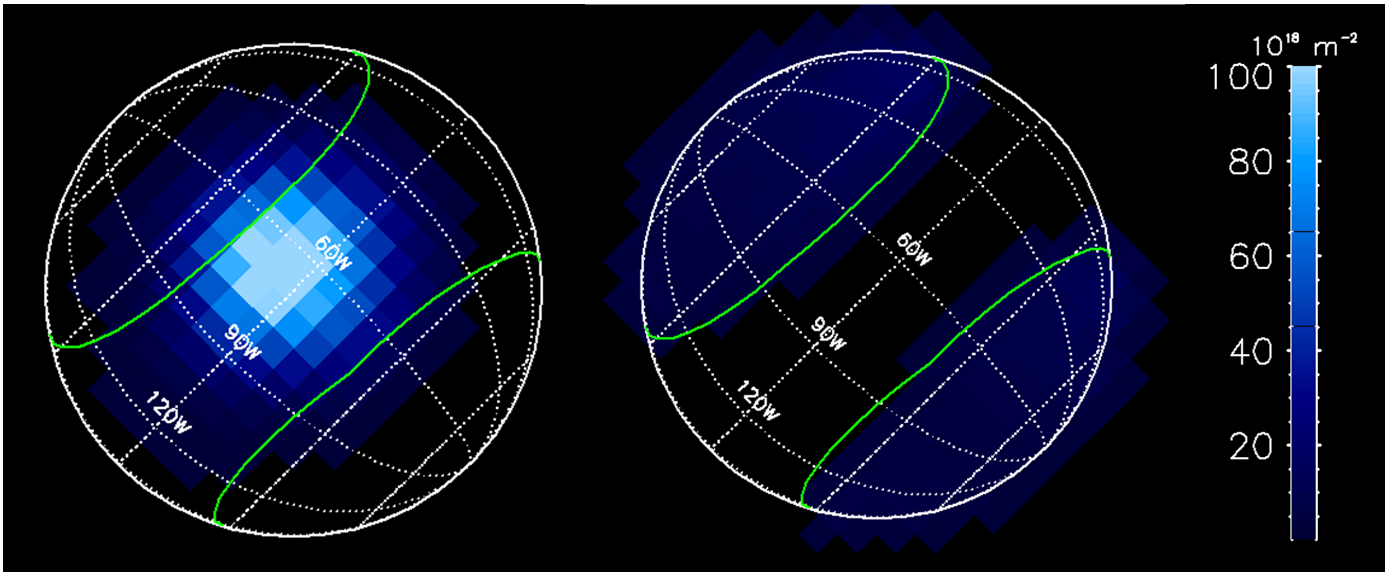}
\caption{Calculated line-of-sight column-density maps of the H$_2$O exosphere of Ganymede (in 10$^{18}$ m$^{-2}$). The green lines display the boundary between open and close field lines (OCFB). Left panel: sublimation only source. Right panel: sputtering only source with $f_c$ = 382. The H$_2$O areal surface fraction is set to $q_{\rm H_2O}$ = 20\%.  The subsolar point is at 2.6$^{\circ}$N, 82$^{\circ}$W. \label{fig:H2O-model}}
\end{figure*}

\begin{figure*}[ht!]
\centering
\includegraphics[scale=0.5]{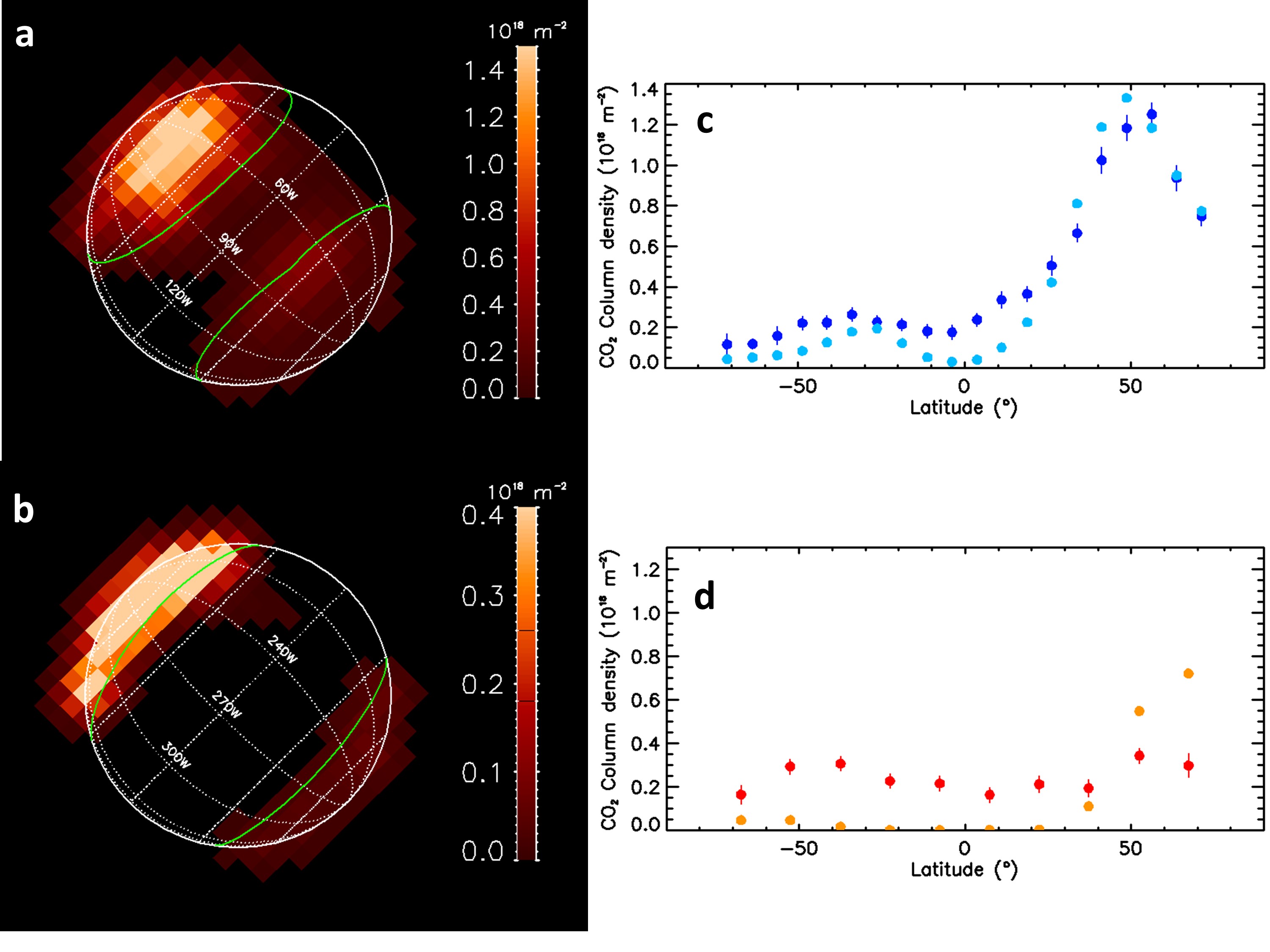}
\caption{Simulation of the entire CO$_2$ exosphere formed by sputtering only. No sputtering is simulated inside the closed-field-line region. Input parameters are given in the text.  Panel a: 2D line-of-sight column density map of CO$_2$ (in 10$^{18}$ CO$_2$/m$^2$), Ganymede-leading side. Panel b: same as panel a but for the Ganymede-trailing side. Panel c: latitudinal variation of the longitudinal average column density for the Ganymede-leading side (for bins of 7.5$^\circ$ in latitude). Panel d: same as for panel c but for the Ganymede-trailing side and bins in latitude of 15$^\circ$. Dark blue symbols: JWST observations. Light blue symbols: EGM simulation.  Red symbols: JWST observations. Orange symbols: EGM simulation.  \label{fig:whole_surface_sputtering}}
\end{figure*}

\subsection{H$_2$O exosphere}
\label{app:L:3}
EGM simulations of Ganymede's H$_2$O exosphere were already performed by \cite{Leblanc2017,Leblanc2023}, but the used surface temperature model did not consider surface roughness, unlike the present calculations. Figure~\ref{fig:H2O-model} displays the H$_2$O column density as seen from JWST for the sublimation (left panel) and sputtering (right panel) cases. The H$_2$O ice  areal surface fraction $q_{\rm H_2O}$ is set to 20\% all over the surface, a value which is consistent with measured water ice abundances on the leading side for latitudes $\leq$ 30--40$^{\circ}$ \cite{ligier2019}. For latitudes of 40--50$^{\circ}$N, values of 40--50\% would be more appropriate \cite{ligier2019}.  As expected the sublimation of H$_2$O follows the surface temperature distribution. A North/South asymmetry can be clearly seen which is driven by the small asymmetry in surface temperature associated with the positive subsolar latitude (see Fig.~\ref{fig:temp-distributiom}). A dawn to dusk asymmetry with larger H$_2$O column densities towards the dusk is also present. A similar dawn/dusk asymmetry, related to surface thermal inertia, is present for the CO$_2$ simulated exosphere (Fig.~\ref{fig:model-exosphere}, left panel). 

The calculated line-of-sight H$_2$O column densities in the leading subsolar region (solar zenith angle SZA $<$ 15$^{\circ}$) and so called "Leading CO$_2$ source" are given in Table~\ref{table:model}. They are below the upper limits set by JWST (7--17$\times$10$^{19}$ m$^{-2}$ for subsolar,  2$\times$10$^{19}$ m$^{-2}$ for CO$_2$ source region,  Table~\ref{tab:H2O}), summing the contributions from sputtering and sublimation.  When considering a more appropriate value of $q_{\rm H_2O}$ = 40--50\% for the northern region of the leading hemisphere, the modelled column density for this region is of the order of the JWST upper limit. The value obtained for the subsolar region (4.1 10$^{19}$ m$^{-2}$) is a factor of 7 above the minimum value derived from HST OI data for the leading hemisphere (6$\times$10$^{18}$ m$^{-2}$, \cite{Roth2021}). 

We stress that calculated H$_2$O sputtered fluxes are under the hypothesis that H$_2$O is the major species released by sputtering of H$_2$O ice \cite{Cassidy2013}. However, the major mass loss from water ice by particle bombardment might not be H$_2$O, as assumed here, but rather via the ejection of O$_2$ and H$_2$ (or at least in significant proportion)  according to laboratory experiments \cite{Teolis2017}. Hence, H$_2$O sputtered fluxes given in Table~\ref{table:model} can be considered as upper limits. This reinforces our conclusion that JWST H$_2$O upper limits measured in the north polar cap of the leading hemisphere are fully consistent with expectations.

\subsection{Simulation of CO$_2$ sputtering on entire Ganymede's surface}
\label{app:L:4}
We also investigated CO$_2$ sputtering over the southern open-field-line region of the leading hemisphere and over the southern and northern open-field-line regions of the trailing hemisphere. This approach allowed us to illustrate how the sputtering ejection rate would need to be changed from a region to another to explain the JWST observations. The surface temperature of the trailing side was calculated using the same approach as for leading side (see parameters of the thermal model in Appendix~\ref{app:K}). Figure~\ref{fig:whole_surface_sputtering} provides one example of results obtained by using the same $q_{\rm CO_2}$ and $f_c$ values for the two hemispheres, the only difference being the location of the OCFBs. The distribution of sputtered CO$_2$ over the whole surface is described according to:
\begin{itemize} 
\item a region (1) of sputtering above a latitude of  50$^{\circ}$N, with a flux of sputtered CO$_2$ molecules determined as described in section I, with $f_c$= 640 and $q_{\rm CO_2}$ = 1\%, 
\item a region (2) in the northern hemisphere open field lines region with $f_c$= 80 and $q_{\rm CO_2}$ = 1\% (describing e.g. regions in between northern OCFB and $>$ 50$^{\circ}$N polar cap on the leading hemisphere),   
\item a region (3) in the southern hemisphere open field lines region with $f_c$= 60 and $q_{\rm CO_2}$ = 1\%,
\item a region (4) in close field lines region with $f_c$= 0.
\end{itemize}

Assuming the same $f_c$ = 640 value for all regions, would be equivalent to adopting $q_{\rm CO_2}$ = 1\%, 0.13\%,  0.09 \%, and 0\% for regions (1), (2), (3), (4) respectively. With these assumptions, the shape of the latitudinal distribution observed on the leading side is reproduced but not that of the trailing side (Fig.~\ref{fig:whole_surface_sputtering}). Reproducing at least approximately Ganymede's CO$_2$ exosphere with the sputtering mechanism might be possible by adjusting $f_c$ or $q_{\rm CO_2}$ parameters geographically over the surface. However, this model cannot explain the exospheric excess observed at equatorial to mid latitudes in the southern regions of the trailing hemisphere (Fig.~\ref{fig:exosphere-combined}) taking into account that, at the time of the trailing-side observation, Ganymede was inside the plasma sheet of Jupiter’s magnetosphere, so southern polar regions were not over-exposed to plasma bombardment with respect to northern regions (Appendix~\ref{app:J}). 

\section{PSF deconvolution of CO$_2$ gas map}
\label{app:M}
\begin{figure*}[ht!]
\includegraphics[width=16cm]{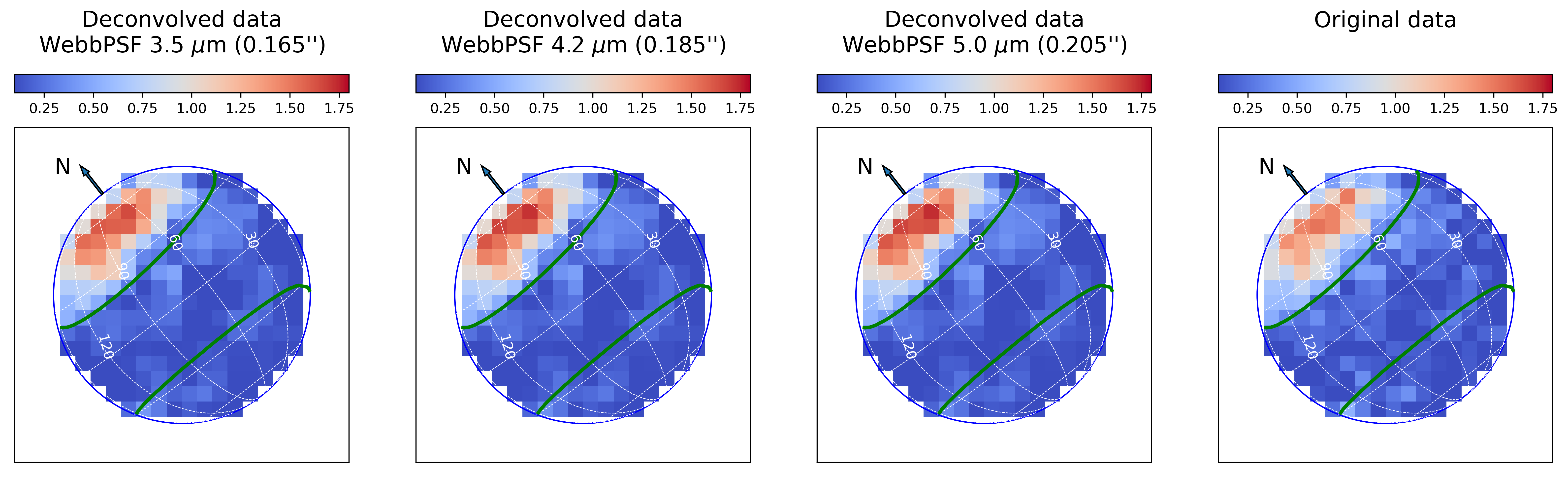}
\includegraphics[width=12cm]{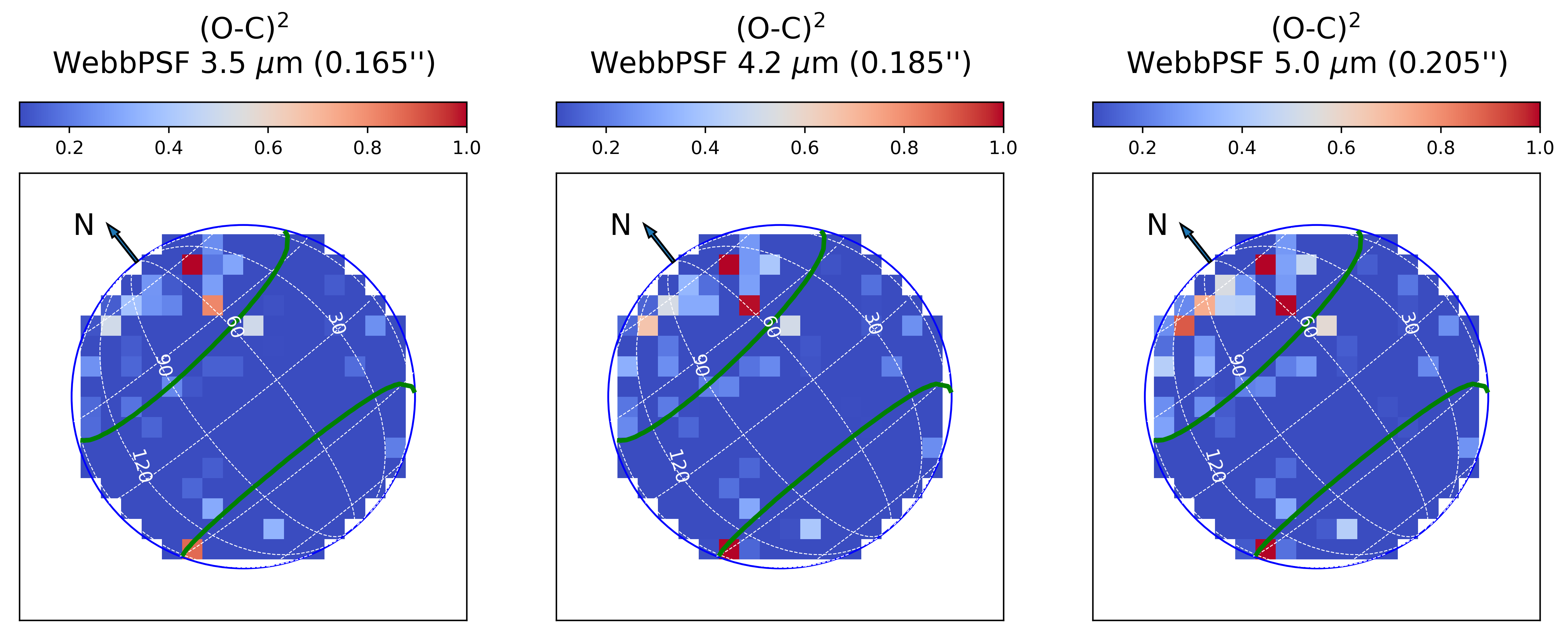}
\caption{Top: deconvolved CO$_2$-gas maps (leading hemisphere) obtained with the AIDA algorithm using NIRSpec PSFs calculated with WebbPSF at 3.5 $\mu$m (FWHM = 0.165''), 4.2 $\mu$m (FWHM = 0.185'') and  5 $\mu$m (FWHM = 0.205''). Bottom: residuals of the deconvolution for each calculated PSF; plotted are $(O-C)^2$, where $O$ is the original CO$_2$ column-density map (leading hemisphere, top-right plot), and $C$ is the convolution of the deconvolved $O$ map. Color bars are in unit of 10$^{18}$ m$^{-2}$. \label{sup:figPSF1}}
\end{figure*}

In order to analyse the spatial distribution of the CO$_2$ exosphere on Ganymede's leading side in more detail, we applied a deconvolution procedure.  The deconvolution process was done using the AIDA algorithm in classical mode \citep{hom2007} that requires science and PSF data files. Reported Full Width at Half Maximum (FWHM) measured at 4.25~$\mu$m from JWST point-source observations range from 0.14 to 0.17'' \citepalias[][]{}{Deugenio2023} and are expected to vary with the observational mode (e.g., number of dithers). Unfortunately, no reference star was observed during the observations of Ganymede, so we used the WebbPSF software (\url{https://www.stsci.edu/jwst/science-planning/proposal-planning-toolbox/psf-simulation-tool}) which can calculate monochromatic PSFs for NIRSPec in spectroscopic mode. PSFs were generated at various wavelengths to explore how deconvolved images vary with the PSF FWHM. 

We show in Fig.~\ref{sup:figPSF1} the deconvolution of the CO$_2$ gas map above the leading hemisphere for PSFs with FWHMs of 0.165, 0.185 and 0.205'' corresponding to  WebbPSF outputs at 3.5, 4.2 and 5 microns, respectively.
For this purpose, we used the CO$_2$ column density map derived for the full 37$\times$ 43 spaxels IFU frame (i.e., including results outside Ganymede disk).  
The deconvolved maps were then reconvolved with the PSFs used for the deconvolution. Residuals with respect to the original data are shown in Fig.~\ref{sup:figPSF1} (bottom row) for the three assumed PSFs, and do not differ much. The deconvolved CO$_2$ gas distribution confirms that the northern CO$_2$ excess is confined in longitude and latitude, and that the decrease of the column density above latitudes of 60$^{\circ}$N is real. The deconvolved map also clearly shows the excess nearby the southern OCFB.

\section{Spatial variations of surface CO$_2$ band depth}
\label{app:N}
\begin{figure*}[ht!]
\centering
\begin{tabular}{l@{\hspace{0.3cm}}l@{\hspace{0.3cm}}l@{\hspace{0.3cm}}}
\includegraphics[scale=0.4,valign=b]{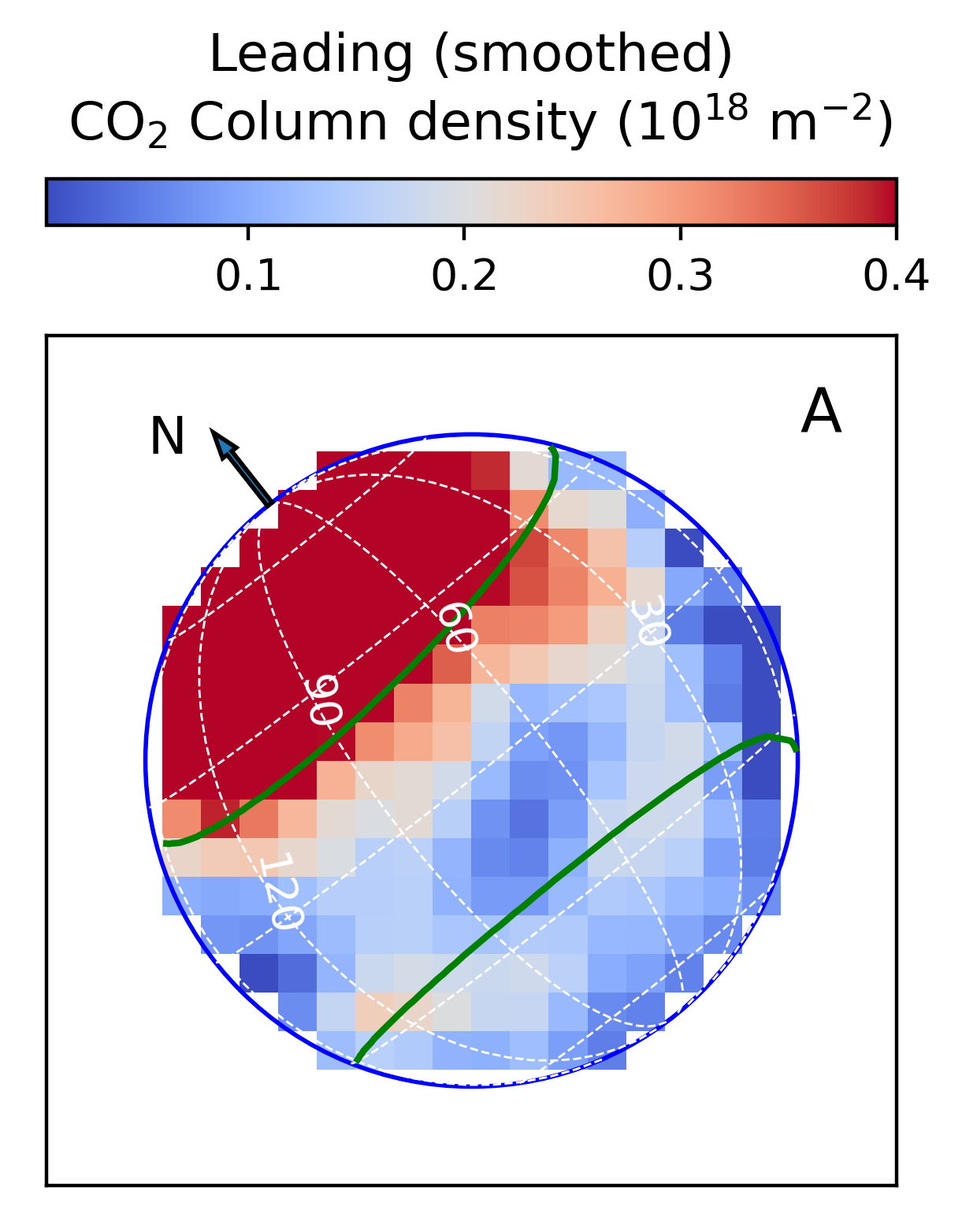}  & \includegraphics[scale=0.4,valign=b]{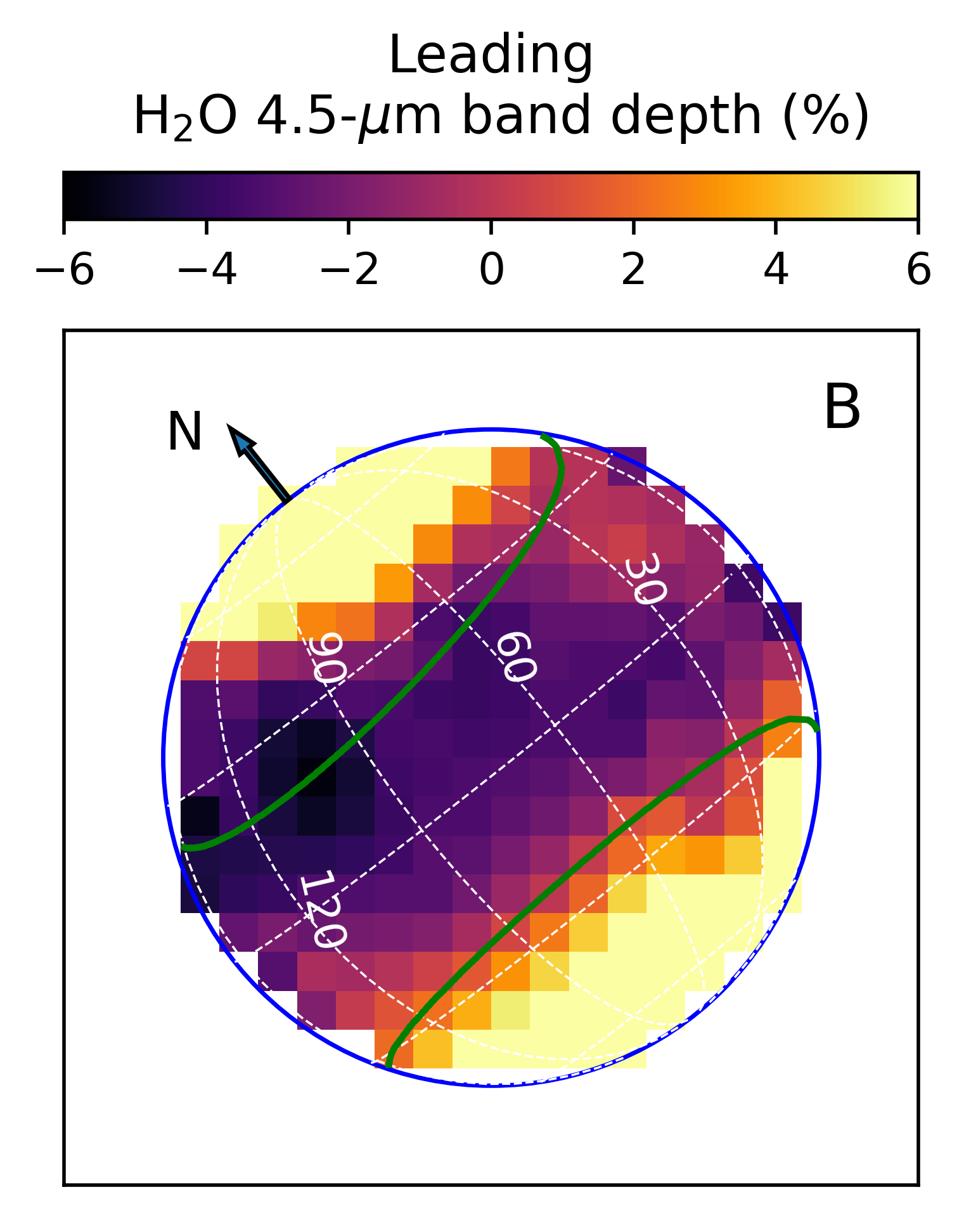} & \includegraphics[scale=0.4,valign=b]{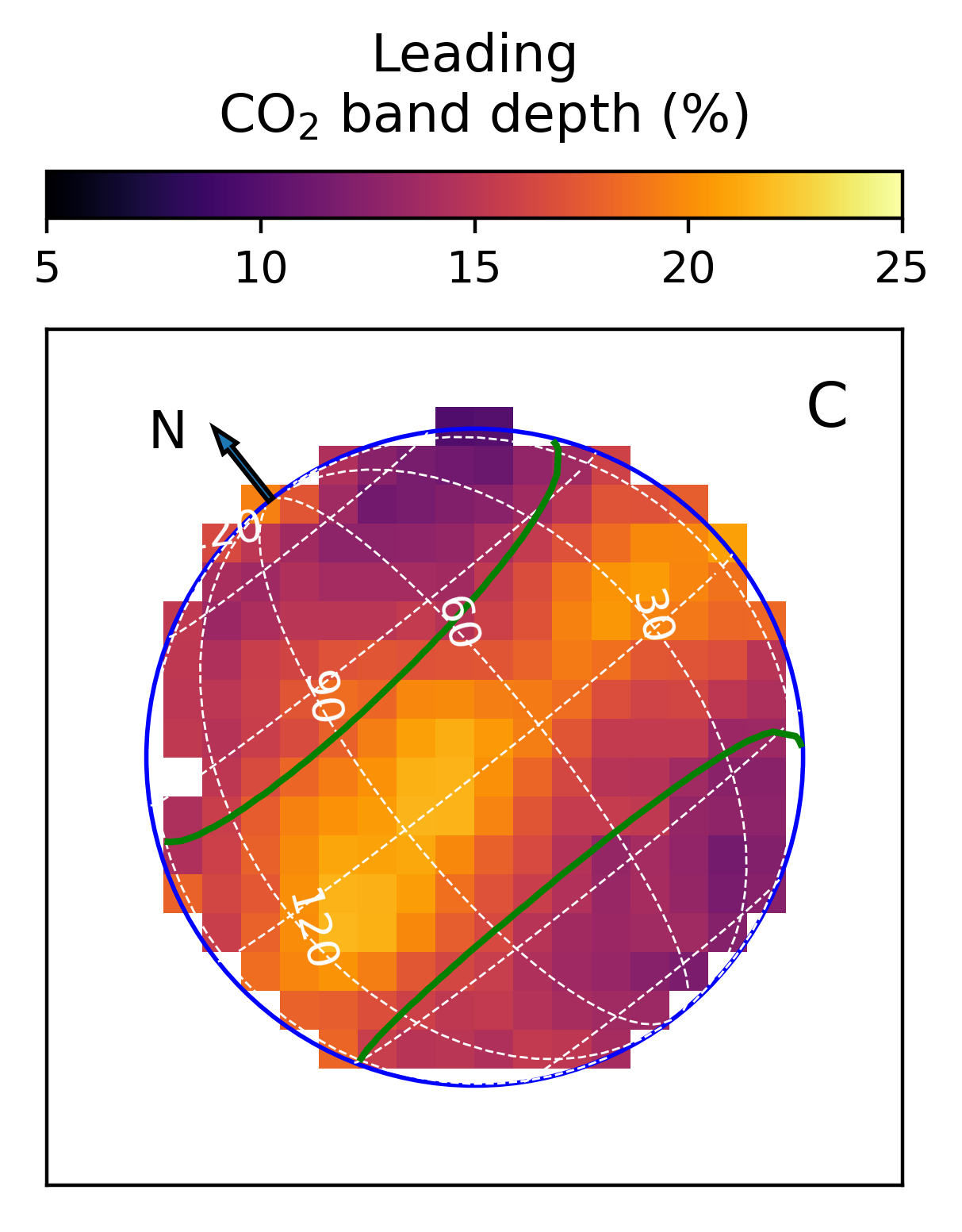}  \\\includegraphics[scale=0.4,valign=b]{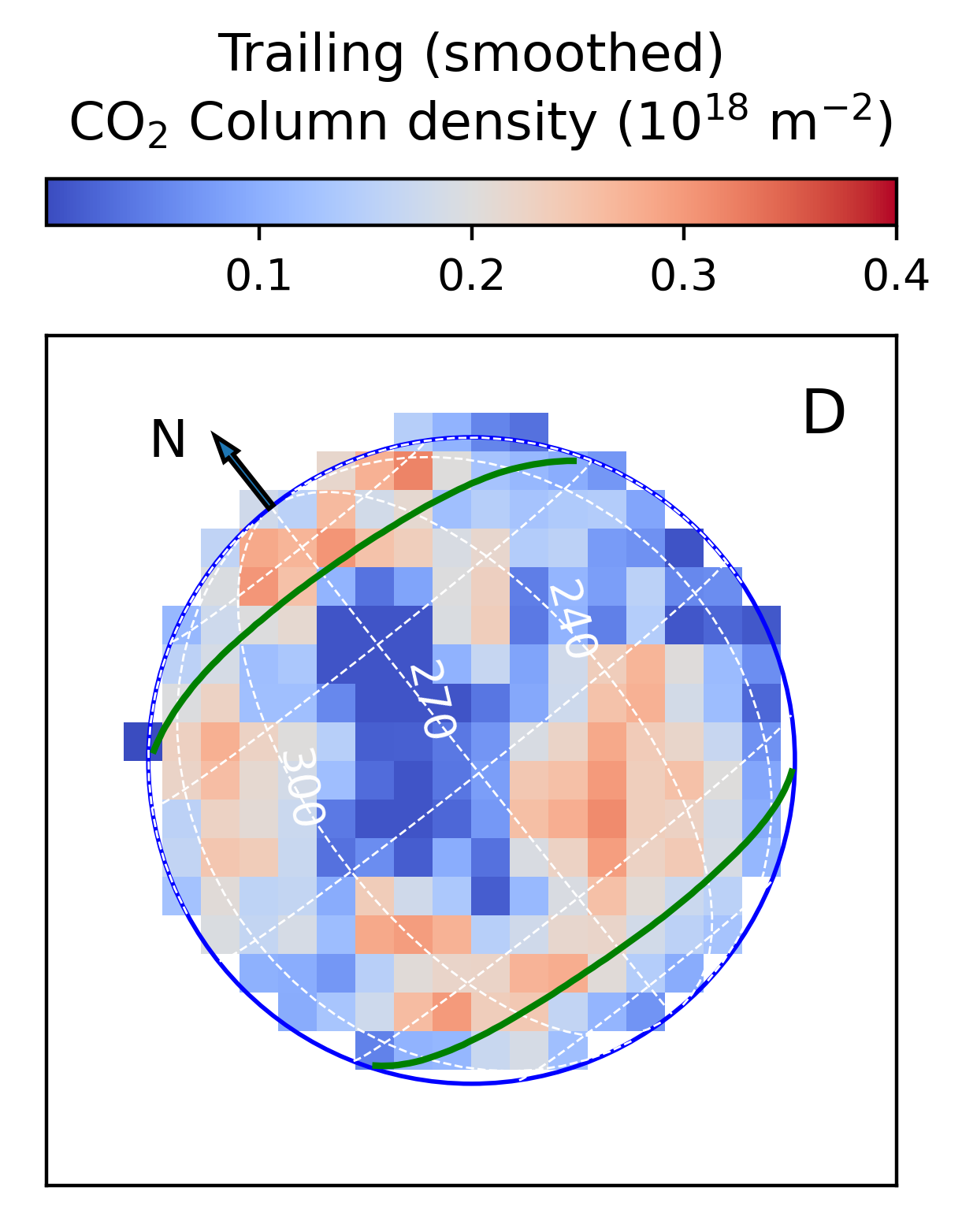} & 
\includegraphics[scale=0.4,valign=b]{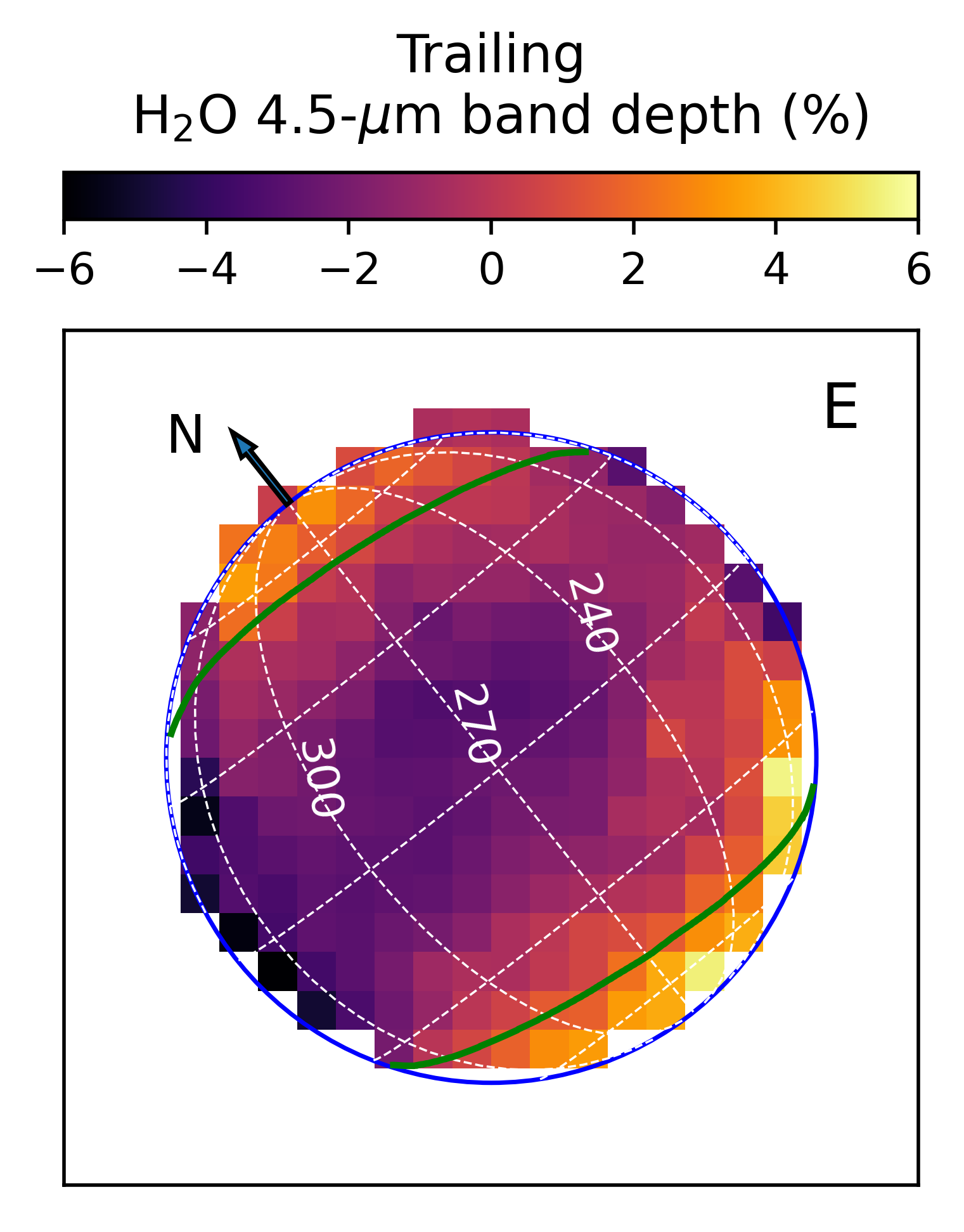} 
&\includegraphics[scale=0.4,valign=b]{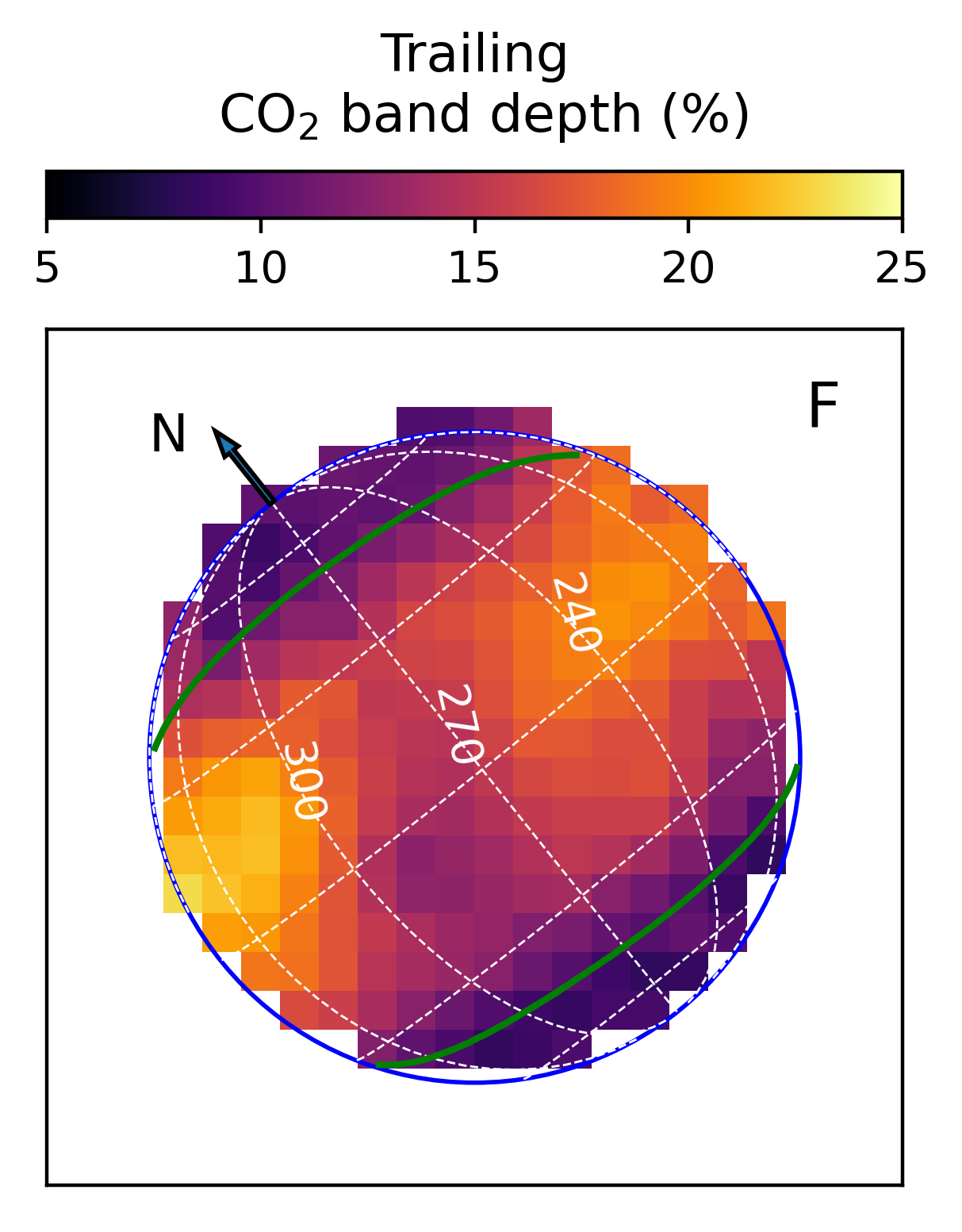} \\
\end{tabular}
\caption{Comparing CO$_2$ exosphere to H$_2$O and CO$_2$ distribution on Ganymede's surface. Top and bottom rows are for the leading and trailing hemispheres, respectively. A, D: CO$_2$ gas column density maps at a 3$\times$3 smoothed resolution (this work) ; B, E: H$_2$O band depth at 4.5 $\mu$m (Paper I); C, F: CO$_2$-solid 4.3-$\mu$m band depth (Paper I). In these plots, unlike in Figs~\ref{fig:exosphere-combined} and \ref{sup-gas-distribution}, the same color scales and boxcar smoothing are used for leading and trailing. \label{leadingtrailing-surface} }
\end{figure*}

\begin{figure*}[ht!]
\centering
\includegraphics[width=10cm]{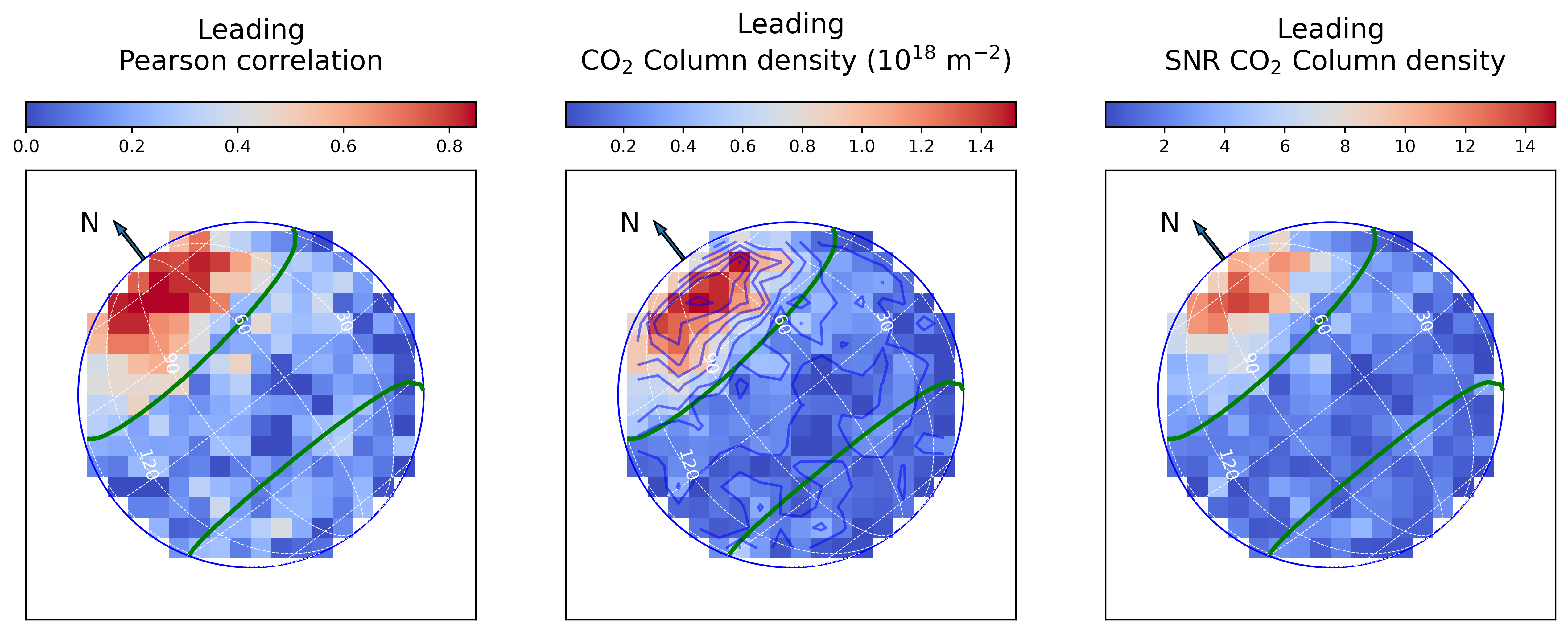}
\includegraphics[width=10cm]{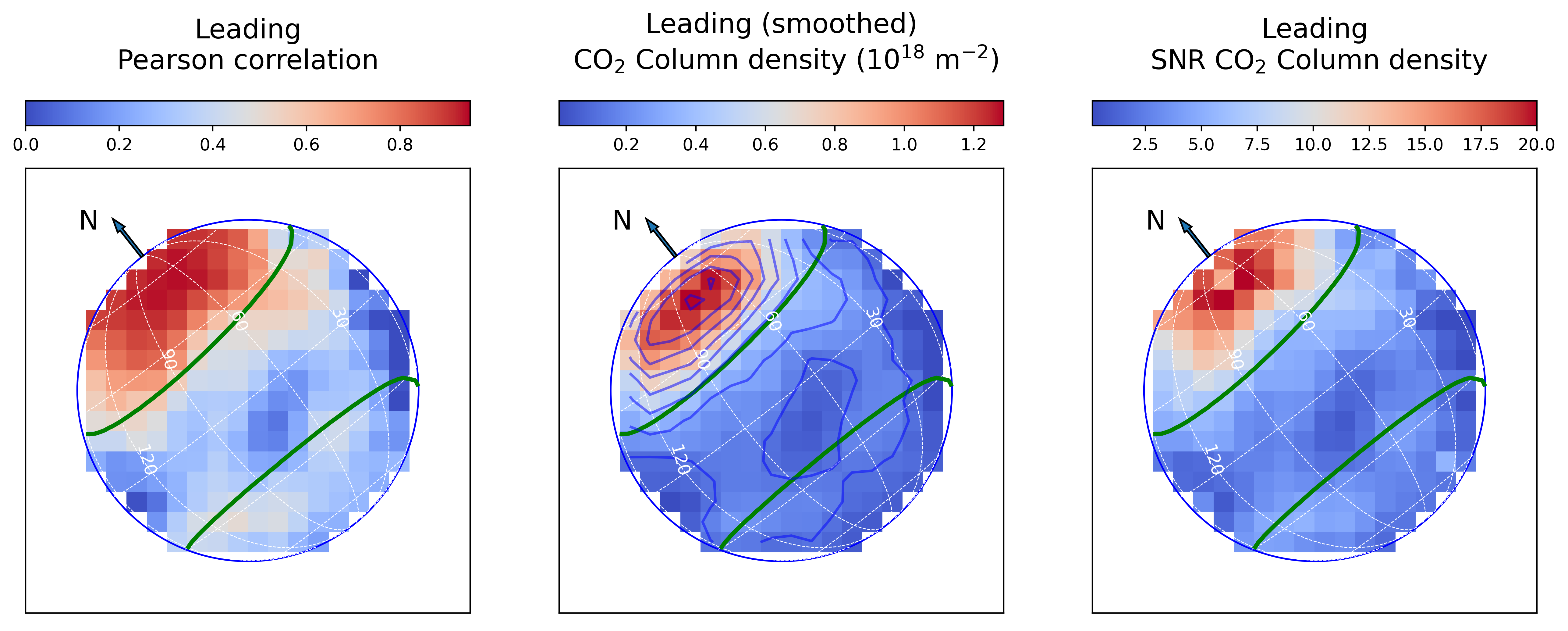} 
\includegraphics[width=10cm]{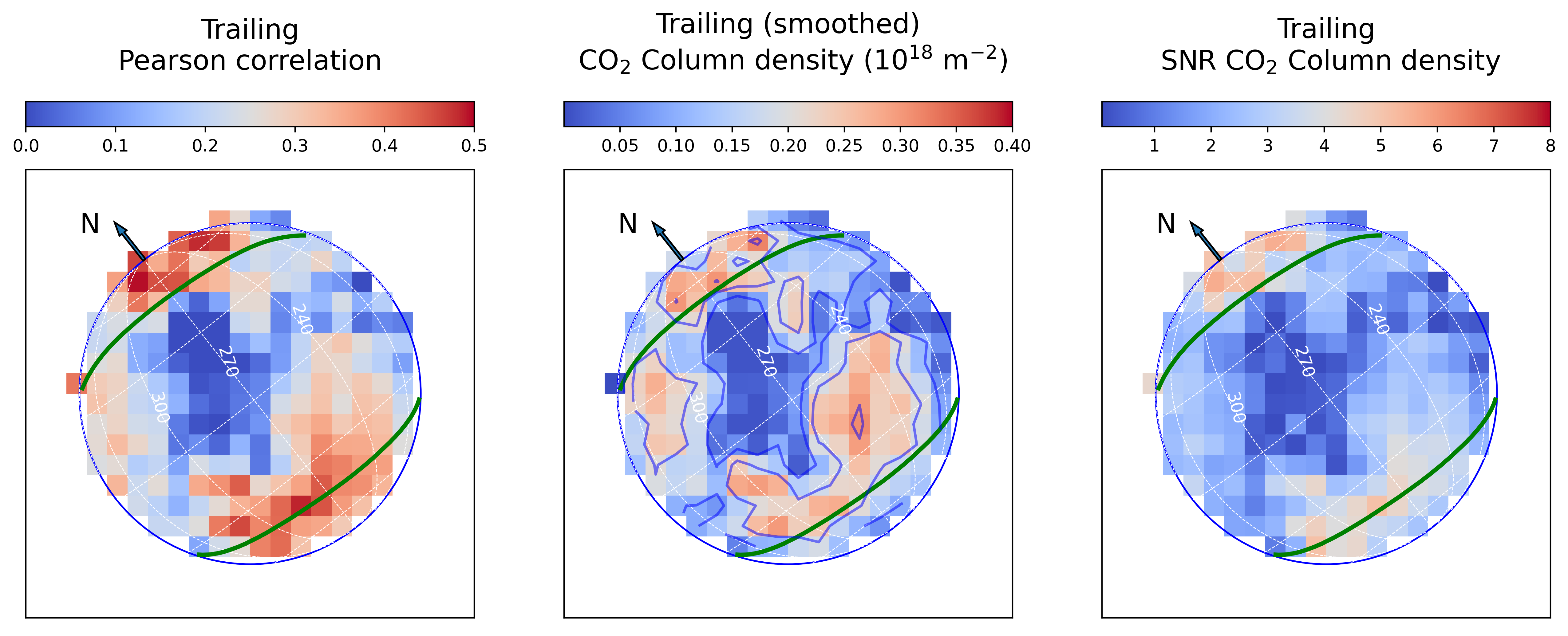}
\caption{CO$_2$ gas distribution above Ganymede surface. First and second rows are for the leading side, at the original and 3$\times$3 boxcar smoothed (for higher SNR) resolutions, respectively. Bottom row is for the trailing side at 3$\times$3 smoothed resolution. Plots on the first column show the Pearson correlation coefficient between continuum-filtered residual CO$_2$ gas emission and a forward CO$_2$ fluorescence model at 105 K (Fig.~\ref{fig:sup-Trot}). Plots on the second and third column show the line-of-sight CO$_2$ column density and SNR inferred by fitting synthetic CO$_2$ fluorescence spectra (Appendix~\ref{app:B}). The color scales for the leading and trailing sides are different, and indicated above the plots. Pixel sizes are 0.1$\times$0.1''. The green lines show the open-closed-field line boundary \cite{Duling2022} \label{sup-gas-distribution}  }
\end{figure*}

\noindent
The CO$_2$ absorption band around 4.26 $\mu$m is widespread over the surface of Ganymede. Globally, the CO$_2$ band depth appears anti-correlated with bond albedo and water ice absorption band depths, with the maximum CO$_2$ band depth on the equatorial regions but much lower values on regions poleward of 30°N (Fig.~\ref{leadingtrailing-surface}, \cite{hibbitts2003}, Paper I). The CO$_2$ band center (Fig.~\ref{leading-surface}C) and shape show stronger relationship with surface brightness for both hemispheres, red-shifting and getting narrower/asymmetric as we reach polar latitudes, respectively. As discussed in Paper I these changes of band position and shape may be due to contributions of CO$_2$ under varying physical states/matrices depending on the latitude: adsorbed on minerals or salts at the equatorial latitudes, and possibly mixed in amorphous water ice at the poles (Fig.~\ref{leading-surface}C). But why is the CO$_2$ band depth weaker at the polar regions, where CO$_2$ gas appears to be released from the surface? Actually, the poles of Callisto also have weaker CO$_2$ band depths and finer-grained water ice than the equatorial regions, just like Ganymede. From Galileo/NIMS data, this was interpreted as fine-grained ice physically covering and spectrally masking the CO$_2$, although such a masking effect was not demonstrated numerically or experimentally \citep{mccord1998, hibbitts2000, hibbitts2003}. JWST/NIRSpec data of Ganymede also show some anti-correlation between the spatial distribution of the CO$_2$ band depth and that of the H$_2$O 4.5-$\mu$m band depth (Fig.~\ref{leadingtrailing-surface}). However, the lower CO$_2$ band depth at the polar regions may not be (only) due to a putative spectral-masking effect, but to other factors. Maybe the surface areal abundance of CO$_2$ is lower within the open-field-line areas because the irradiation releases it more efficiently from its mineral association. As a result, CO$_2$ may be less concentrated over non-ice mineral-rich terrains and only present on bright ice-rich patches (the coldest surfaces), reducing its geographically averaged band depth and making its band shape and position more compatible to CO$_2$ mixed in water ice. Another explanation might also be that the absorption coefficient of CO$_2$ decreases at the poles because CO$_2$ is in a different state/matrix and/or at a different temperatures than at equatorial regions.

Despite the lack of correlation between the global distribution of CO$_2$ gas column density and solid CO$_2$ band depth, there are some local areas where both are minimum (in the south polar regions and in an area at about 260-300$^{\circ}$W and 30$^{\circ}$S-50$^{\circ}$N, Fig.~\ref{leadingtrailing-surface}), and some spaxels at the extreme north of the leading hemisphere show an enhanced CO$_2$ band depth (Fig.~\ref{leading-surface}).

\section{Additional figures}
\label{app:O}

\begin{figure*}[ht]
\centering
\includegraphics[width=7cm]{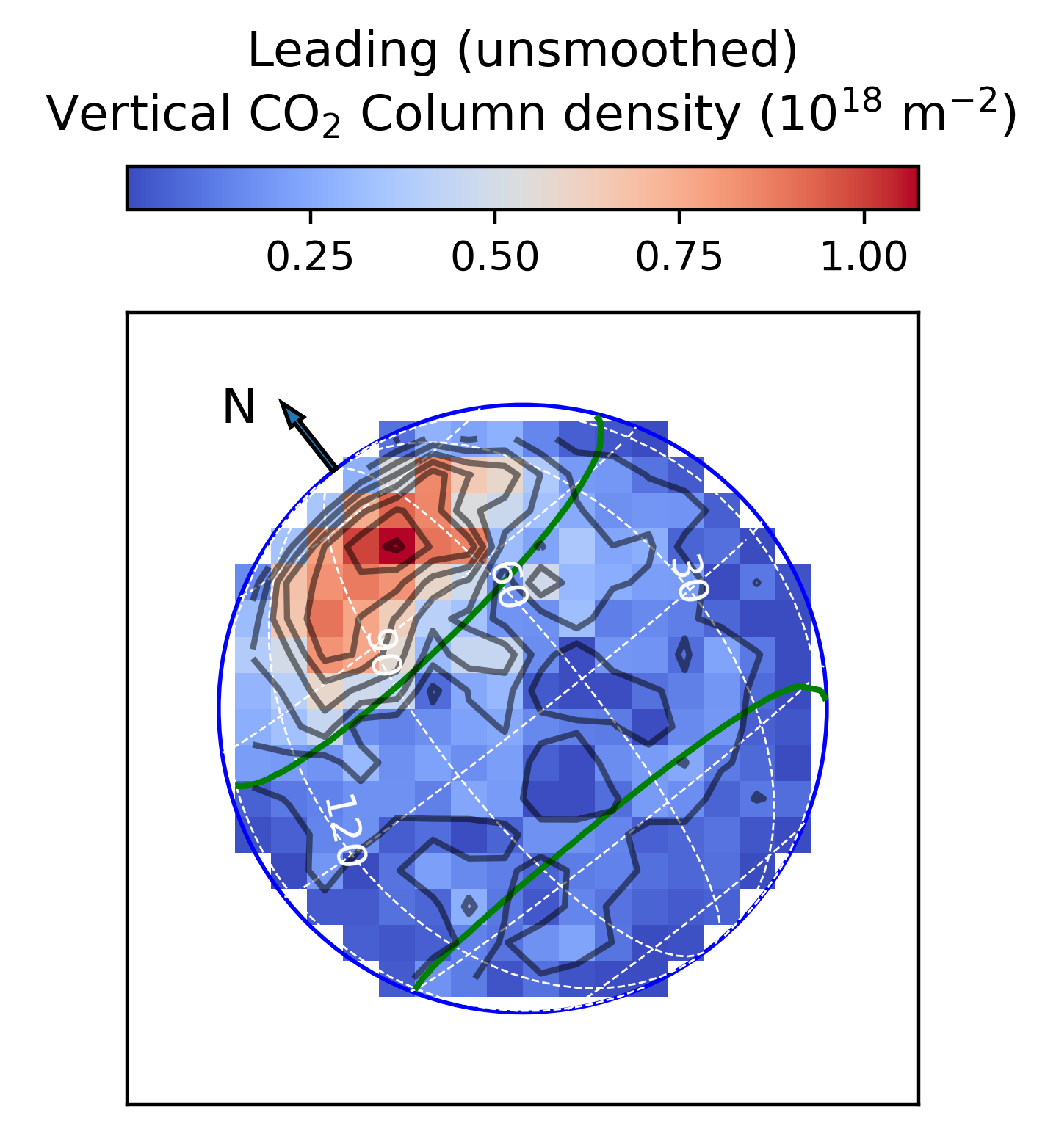}
\includegraphics[width=7cm]{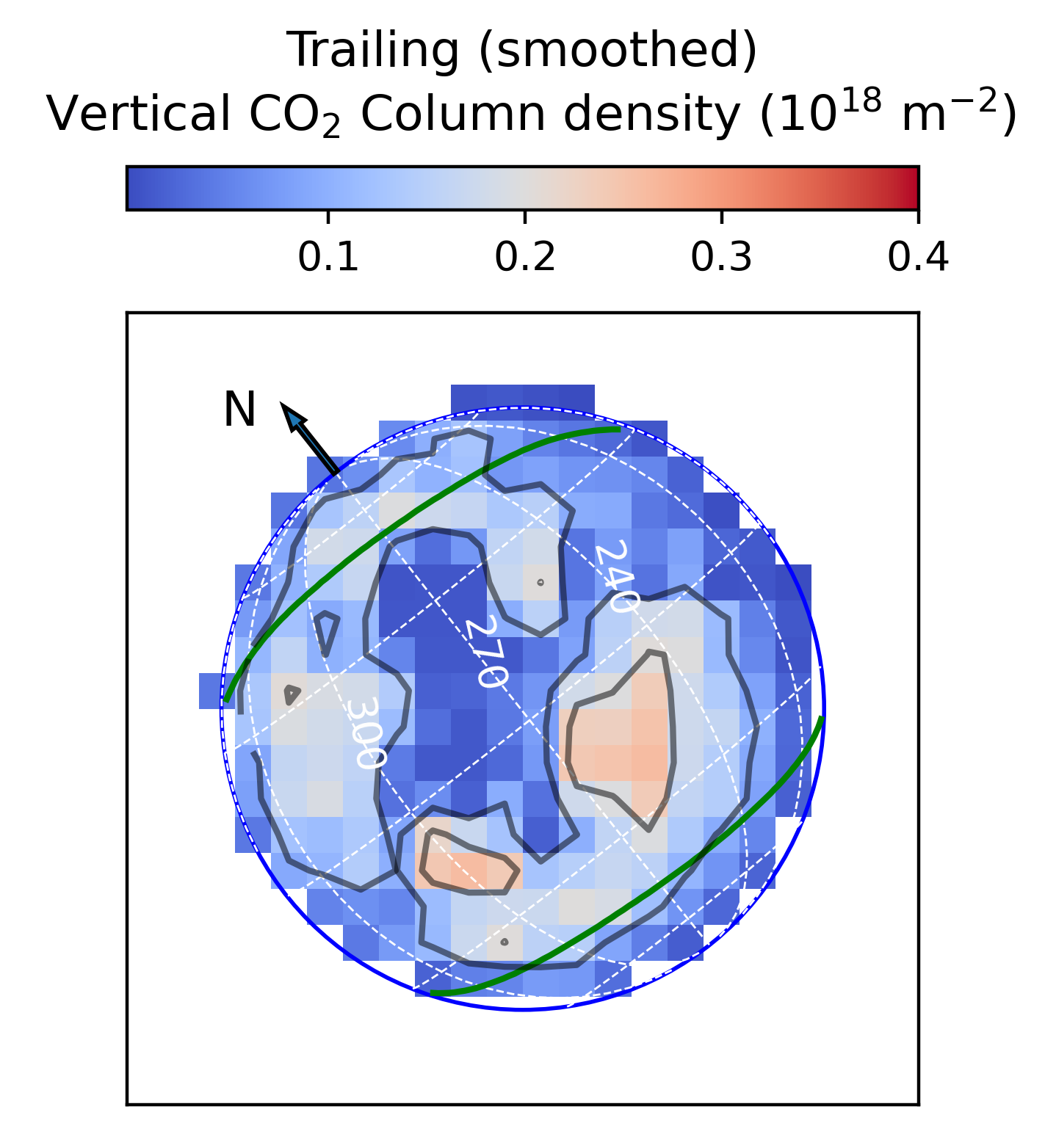}
\caption{CO$_2$ vertical column density maps of the leading (left) and trailing (right) hemispheres. They were deduced from the line-of-sight column density maps, by multiplying them by the cosine of the angle between local zenith and line of sight directions. For the leading hemisphere, the central contour for the north excess is at about 72$^{\circ}$W (12.6 h local time), 45$^{\circ}$N. Trailing data were smoothed using a 3$\times$3 boxcar filter.
\label{fig:vertical-colum-CO2} }
\end{figure*}

\begin{figure*}[ht]
\centering
\includegraphics[width=14cm]{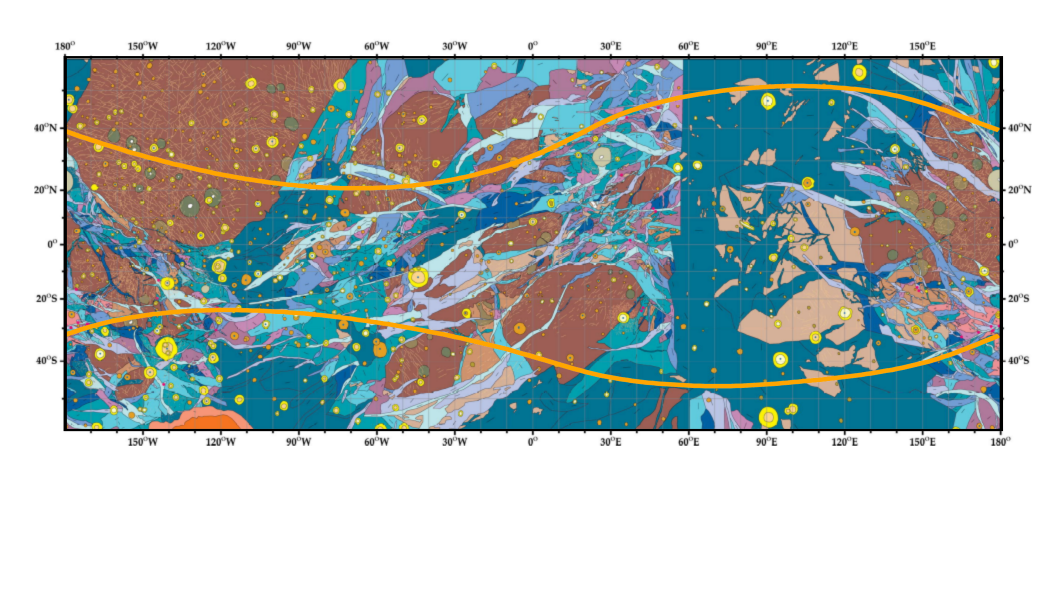}
\caption{Geological map of Ganymede (Plate 2 of \cite{Patterson2010}). The green lines show the open-closed-field line boundary at the time of the JWST observations \cite{Duling2022}. \label{fig:sup-Patterson}  }
\end{figure*}

\end{appendix}
\end{document}